%% file: 2D-Review.tex
\documentclass{ws-rv9x6}
\usepackage{subfigure}     
\usepackage{ws-rv-van}     
\makeindex

\usepackage{graphicx}
\usepackage{amssymb}
\usepackage{amsmath}
\usepackage{color}
\usepackage{comment}
\usepackage{bm}
\usepackage{hyperref}

\newcommand{\vect}[1]{{\mathbf #1}}

\newcommand{\kb}{k_{\rm{B}}}
\newcommand{\as}{a_s}
\newcommand{\eb}{\varepsilon_b}
\newcommand{\ad}{a_{2\rm{D}}}
\newcommand{\add}{a_{\rm{dd}}}
\newcommand{\aad}{a_{\rm{ad}}}
\newcommand{\sad}{s_{\rm{ad}}}
\newcommand{\mad}{m_{\rm{ad}}}

\newcommand{\ecoll}{E_{\rm{coll}}}
\newcommand{\up}{\uparrow}
\newcommand{\down}{\downarrow}
\renewcommand{\k}{{\bf k}}
\newcommand{\p}{{\bf p}}
\newcommand{\q}{{\bf q}}
\newcommand{\0}{{\bf 0}}
\newcommand{\x}{\hat {\bf x}}

\newcommand{\R}{\bf R}
\renewcommand{\r}{{\bf r}}
\newcommand{\ef}{\varepsilon_F}

\newcommand{\kf}{k_F}
\newcommand{\kfup}{k_{F\up}}
\newcommand{\kfdown}{k_{F\down}}

\newcommand{\ek}{\epsilon_{\k}}
\newcommand{\ekup}{\epsilon_{\k\up}}

\newcommand{\epup}{\epsilon_{\k\up}}

\newcommand{\equp}{\epsilon_{\q\up}}

\newcommand{\nn}{\nonumber}
\newcommand{\op}{\omega_z}
\newcommand{\operp}{\omega_\perp}
\newcommand{\T}{{\cal T}}
\newcommand{\F}{{\cal F}}

\newcommand{\bra}[1]{\langle\left.{#1}\right|}
\newcommand{\ket}[1]{\left|{#1}\right.\rangle}
\newcommand{\fs}{\ket{FS}}

\newcommand{\req}[1]{Eq.~\eqref{#1}}
\newcommand{\sch}{{Schr{\"o}dinger }}

\def\lba{\left(}    \def\rba{\right)}
\def\lbc{\left[}    \def\rbc{\right]}
    
\def \del{\partial}    

\mathchardef\mhyphen="2D

\begin{document}

\chapter[Strongly interacting two-dimensional Fermi gases]
{Strongly interacting two-dimensional Fermi gases}
\label{ra_ch1}

\author[J. Levinsen and M. M. Parish]{Jesper
  Levinsen\footnote{\url{jfle@aias.au.dk}}} \address{ Aarhus Institute
  of Advanced Studies, Aarhus University,\\ DK-8000 Aarhus C, Denmark}

\author[J. Levinsen and
M. M. Parish]{Meera~M.~Parish\footnote{\url{meera.parish@ucl.ac.uk}}}
\address{London Centre for Nanotechnology, University College London,\\
  Gordon Street, London, WC1H 0AH, United Kingdom }

\begin{abstract}
  We review the current understanding of the uniform two-dimensional
  (2D) Fermi gas with short-range interactions.  We first outline the
  basics of two-body scattering in 2D, including a discussion of how
  such a 2D system may be realized in practice using an anisotropic
  confining potential.  We then discuss the thermodynamic and
  dynamical properties of 2D Fermi gases, which cold-atom experiments
  have only just begun to explore.  Of particular interest are the
  different pairing regimes as the interparticle attraction is varied;
  the superfluid transition and associated finite-temperature
  phenomenology; few-body properties and their impact on the many-body
  system; the ``Fermi polaron'' problem; and the symmetries underlying
  the collective modes.  Where possible, we include the contributions
  from 2D experiment.  An underlying theme throughout is the effect of
  the quasi-2D geometry, which we view as an added richness to the
  problem rather than an unwanted complication.
\end{abstract}


\body

\input{intro.tex}

\input{2dscat.tex}

\input{fewbody.tex}

\input{manybody.tex}

\input{temp.tex}

\input{polaron.tex}

\input{dynamics.tex}

\input{conclusion.tex}

\section*{Acknowledgements}
We gratefully acknowledge our collaborators on 2D Fermi gases and
related subjects for many illuminating discussions. In particular, we
thank Marianne Bauer, Stefan Baur, Georg Bruun, Nigel Cooper, Tilman
Enss, Andrea Fischer, Peter Littlewood, Francesca Marchetti, Pietro
Massignan, Vudtiwat Ngampruetikorn, Dmitry Petrov, and Gora
Shlyapnikov. Michael K{\"o}hl is thanked for several very useful
discussions on experiments in 2D Fermi gases, and for sharing the data
of Refs.~\refcite{Koschorreck2012,Vogt2012,Koschorreck2013}. We also
wish to thank Stefan Baur, Georg Bruun, Pietro Massignan, and Vudtiwat
Ngampruetikorn for helpful feedback on the manuscript. Stefan Baur and
Vudtiwat Ngampruetikorn are also thanked for help with figures. We
thank Johannes Hofmann for sharing the data of
Ref.~\refcite{Hofmann2012}, Jonas Vlietinck for sharing the data of
Ref.~\refcite{Vlietinck2014}, and Peter Kroi{\ss} for sharing the data
of Ref.~\refcite{Kroiss2014}. This work was supported in part by the
National Science Foundation under Grant No.~PHYS-1066293 and the
hospitality of the Aspen Center for Physics.  MMP acknowledges support
from the EPSRC under Grant No.\ EP/H00369X/2.

\bibliographystyle{ws-rv-van}
\bibliography{review}

\end{document}

%% file: intro.tex
\section{Introduction}\label{sec:intro}

Following the successful realisation of strongly interacting atomic
Fermi gases in three dimensions (3D), attention has now turned to
Fermi systems that have, in principle, even stronger correlations,
such as low-dimensional gases and fermions with long-range dipolar
interactions.  Model two-dimensional (2D) systems are of particular
interest, since they may provide insight into technologically
important, but complex, solid-state systems such as the
high-temperature superconductors~\cite{Norman2011}, semiconductor
interfaces~\cite{SmiM90}, and layered organic
superconductors~\cite{SinM02}. Moreover, 2D gases pose fundamental
questions in their own right, being in the so-called marginal
dimension where particle scattering can be strongly energy dependent,
and quantum fluctuations are large enough to destroy long-range order
at any finite temperature~\cite{MW,Hohenberg}.

In this review, we focus on the uniform 2D Fermi gas with short-range
interactions, since this has already been successfully realised
experimentally~\cite{gunter2005,Martiyanov2010,Frohlich2011,Dyke2011,Feld2011,Sommer2012,Zhang2012,Baur2012,Koschorreck2012,Frohlich2012,Vogt2012,Makhalov2014}.
Here, two species of alkali atom are confined to one or more layers
using a 1D optical lattice or a highly anisotropic trap.  The
interspecies interactions may then be tuned using a Feshbach
resonance, making cold atomic gases ideal for studying the behavior of
fermions in low dimensions.  While the cold-atom system is clearly
much simpler than solid-state systems, where the long-range Coulomb
interactions are difficult to treat and there are often complex
crystal structures, the usual toy models for such systems neglect the
long-range interactions and consider simple contact interactions like
the ones described here in this review.  In particular, the attractive
2D Fermi gas provides a basic model for understanding pairing and
superconductivity in 2D~\cite{Randeria1989,Randeria1990,SR1989}.
Here, by varying the attraction, one can investigate the crossover
from BCS-type pairing to the Bose regime of tightly bound dimers. In
the interests of space, we do not consider further extensions such as
dipolar interactions, spin-orbit coupling, or any lattice within the
plane.  Indeed, we note that a degenerate 2D dipolar Fermi gas has yet
to be achieved experimentally, while the pursuit of the 2D Hubbard
model is still ongoing.

The investigation of strongly interacting 2D Fermi gases, as described
in the following, may be encompassed within several broad themes.
Firstly, there is the interplay between Bose and Fermi behavior as the
attraction is varied. This is particularly apparent at finite
temperature where the normal state evolves from a Fermi to a Bose
liquid, and one has the possibility of the so-called pseudogap regime.
Potentially even richer behavior may be derived from Fermi-Fermi
mixtures with unequal masses and/or imbalanced ``spin''
populations. While attempts to confine mass-imbalanced mixtures to 2D
are still underway, experiments with equal masses have already
realized the regime of extreme spin imbalance~\cite{Koschorreck2012},
corresponding to a single impurity problem.  Here, it has emerged that
even the strongly interacting impurity can be well described by wave
functions that only contain two- and three-body correlations.  A
related theme is the importance of few-body phenomena in the many-body
system.  As well as being relevant to high temperatures, where the
thermodynamic properties are well described by the behavior of
few-body clusters (i.e., the virial expansion), few-body properties
are also required to properly describe the Bose regime of the pairing
crossover. Turning to themes unique to the 2D system, we have the
existence of classical scale invariance and its impact on the
collective modes in a harmonic trap.  Finally, there is the question
of how 2D experiments really are, since in practice there is always a
finite transverse ``size'' of the quasi-2D geometry.  To be in the 2D
limit, we require the length scales associated with the gas (e.g., the
dimer size) to be much larger than the confinement length.
Ultimately, it would be interesting to understand how the gas evolves
from 2D to 3D.

The review is organized as follows: Section~\ref{sec:2d} surveys the
basic properties of two-body scattering in a two-dimensional geometry
--- since the literature offers multiple different definitions in the
2D scattering problem, this may be thought of as a reference section
for the remainder of the review. We also present here an alternative
formulation of the scattering problem in a quasi-2D geometry, and
discuss the issue of confinement induced
resonances. Section~\ref{sec:few} focuses on recent advances in the
understanding of few-body physics. We discuss elastic scattering
properties, as well as the bound trimer and tetramer states that are
predicted to occur in the heteronuclear Fermi gas, for a sufficiently
large mass imbalance. Turning to the many-body physics in a 2D Fermi
gas, Sec.~\ref{sec:bcsbec} reviews the properties of the BCS-BEC
crossover, including the mean-field approach and the equation of state
at zero temperature.  Section~\ref{sec:temp} considers the behavior of
the gas at finite temperature, which includes an outline of the
high-temperature virial expansion, a sketch of the phase diagram for
superfluidity, and a discussion of the existence of the pseudogap.
Section~\ref{sec:polarized} discusses the recent experimental and
theoretical advances in the 2D Fermi polaron problem, with both
metastable states and the nature of the ground state being considered.
In Sec.~\ref{sec:dynamics}, dynamical quantities such as collective
modes and spin diffusion are reviewed, as well as the breakdown of
classical scale invariance in the interacting quantum system --- the
so-called quantum anomaly.  Finally, Sec.~\ref{sec:outlook} provides
an outlook into future investigations of strongly interacting 2D Fermi
gases.

%% file: 2dscat.tex
\section{Basics of the two-dimensional system}\label{sec:2d}

\subsection{General properties of scattering in two dimensions \label{sub:scat}} 

We now summarize several properties of two-body scattering in two
dimensions that are relevant to the results presented in this
review. In the following discussion, we mostly follow
Refs.~\refcite{LL,Adhikari1986}.  The starting point is the 2D \sch
equation for two particles interacting via a short-range local
potential $V(\r)$ at energy $E$ in the center-of-mass frame:
\begin{align}
  -\frac{\hbar^2\nabla^2}{2m_r}\psi(\r)+V(\r)\psi(\r)=E\psi(\r).
\end{align}
Here, the reduced mass is defined in terms of the masses of particle 1
and 2 as $m_r=m_1m_2/(m_1+m_2)$, $\r=(m_1\r_1-m_2\r_2)/m_r$ is the
relative coordinate, and $\nabla$ is the 2D gradient. We further
assume that the potential only depends on $r \equiv |\r|$;
then the \sch equation is separable, the wavefunction may be written
as $\psi(\r)=R(r)T(\theta)$, and the equation for the radial part takes the
form
\begin{align}
  -\frac{\hbar^2}{2m_r}\frac
  1r\frac{d}{dr}\left(r\frac{dR}{dr}\right)
+  \frac{\hbar^2\ell^2}{2m_rr^2}R
+V(r)R=ER.
\end{align}
The quantum number $\ell$ is determined from the azimuthal equation
$d^2T/d\theta^2=-\ell^2T$ and corresponds to the angular momentum in
the plane. In order for the wavefunction to be single valued we must
have $T_\ell(\theta)\propto e^{i\ell\theta}$ with $\ell$ integer. Thus
we have one $s$-wave component ($\ell=0$) but two of all higher
partial wave components ($p$, $d$, etc.\ corresponding to
$\ell=\pm1,\pm2$, etc.). This may be thought of as clockwise and
anti-clockwise rotation and should be compared with the degeneracy
factor $2\ell+1$ in 3D\cite{LL}.

In the asymptotic limit, we write the wavefunction as a sum of an
incident plane wave along the $\x$ direction and an outgoing circular
wave
\begin{align}
\psi(\r)\underset{r\to\infty}{\rightarrow}
e^{ikx}-\sqrt{\frac{i}{8\pi k r}}f(\k)e^{ikr},
\end{align}
with the incident relative wavenumber $k$ defined by
$E=\hbar^2k^2/2m_r$. The vector $\k\equiv k\hat \r$ is defined in the
direction of the scattered wave at an angle $\theta$ with respect to
the incident wave. The dimensionless scattering amplitude $f(\k)$ may
then be expanded in the partial waves as
\begin{align}
f(\k)=\sum_{\ell=0}^\infty(2-\delta_{\ell0})\cos(\ell\theta)f_\ell(k),
\end{align}
where the Kronecker delta takes account of the degeneracy within the partial
wave.

The scattering amplitude gives access to the differential elastic
cross section $\frac{d\sigma}{d\theta}=\frac{|f(\k)|^2}{8\pi k}$, and
to both the total and elastic cross sections:
\begin{align}
\sigma^{\rm{tot}}_\ell(E)&=-\frac1k\mbox{Im}[f_\ell(k)](2-\delta_{\ell0}),\\
\sigma^{\rm{el}}_\ell(E)&=\frac{\left|f_\ell(k)\right|^2}{4k}(2-\delta_{\ell0}),
\end{align}
where the first equation corresponds to the well-known optical theorem. For both
cross sections we use the partial wave expansion
$\sigma(E)=\sum_{\ell=0}^\infty\sigma_\ell(E)$, noting that the
partial waves decouple in the cross section.  The inelastic cross
section simply follows as
$\sigma^{\rm{inel}}(E)=\sigma^{\rm{tot}}(E)-\sigma^{\rm{el}}(E)$. Note
that in 2D the cross section has dimensions of length.

The scattering amplitude may be related to the phase shift experienced
by the scatterers at distances outside the range of the potential:
\begin{align}
f_\ell(k)=\frac{-4}{\cot\delta_\ell(k)-i}.
\label{eq:fell}
\end{align}
The phase shifts are real for elastic scattering and have the low
energy behavior (see, e.g., Ref.~\refcite{Randeria1990})
\begin{align}
&\cot\delta_s(k)=-\frac2\pi\ln(1/ka)+{\cal O}(k^2),
\\
&k^2\cot\delta_p(k)=-s^{-1} +{\cal O}(k^2\ln k),
\end{align}
where we denote the phase shifts $\delta_s\equiv \delta_0$,
$\delta_p\equiv\delta_1$, etc. Here, $a>0$ is a 2D scattering length,
while $s$ is a 2D scattering surface (of unit length
squared). Interestingly, we see that $\cot\delta_s$ diverges
logarithmically at low energies, and thus the definition of the
scattering length is ambiguous (indeed several conventions are used in
the literature). The logarithmic divergence means that the scattering
amplitude goes to zero at zero collision energy; this is manifestly
different from the 3D behavior, where the scattering amplitude at zero
energy equals minus the scattering length.  While the $p$-wave
amplitude also goes to zero in this limit, we see that it does so much
faster than $f_s$. Indeed, while the $s$-wave cross section is seen to
diverge at zero energy, the $p$-wave cross section $\sigma_p\to0$ in
this limit. The low-energy behavior has important consequences in both
few- and many-body physics of the 2D gas with short-range
interactions.

\subsection{Scattering with a short-range potential
\label{sub:short}
}

We now specialize to the typical interactions occuring in the
two-component Fermi gas in 2D. We use a spin notation for the two
components, $\sigma=\up,\down$; the spin indices may denote different
hyperfine states of the same atom or, in the case of a heteronuclear
mixture, single hyperfine states of two different atomic species. The
atomic interaction is characterized by a van der Waals range $R_e$
much shorter than both the average interparticle spacing and the
thermal wavelength. Thus we may consider the interaction to be
effectively a contact, $s$-wave interaction, and model the two-body
problem with the following Hamiltonian
\begin{align}
\label{eq:H}
{\cal H} = & \sum_{\k} \frac{\hbar^2 k^2}{2m_r} \ket{\k} \bra{\k} +
\frac{1}{A} \sum_{\k,\k'} g(\k,\k') \ket{\k} \bra{\k'}.
\end{align}
Here, $A$ is the system area and in the following we set $A = \hbar =
1$. The attractive contact interaction $g(\k,\k')\equiv \bra{\k}\hat
g\ket{\k'}$ has strength $g<0$ and is taken constant up to a large
ultraviolet cutoff $\Lambda\sim1/R_e$. The reduced mass in this
two-component system is $m_r=m_\up m_\down/(m_\up+m_\down)$. As we are
considering low-energy $s$-wave scattering, interactions between the
same species of fermion are suppressed by Pauli exclusion.

\begin{figure}
\centering
\includegraphics[width=0.6\linewidth]{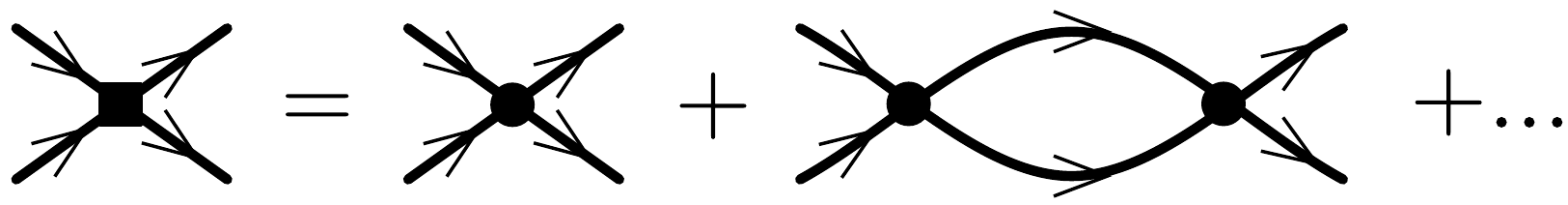}
\caption{The sum of all possible repeated scattering processes of two atoms,
  resulting in the $T$ matrix (black square). The circles represent
  the interaction $\hat g$.
  \label{fig:T}}
\end{figure}

The interaction between two atoms is conveniently described in terms
of a $T$ matrix, illustrated in Fig.~\ref{fig:T}, which describes the
sum of repeated scattering processes between two atoms. In the center
of mass frame, with incoming (outgoing) momenta of $\pm \k_i$ ($\pm
\k_f$), the $T$ matrix takes the form
\begin{align}
  \bra{\k_f}\hat T(E+i0)\ket{\k_i}=\bra{\k_f}\hat g+\hat
  g\frac1{E-\hat H_0+i0}\hat g +\dots\ket{\k_i} =
  \frac1{g^{-1}-\Pi(E)},
\end{align}
where the notation $+i0$ indicates an infinitesimal positive imaginary
part.  Here $\hat H_0$ is the non-interacting part of the
Hamiltonian. The one loop polarization bubble takes the form
\begin{align}
\Pi(E)=\sum_\q^\Lambda\bra{\q}\frac1{E-\hat H_0+i0}\ket{\q}=
\sum_\q^\Lambda \frac1{E-q^2/2m_r+i0}.
\end{align}

Considering scattering at negative energies, it is immediately clear
that the attractive contact interaction in 2D always admits a bound
diatomic molecule (dimer) state in contrast to the 3D case. The energy
of the bound state, $-\eb$ (we define $\eb$ positive), is determined
through the pole of the $T$ matrix, {\em i.e.}
\begin{align}
\frac1g=\Pi(-\eb).
\label{eq:renorm}
\end{align}
This relation acts to renormalize the interaction: the integral
logarithmically diverges at fixed $\eb$ if we take $\Lambda\to\infty$,
however, the physics beyond the two-body problem becomes independent
of $\Lambda$ once Eq.~(\ref{eq:renorm}) is used to replace $g$ with
the binding energy. Thus we arrive at the renormalized $T$ matrix
\begin{align}
T(E)\equiv\bra{\k_f}\hat T(E+i0)\ket{\k_i} = \frac1{\Pi(-\eb)-\Pi(E)}
=\frac{2\pi}{m_r}\frac{1}{\ln(\eb/E) +i\pi}.
\label{eq:2dT}
\end{align}
As the $T$ matrix does not depend on incoming momenta in the center of
mass frame, we will simply denote it $T(E)$.

The on-shell scattering of two atoms at momenta $\pm \k_i$ into
momenta $\pm \k_f$ with $k = |\k_i|=|\k_f|$ yields the scattering
amplitude through the relation $f(\k)=2m_r\bra{\k_i}\hat
T(k^2/2m_r)\ket{\k_f}$. Then, using Eq.~(\ref{eq:fell}), we find that
the two-body phase shift with this contact interaction takes the form
$\cot\delta_s(k)=-\frac2\pi\ln(1/k\ad)$, which defines the 2D
atom-atom scattering length $\ad$.  \footnote{In the literature, the
  alternative definition $2e^{-\gamma}\ad$ of the 2D scattering length
  is often employed, with $\gamma$ the Euler gamma constant. This
  definition arises naturally when considering scattering from a hard
  disc of radius $a_c$, in which case $\ad=(e^\gamma/2)a_c$.}
The relation between the binding energy and the 2D scattering length
is then simply
\begin{align}
\eb=\frac1{2m_r\ad^2}.
\end{align}

\subsection{Quasi-two-dimensional Fermi gases}\label{sub:q2d}

Under realistic experimental conditions, the extent of the gas
perpendicular to the plane is necessarily finite.  The
quasi-two-dimensional (quasi-2D) regime occurs when the confinement
width is much smaller than both the interparticle spacing and the
thermal wavelength, such that transverse degrees of freedom are frozen
out. However, the length scale associated with the confinement to the
quasi-2D geometry is necessarily much larger than the range of the van
der Waals type interactions, and thus at short distances the two-body
interactions are unaffected by the confinement. The relationship
between the 2D scattering theory detailed above, and the realistic
interatomic potential was considered in detail in
Ref.~\refcite{petrov2001}. Here we present an alternative derivation
of the quasi-2D scattering amplitude, and arrive at a form which is
closer to that in Ref.~\refcite{Bloch:2008vn}.

We thus consider the experimentally relevant harmonic confinement
$V_\sigma(z)=\frac12m_\sigma \omega_z^2z^2$ acting in the direction
perpendicular to the 2D plane. While for the heteronuclear gas the
confining frequency $\omega_z$ is not necessarily the same for both
species, this choice in general allows a separation of the center of
mass from the relative motion and provides a major simplification of
the formalism. In relative coordinates, the non-interacting two-body
problem in the $z$ direction reduces to the harmonic oscillator
equation
\begin{align}
  \lba - \frac{1}{2 m_r} \frac{d^2}{dz^2} + \frac{1}{2} m_r \op^2 z^2
  \rba \phi_n(z) = \lba n + \frac{1}{2} \rba \op \phi_n(z).
\end{align}
Here, the motion along the $z$ direction is clearly quantized, with a
constant spacing $\op$ between energy levels.  The non-interacting
part of the quasi-2D Hamiltonian is thus
\begin{align}
  \hat{H}_0 = \sum_{\k n} \lbc \frac{k^2}{2m_r} + \lba n +
  \frac{1}{2} \rba \op \rbc \ket{\k n}\bra{\k n},
\end{align}
where $n$ is the harmonic oscillator quantum number for the $z$
direction. The gas is considered to be kinematically 2D if motion is
restricted to the $n=0$ level.

To investigate two-body scattering in the quasi-2D geometry, we need
to consider the bare interaction in three-dimensional space.  For
convenience, {\em in this section only}, we consider a separable 3D
interaction of the form
\begin{align}
  g(\k_{\rm 3D},\k'_{\rm 3D})= \bra{\k_{\rm 3D}}\hat g\ket{\k'_{\rm
      3D}}\equiv g e^{-(k^2+k'^2+k_z^2+k'^2_z)/\Lambda^2}.
\label{eq:intq2d}
\end{align}
where $k_z$ is the $z$-component of the 3D momentum and $k$ is the
magnitude of the inplane momentum $\k$ as above.  Letting the incoming
(outgoing) atoms have momenta $\pm\k_i$ ($\pm\k_f$) in the plane and
relative motion in the harmonic potential described by the index $n_i$
($n_f$), the matrix elements of the 3D interaction in the quasi-2D
basis are
\begin{align} \notag \bra{\k_fn_f}\hat g\ket{\k_in_i}&=\sum_{\q_{\rm
      3D}\q'_{\rm 3D}}\langle\k_fn_f|\q_{\rm 3D}\rangle\bra{\q_{\rm
      3D}}\hat g \ket{\q'_{\rm 3D}}\langle\q'_{\rm 3D}|\k_in_i\rangle
  \\ & =gf_{n_f}f_{n_i}e^{-(k_i^2+k_f^2)/\Lambda^2},
\end{align}
where $f_n\equiv \sum_{q_z} \tilde{\phi}_n(q_z)e^{-q_z^2/\Lambda^2}$
and $\tilde{\phi}_n(q_z)$ is the Fourier transform\footnote{ The
  harmonic oscillator wave function is
\begin{align*}  \phi_n(z) = \sqrt{\frac{1}{2^n n!}} \lba \frac{m_r\op}{\pi}
  \rba^{\frac{1}{4}} \exp\lba-\frac{m_r\op z^2}{2}\rba H_n\lba
  \sqrt{m_r\op} \ z\rba,
\end{align*}
where $H_n(x)$ are the Hermite polynomials.  $\phi_n(z)$ also happens
to be an eigenfunction of the Fourier transform, so in momentum space
it is simply
\begin{align*}
  \tilde{\phi}_n(k_z) = (-i)^n \sqrt{\frac{2}{2^n n!}} \lba
  \frac{\pi}{m_r\op} \rba^{\frac{1}{4}} \exp\lba-\frac{
    k_z^2}{2m_r\op}\rba H_n\lba \sqrt{\frac{1}{m_r\op}}
  k_z\rba.
\end{align*}} of the harmonic oscillator wave function. For the $f$
coefficients, we then find
\begin{align} \label{eq:funcn}
  f_{2n}=(-1)^n\frac1{(2\pi l_z^2)^{1/4}}\frac{\sqrt{(2n)!}}{2^n n!}
  \frac1{\sqrt{1+\lambda}}\left(\frac{1-\lambda}{1+\lambda}
  \right)^n,
\end{align}
and $f_{2n+1}=0$. Here $l_z\equiv 1/\sqrt{2m_r\omega_z}$ is the
harmonic oscillator length.\footnote{For equal masses, $l_z$ reduces
  to the usual harmonic oscillator length for the motion of the
  individual atoms. For a general mass ratio it differs by a factor
  $\sqrt2$ from the usual definition of the harmonic oscillator length
  of the relative motion.} $\lambda\equiv 1/(\Lambda l_z)^2$ is the
(squared) ratio between the length scale of the short distance physics
and the harmonic oscillator length, and is very small in typical
experiments. Indeed, our approach of using a 3D interaction would be
invalid if this were not the case.

We then evaluate the $T$ matrix in a manner similar to the 2D case
above:
\begin{align}
  \bra{\k_fn_f}\hat T(E+i0)\ket{\k_in_i}=& \bra{\k_fn_f}\hat g +\hat g
  \frac1{E-\hat H_0+i0}\hat g +\dots\ket{\k_in_i}
  \nn \\
  =&e^{-(k_i^2+k_f^2)/\Lambda^2}f_{n_i}f_{n_f}\frac1{g^{-1}-\Pi_{\mbox{\tiny
        Q2D}}(E)}.
\end{align}
The quasi-2D polarization bubble takes the form
\begin{align}
\Pi_{\mbox{\tiny Q2D}}(E)
=\sum_{\q,n}|f_n|^2\frac{e^{-2q^2/\Lambda^2}}
{E-(n+1/2)\omega_z-q^2/2m_r+i0}.
\end{align}
The sum over $n$ may be evaluated by changing variables to
$u=-2\lambda\frac{q^2/2m_r}{E-(n+1/2)\op}$ and using\footnote{While
  formally this approach is valid only for $-1/4\leq x<1/4$, by the
  analytic continuation $E\to E+i0$ the result can be extended to all
  energies.}
\begin{align}
 \sum_{n=0}^\infty \frac{(2 n)!}{(n!)^2}  x^n = \frac{1}{\sqrt{1-4x}}.
\end{align}
We then find
\begin{align}
\Pi_{\mbox{\tiny Q2D}}(E)=-\frac{m_r}{(2\pi)^{3/2}l_z}\int_0^\infty
\frac{du}{u+2\lambda}\frac{e^{-(-E/\op+1/2)u}}{\sqrt{(1+\lambda)^2-
(1-\lambda)^2e^{-2u}}}.
\end{align}

Finally, we relate this result back to the 3D physics: The interaction
(\ref{eq:intq2d}) is renormalized using the relationship\footnote{The
  3D scattering length is related to the $T$ matrix at vanishing
  energy by $\as=(m_r/2\pi)\langle
  \0|T(0)|\0\rangle=(m_r/2\pi)/(g^{-1}+m_r\Lambda/(2\pi)^{3/2})$.}
between the $T$ matrix at vanishing energy and the 3D scattering
length, $\as$. Thus we arrive at the $T$ matrix
\begin{align}
  \bra{\k_fn_f}\hat T(E+i0)\ket{\k_in_i}=&
  e^{-(k_i^2+k_f^2)/\Lambda^2}f_{n_i}f_{n_f}\frac{2\pi l_z}{m_r}
  \frac1{\frac{l_z}{\as}-{\cal F_\lambda}(-E/\op+1/2)},
\label{eq:Tq2d}
\end{align}
with
\begin{align} {\cal F_\lambda}(x)=\int_0^\infty
  \frac{du}{\sqrt{4\pi(u+2\lambda)^3}}\left[1-\frac{e^{-xu}}{\sqrt{[(1+\lambda)^2-
        (1-\lambda)^2e^{-2u}]/(2u+4\lambda)}}\right].
\label{eq:Fcali}
\end{align}
This expression reduces to that of Ref.~\refcite{Bloch:2008vn} in the
limit $\lambda\to0$. In this case, the $T$ matrix only depends on $E$
and the quantum numbers in the harmonic potential, and we write
\begin{align} \label{q2D-T}
  \bra{\k_fn_f}
\hat T(E+i0)\ket{\k_in_i} \equiv \sqrt{2\pi}
  l_zf_{n_i}f_{n_f}\T(E),
\end{align}
where $\T(E)\equiv \frac{\sqrt{2\pi} }{m_r}
\frac1{\frac{l_z}{\as}-{\cal F}_0(-E/\op+1/2)}$ contains the entire
energy dependence.

\subsubsection{Low-energy scattering}
At energies close to the scattering threshold, the function ${\cal F}$
may be expanded. Specializing to the case $\lambda=0$, this results in
\begin{align}
{\cal F}_0(x)\approx \frac1{\sqrt{2\pi}}\ln(\pi
x/B)+\frac{\ln2}{\sqrt{2\pi}}x-
\frac{\pi^2-12\ln^22}{48\sqrt{2\pi}}x^2
+{\cal O}(x^3), \hspace{5mm} |x|\ll1,
\label{eq:fcalapp}
\end{align}
with $B\approx 0.905$; see
Refs.~\refcite{petrov2001,Bloch:2008vn}. This in turn yields the 2D
scattering length
\begin{align}
\ad=l_z\sqrt{\pi/B}\exp(-\sqrt{\pi/2}\,l_z/\as).
\label{eq:ad}
\end{align}
We emphasize that this result is valid across the 3D resonance, as it
only requires the scattering energy to be negligible compared with
the strength of the confinement. In particular, if $|l_z/\as|\gg1$,
then for a large range of energies in the continuum close to the
threshold, {\em i.e.} for $|E-\op/2|\ll \op$, the scattering amplitude
may simply be approximated by
\begin{align}
f(\k)\approx 2\sqrt{2\pi}\,\as/l_z.
\end{align} 
Thus, in this regime, the two-body interaction is approximately
independent of energy, and the system may be considered scale
invariant. This can have important consequences for the many-body
system. 

\subsubsection{Bound state}
\begin{figure}
\centering
\includegraphics[width=0.7\linewidth]{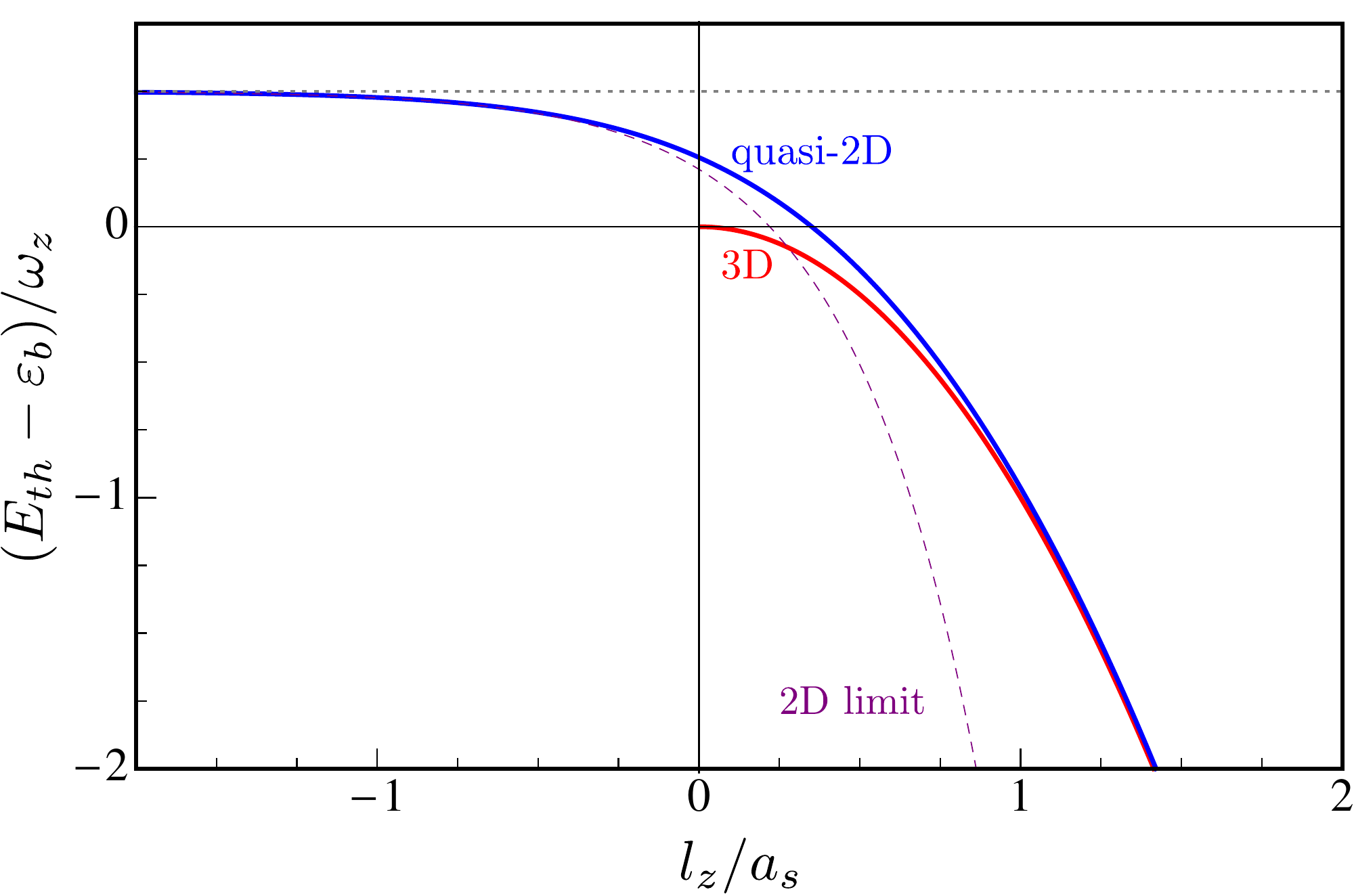}
\caption{The binding energy of the quasi-2D dimer with $\lambda=0$
  (blue, solid). Also shown is the 3D dimer (red, solid), and the 2D
  expression $\eb=\frac1{2m_r\ad^2}$ (dashed). The threshold energy
  $E_{th}$ is 0 in the 3D case, and $\op/2$ in quasi-2D.
  \label{fig:eb}}
\end{figure}

The binding energy of the dimer is the solution of
\begin{align}\label{eq:q2d-bound}
\frac{l_z}{\as}={\cal F_\lambda}(\eb/\op),
\end{align}
where we measure the binding energy from the threshold of free
relative motion of the two atoms. In contrast to the situation in 3D,
where a bound state only exists for $\as>0$, under a harmonic
confinement a bound state exists for a zero-range interaction of
arbitrary strength. In this sense, for negative 3D scattering length,
the dimer in the quasi-2D geometry is confinement induced. This may be
viewed as resulting from the increase of the continuum by
$\frac12\op$, as illustrated in Fig.~\ref{fig:eb}.  In the following,
we focus on the case $\lambda = 0$, but similar behavior should hold
for $\lambda \ll 1$.

For small positive scattering length, $l_z/\as\gg 1$, the 3D dimer
with size $\sim \as$ fits well within the confining potential and is
only weakly perturbed by the harmonic confinement, as illustrated in
Fig.~\ref{fig:eb}. As the scattering length is increased, eventually
the dimer energy becomes strongly modified; for instance at the 3D
resonance the binding energy takes the universal
value\cite{petrov2001} $\eb=0.244\op$. On the other hand, in the limit
of a small negative 3D scattering length, $l_z/\as\ll -1$, the dimer
spreads out in the 2D plane and the binding energy follows from the
expansion \req{eq:fcalapp}. Taking only the first term, the result is
seen to match the 2D expression $\eb=\frac1{2m_r\ad^2}$, {\em i.e.}
\begin{align}
\eb\approx
\op\frac{B}\pi\exp(\sqrt{2\pi}l_z/\as), \hspace{5mm} l_z/\as \ll -1
\end{align}
as is also seen in Fig.~\ref{fig:eb}. However, this expression breaks
down when $l_z/\as>-1$. The fact that in general $\eb\neq
\frac{\hbar^2}{2m_r\ad^2}$, should not come as a surprise. It simply
follows from the introduction of an extra length scale, $l_z$, into
the problem, and is analogous to the problem of a narrow Feshbach
resonance in the 3D gas~\cite{petrov-fewatom}.

\begin{figure}
\centering
\includegraphics[width=0.49\linewidth]{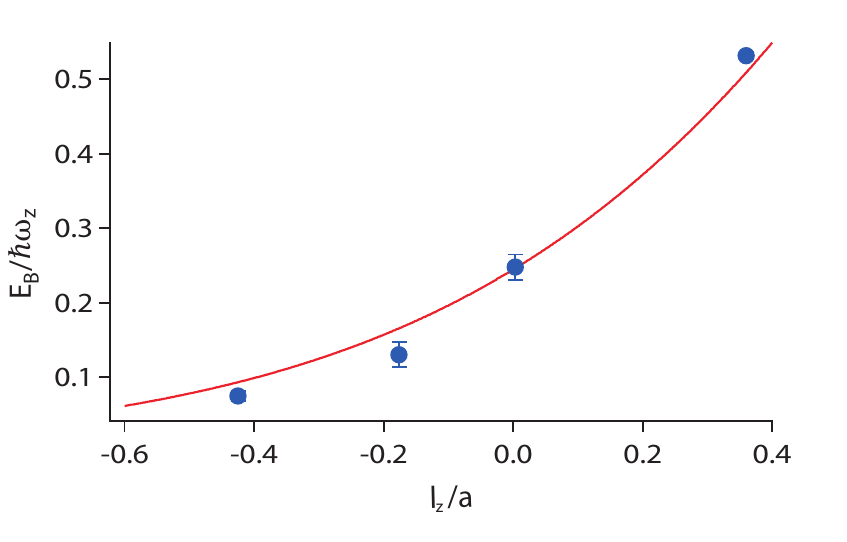}
\includegraphics[width=0.49\linewidth]{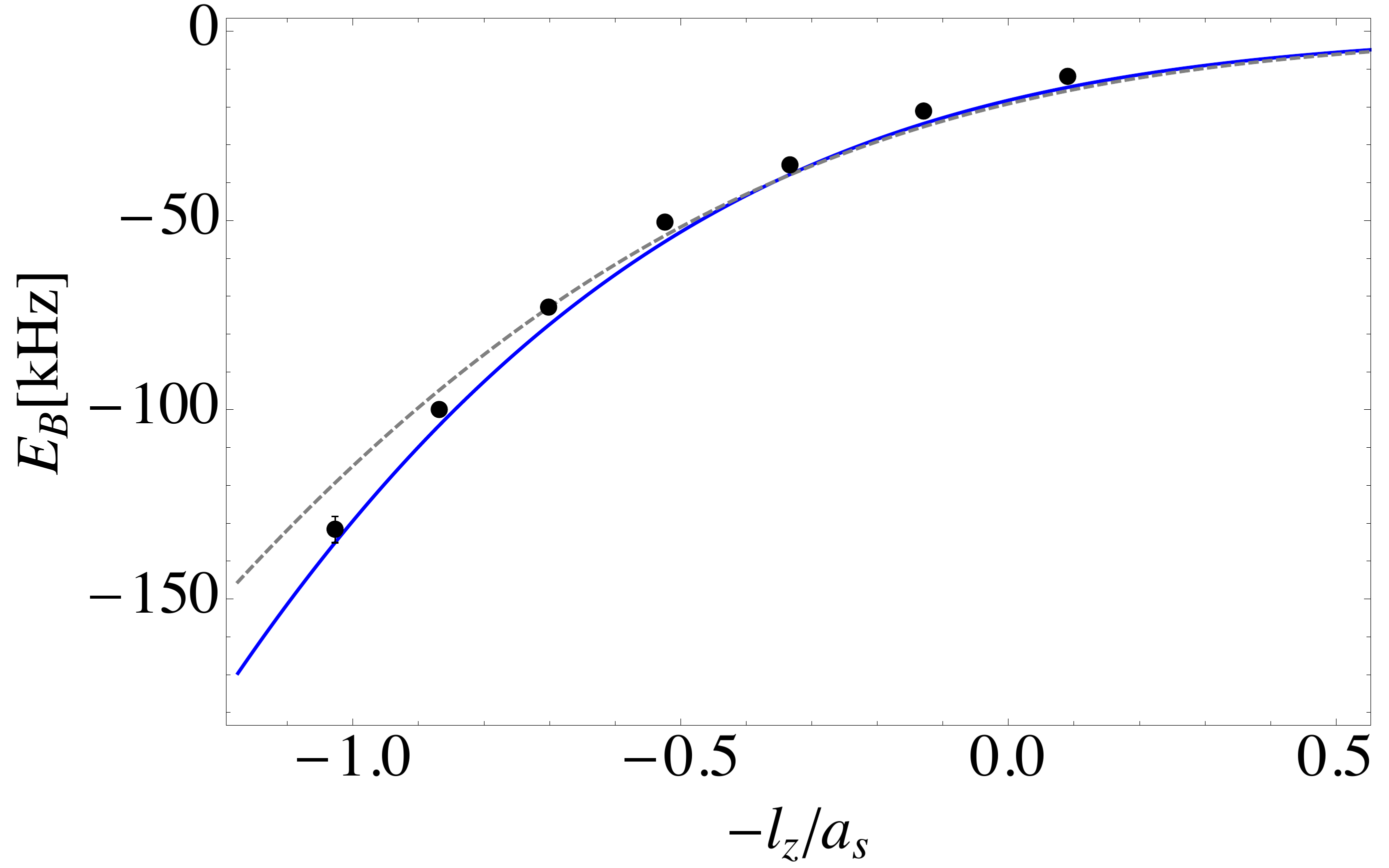}
\caption{(a) Experimental (blue circles) and theoretical (solid line)
  binding energy of a fermion pair in a gas of $^6$Li atoms
  \cite{Sommer2012}. (b) Binding energy in a $^{40}$K gas\cite{Baur2012} at a
  confinement of $\op=2\pi\times75$kHz. The
  experimental result (black circles) is compared with the theoretical
  prediction at zero effective range (gray, dashed), and finite
  effective range (blue, solid), according to the equation
  for the binding energy modified by the 3D effective range
  $r_{\mbox{\tiny eff}}$, $\frac{l_z}{\as}+\frac{r_{\mbox{\tiny
        eff}}}{2l_z}(\eb/\op-1/2)={\cal F}_0(\eb/\op)$. \newline\hspace{\textwidth}
  \tiny{Reprinted figure in (a) with permission from: A. T. Sommer, L. W. Cheuk, M. J. H. Ku, W. S. Bakr, and M. W. Zwierlein, 
   \textit{Phys.\ Rev.\ Lett.} \textbf{108}, 045302 (2012). Copyright 2012 by the American Physical Society.}
  \label{fig:ebexpt}}
\end{figure}

The dimer binding energy has been measured using radio-frequency (RF)
spectroscopy in experiments on ultracold $^6$Li
(Ref.~\refcite{Sommer2012}) and $^{40}$K (Ref.~\refcite{Baur2012})
atoms subjected to a tight optical confinement. The results are shown
in Fig.~\ref{fig:ebexpt}, and both experiments agree well with
theory\cite{petrov2001,Bloch:2008vn} across the Feshbach resonance.

\subsection{Confinement induced resonances}

One consequence of confining the gas to lower dimensions 
is the appearance of so-called \emph{confinement induced resonances}.
At the simplest level, these refer to any region of resonantly
enhanced two-body scattering resulting from the confinement.  However,
the situation in 2D is slightly more subtle, given that the purely 2D
system (where we effectively have $l_z=0$) already exhibits an
enhancement of the scattering amplitude for energy $E \sim \eb$, as
can be seen from the $T$ matrix $ T(E) =
\frac{2\pi}{m_r}\frac{1}{\ln(\eb/E) +i\pi}$.  Thus, it is important to
make a distinction between enhanced scattering that can arise from 2D
kinematics, and resonances that only result from the finite extent of
the gas in the confined direction.

An example of the latter case is the confinement induced resonance
associated with quasi-1D systems~\cite{Olshanii1998,Bergeman2003}.
Here, a resonance occurs when a virtual bound state (arising from the
excited levels of the transverse confinement) crosses the 1D atom-atom
scattering threshold. This process can be captured with a simplified
two-channel model~\cite{Bergeman2003}
\begin{align} \label{eq:q1d} {\cal H}_{\mbox{\tiny Q1D}} = & \sum_{k}
  \frac{k^2}{2m_r} \ket{k} \bra{k} + \nu \ket{b} \bra{b} + \alpha
  \sum_{k} \lba \ket{k} \bra{b} + \ket{b} \bra{k} \rba,
\end{align}
where $\ket{b}$ corresponds to the virtual ``closed channel'' bound
state associated with the excited states in the harmonic confinement,
$\nu$ is the energy of this state with respect to the continuum
threshold, and $\alpha$ is the coupling between $\ket{b}$ and the
scattering states $\ket{k}$ in 1D.  Here, we neglect the interactions
between the 1D scattering states and we take $\alpha$ and $\nu$ to be
independent parameters.\footnote{In the real quasi-1D system, $\alpha$
  and $\nu$ are not independent, as is apparent from the two-channel
  model in Ref.~\refcite{Bergeman2003}.}  In general, this model leads
to energy-dependent interactions, but for zero-energy scattering
we have an effective 1D contact potential $g_{1\rm{D}} \delta(x)$ with
interaction strength $g_{1\rm{D}} = - \alpha^2/\nu$. Thus, we obtain a
scattering resonance where $g_{1\rm{D}} \to \pm \infty$ when $\nu \to
0^{\mp}$. Note that this 
does not signal the appearance of a two-body bound state like in the
3D case where the scattering length diverges ($1/\as = 0$).  Instead,
Eq.~\eqref{eq:q1d} always yields a two-body bound state with binding
energy $\eb$ satisfying the condition $\sqrt{2}(\nu + \eb)/\alpha^2 =
\sqrt{m_r/\eb}$. Moreover, we see that $\eb$ is finite at the
resonance $\nu=0$, thus illustrating how this resonance is a feature
of confinement that goes beyond the behaviour in a purely 1D system.

On the other hand, such a confinement induced resonance does not exist
in the quasi-2D system. If we consider the two-channel model
\eqref{eq:q1d} in 2D, where we have $\ket{\k}$ instead of $\ket{k}$,
then we obtain the modified $T$ matrix
\begin{align}
T(E) = \frac{2\pi}{m_r} \lbc \ln\lba\frac{1}{2m_r\ad^2 E}\rba 
+ \frac{2\pi E}{m_r\alpha^2} +i\pi \rbc^{-1},
\end{align}
with $\ad = \Lambda^{-1} \exp(\pi\nu/m_r\alpha^2)$.  This is
essentially the quasi-2D $T$ matrix $\T(E)$ expanded up to linear
order in $E/\op$.  Comparing with the terms in the expansion
\eqref{eq:fcalapp} yields $ \op = m_r \alpha^2 \ln(2)/2\pi$.  However,
this modification to the $T$ matrix only shifts the scattering
enhancement away from $E \sim \eb$ (where $T(-\eb)^{-1} = 0$).  The
resonance still remains strongly energy dependent like in the purely
2D case.
However, it can still be characterized experimentally: For a Boltzmann
gas in the 2D limit, the scattering is enhanced for temperature $T
\sim \eb$.  The requirement $T \ll \op$ then implies that $\eb \ll
\op$ and thus the resonance occurs on the attractive side of the 3D
Feshbach resonance, $\as < 0$, according to Fig.~\ref{fig:eb}.


Additional resonances will appear when there are anharmonicities in
the confining potential, as is usually the case in experiments
employing an optical lattice. Any anharmonicity inevitably leads to
coupling between two-body states with different center-of-mass
harmonic quantum numbers $N$. In particular, there will (for instance)
be a coupling between scattering states $\ket{\k}$ with
$N=0$ 
and two-body bound states with $N = 2$, due to the selection rules.
This leads to increased molecule formation when
$\eb$ 
is close to $2 \hbar \omega_z$, which in turn leads to enhanced losses
due to subsequent collisions with other particles. Such an
``inelastic'' confinement induced resonance was recently observed
experimentally in low-dimensional
geometries~\cite{Haller2010,Sala2012,Sala2013}.
In the absence of effective range corrections, the inelastic resonance
in the quasi-2D geometry arising from the above mentioned coupling
occurs when $l_z/\as \simeq 1.2$, in contrast to the resonance derived
from 2D kinematics.

%% file: fewbody.tex
\section{Universal few-body physics in a 2D Fermi gas}\label{sec:few}

Recent years have brought a wealth of experiments exploring few- and
many-body physics in ultracold atomic gases --- see, for instance,
Refs.~\refcite{Bloch:2008vn,petrov-fewatom} and references therein.
In particular, it has enabled the study of \emph{universal} few-body
physics, since the low-energy scattering is insensitive to the details
of the short-range interactions, and this in turn has major
consequences for the many-body system. For instance, in the regime
where a dimer exists, a knowledge of the
dimer-dimer\cite{PetrovPRL2004} and atom-dimer\cite{stm} scattering
lengths is necessary for a complete description of the
balanced~\cite{Levinsen2006} and polarized~\cite{Combescot2010} Fermi
gases.  The properties of few-body inelastic processes furthermore
explains the exceptional stability of 3D Fermi gases close to the
unitary limit~\cite{PetrovPRL2004}. The experimental exploration of 2D
Fermi gases is still ongoing, but few-body physics has already played
an important role in understanding many-body phenomena, as can be seen
in Secs.~\ref{sec:bcsbec} and \ref{sec:temp}.

While experiments have thus far focussed on the equal-mass case,
heteronuclear Fermi-Fermi mixtures with mass imbalance promise to
provide even richer few-body phenomena.  For instance, in 1D it has
been shown that when the mass ratio exceeds one, trimers (bound states
of 1 light atom, 2 heavy atoms) can form~\cite{Malykh2009}. This can
lead to a Luttinger liquid of trimers in the polarized gas, while more
exotic bound states such as tetramers (1,3), pentamers (2,3), etc.,
can also exist at higher mass ratios~\cite{orso2010}. Similarly, in
2D\cite{Pricoupenko2010} and 3D~\cite{trimer}, trimers are predicted
to exist above a mass ratio of 3.3 and 8.2, respectively, which can
lead to a trimer phase in the highly polarized Fermi
gas\cite{Mathy2011}. Additionally, tetramers have been predicted in
1D\cite{orso2010,Mehta2014}, 2D\cite{Levinsen2013}, and
3D\cite{Blume2012}. These bound states all share the property of being
universal, in the sense that their binding energy is a multiple of the
dimer binding energy without the need for additional parameters.  For
a recent review of bound few-body states, we refer the reader to
Ref.~\refcite{Kartavtsev2012}.

Thus far, the only heteronuclear Fermi-Fermi mixture where tunable
short-range interactions have been experimentally
demonstrated\cite{Wille2008,Costa2010,wu2011} is $^{40}$K-$^6$Li with
a mass ratio of 6.64. However, the periodic table offers ample
opportunity for exploring additional mass ratios, while the
possibility to tune the effective mass ratio using optical lattices
also exists. Thus, for the theoretical approach employed in this
section, we may consider the mass ratio to be a free parameter.

Finally, we note that there is a special class of states --- the
well-known Efimov states --- where the 3D physics is manifestly
different from the situation in confined geometries \cite{Efimov}.
Efimov states have been observed experimentally as sharp peaks in the
loss rate in both bosonic~\cite{efimovobservation} and
(three-component) fermionic\cite{Ottenstein2008} systems. In the
heteronuclear fermionic system in 3D, the Efimov effect occurs for
(2,1) trimers above a mass ratio of 13.6 and for (3,1)
tetramers\cite{Castin2010} for mass ratios exceeding 13.4.  On the
other hand, it is known that Efimov's scenario does not occur in the
1D and 2D geometries~\cite{Nishida2011}. Very recently it was shown
that under realistic experimental conditions, if Efimov trimers exist
in 3D, these will impact the few-body physics in a strongly confined
geometry\cite{Levinsen2014}. However, in the following we ignore this
effect and focus either on the idealized 2D scenario and/or on mass
ratios for which the Efimov effect is absent.

\subsection{Equal mass fermions}

\begin{figure}
\centering
\includegraphics[width=0.7\linewidth]{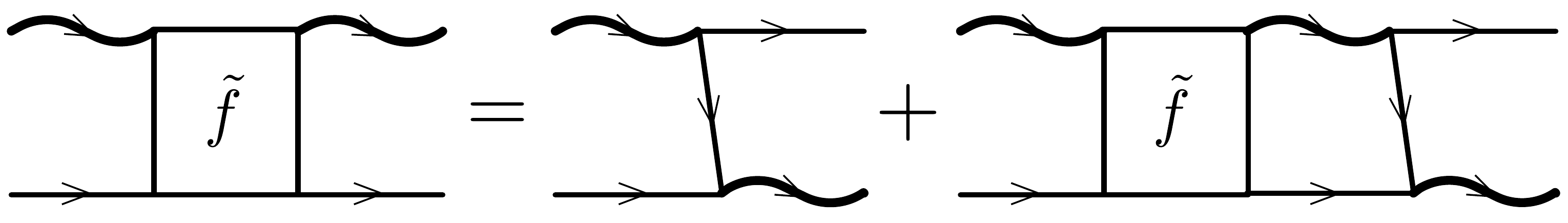}
\caption{Illustration of the Skorniakov--Ter-Martirosian equation
which governs the interaction of an atom (straight line) with a dimer
(wavy line). $\tilde f$ is the atom-dimer scattering amplitude.
\label{fig:stm}}
\end{figure}

We now discuss the three- and four-body problem in a homonuclear
gas. The first of these plays an important role in accurately
determining the energy of an impurity atom immersed in a Fermi sea ---
see Sec.~\ref{sec:polarized}. It is also of practical importance when
considering inelastic processes such as three-body
recombination~\cite{Ngamp2013}, the process whereby three atoms
collide to produce an atom and a dimer. The dimer-dimer scattering
length, on the other hand, is important in describing the many-body
system in the limit of tightly bound pairs, as shown in
Sec.~\ref{sec:bcsbec}.  In the present discussion we confine ourselves
to on-shell scattering properties.

The interaction between a spin-$\up$ atom with an $\up\down$ dimer may
be investigated with the Skorniakov--Ter-Martirosian (STM) equation
introduced in the context of neutron-deuteron scattering~\cite{stm}:
The atom-dimer scattering arises from the repeated exchange between
identical spin-$\up$ atoms of the spin-$\down$ atom, and the STM
integral equation yields the sum of diagrams with any number of such
exchanges as illustrated in Fig.~\ref{fig:stm}. We are interested in
the on-shell scattering amplitude, and thus we let the incoming
[outgoing] atom and dimer have four-momentum $(\k,\ekup)$ and
$(-\k,E-\ekup)$ [$(\p,\epup)$ and $(-\p,E-\epup)$], respectively, with
$E=k^2/2\mad-\eb$ such that the incoming dimer is on shell. Here the
single particle kinetic energy is $\epsilon_{\k,\sigma} =
k^2/2m_\sigma$, where $m_\sigma$ is the mass.  $\mad$ is the
atom-dimer reduced mass $\mad^{-1}=m_\up^{-1}+M^{-1}$, with $M=m_\up
+m_\down$. The on-shell condition $|\k|=|\p|$ is taken at the end of
the calculation. With these definitions, the STM equation takes the
form of an integral equation~\cite{Ngamp2013}
\begin{equation}
\tilde f_\ell(k,p) = - h(k,p)\Big[ g_\ell(k,p)
  - \int  \frac{q\, d q}{2\pi} \,
\frac{g_\ell(p,q)\tilde f_\ell(k,q)}{q^2-k^2-i0}\Big],
\label{eq:stm}
\end{equation}
where we note that the atom-dimer scattering preserves angular
momentum, allowing a decoupling of $\tilde{f}(\k,\p)$ into its partial
wave components.  The scattering amplitude then follows from taking
the on-shell condition $f_\ell(k) = \tilde f_\ell(k,k)$. $g_l$ is the
partial wave projection of the spin-$\down$ propagator and $h$ is
proportional to the two-body $T$ matrix.\footnote{The projection of
  the propagator of the spin-$\down$ fermion onto the $\ell$'th
  partial wave is
\begin{equation}
g_\ell(p,q)=\int_0^{2\pi}
\frac{d\phi}{2\pi}
\frac{\cos(\ell\phi)}{
E-\epup-\equp-\epsilon_{\p+\q\down}+i0},
\end{equation}
with $\phi$ the angle between $\p$ and $\q$. We also define
$h(k,p)\equiv (k^2-p^2)T(E-p^2/2\mad) $ [see \req{eq:2dT}] in order to
separate out the simple pole of the two-particle propagator occuring
at $|\k|=|\p|$.}

\begin{figure}
\centering
\includegraphics[width=\linewidth]{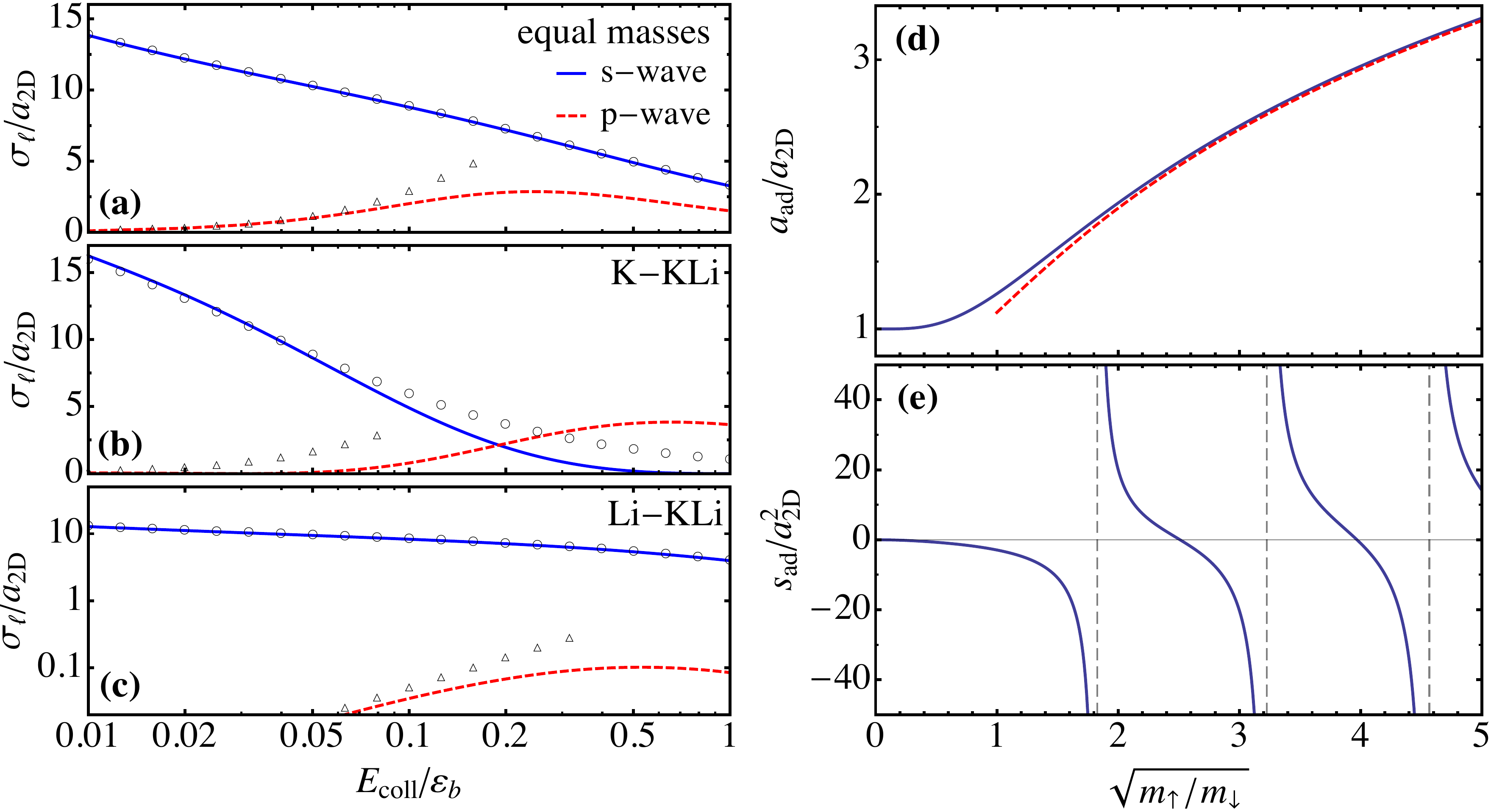}
\caption{
  (a,b,c) $s$- and $p$-wave atom-dimer cross sections as a function of
  collision energy $\ecoll\equiv k^2/2M+k^2/2m_\up$ for equal masses
  and for the K-Li mixture. The circles [triangles] correspond to the
  low-energy expansion, see Sec.~\ref{sub:scat}, using the atom-dimer
  scattering length [area]. Note the log-scale used in (c).  (d)
  Atom-dimer scattering length and (e) area as a function of mass
  ratio. In (d) the dashed line is the asymptotic behavior at large
  mass ratio\cite{Ngamp2013} and in (e) the vertical dashed lines
  indicate the appearance of trimers.  The figure is taken from
  Ref.~\refcite{Ngamp2013}.
  \label{fig:scat}}
\end{figure}

From the resulting scattering amplitude, the low-energy
scattering properties such as the $s$-wave scattering length and the
$p$-wave scattering area (see Sec.~\ref{sub:scat}) may be
extracted. For equal masses, we find:~\cite{Ngamp2013}
\begin{align}
\aad\approx1.26\ad, \hspace{10mm}\sad\approx-2.92\ad^2.
\label{eq:aad}
\end{align}
Thus, the $s$-wave scattering is repulsive at low energies while the
$p$-wave scattering is attractive in this case. The
scattering amplitude furthermore gives access to the partial wave
cross sections, which are shown for $s$- and $p$-waves in
Fig.~\ref{fig:scat}(c). While the $s$-wave cross section is apparently
completely described in terms of the scattering length, interestingly
we note that the $s$- and $p$-wave cross sections are comparable for
collision energies of the order of the binding energy. This is unlike
the 3D case~\cite{Levinsen2009}, and is due to the weaker centrifugal
barrier between identical fermions in 2D. Thus three-body
correlations are likely to be more important in 2D than in 3D.

Both the atom-dimer and the dimer-dimer scattering lengths have been
extracted from a quantum Monte Carlo (QMC) calculation of the
excitation gap in the BEC regime~\cite{Bertaina2011}. The result was
$\aad=1.7(1)\ad$ and $\add=0.55(4)\ad$.  Note that these relations do
not depend on the definition of the scattering length.  The
discrepancy between the exact result in \req{eq:aad} and the QMC
result is likely to be due to the fact that the extraction of $\aad$
from the data relied on an equation of state that was only
logarithmically accurate. On the other hand, the dimer-dimer
scattering length is in perfect agreement with the exact few-body
calculation of Petrov et al.~\cite{Petrov2003}: $\add\approx0.56\ad$.

\subsection{Heteronuclear Fermi gas}

The three-body problem in a heteronuclear 2D Fermi gas with
short-range interparticle interactions was first studied by
Pricoupenko and Pedri~\cite{Pricoupenko2010}. Remarkably, even in the
absence of Efimov physics, they still found that two heavy fermionic
atoms and a light atom can form an ever increasing number of trimers
as the mass ratio is increased. However, at any given mass ratio, the
number of trimers in the spectrum was found to be finite. In
Refs.~\refcite{Bellotti2013,Ngamp2013} it was argued that the
appearance of trimers was due to an effective $1/R$ potential in odd
partial waves between heavy atoms at a separation $R$, mediated by the
light atom.\footnote{As the heavy atoms are identical fermions, the
  wavefunction of the light atom is necessarily antisymmetric
  (symmetric) for scattering in even (odd) partial waves. The
  antisymmetric state suppresses tunneling of the light atom between
  heavy atoms, and as a result the effective potential between the
  heavy atoms is repulsive (attractive) in even (odd) partial
  waves. Consequently, trimers only form in odd partial waves.}
Consequently, at large mass ratios, the spectrum of bound states is
hydrogen-like.

Signatures of trimer formation are clearly seen in the atom-dimer
scattering properties, Fig.~\ref{fig:scat}. Here the $p$-wave
scattering surface diverges at the mass ratios~\cite{Pricoupenko2010}
$m_\up/m_\down=3.33$, $10.41$, etc., when a trimer state crosses the
atom-dimer scattering threshold. Thus, the $p$-wave interaction
becomes resonant at the crossing, and while the K-Li mass ratio of
6.64 is in-between the appearance of bound states in 2D, we still
observe that the $p$-wave cross section in K-KLi scattering dominates
at a collision energy comparable to $\eb$. On the other hand, the
scattering length increases monotonically with mass ratio.

The strong atom-dimer scattering in higher partial waves due to the
proximity of trimers may be investigated using a mixture of heavy
atoms and heavy-light dimers, as in a recent experiment using a K-Li
mixture in 3D~\cite{Jag2014}.  In such a mixture, the energy shift of
an atom due to the interaction with dimers is proportional to the
real part of the atom-dimer scattering amplitude, and may be directly
accessed by radiofrequency spectroscopy.

\begin{figure}
\centering
\includegraphics[width=0.7\linewidth]{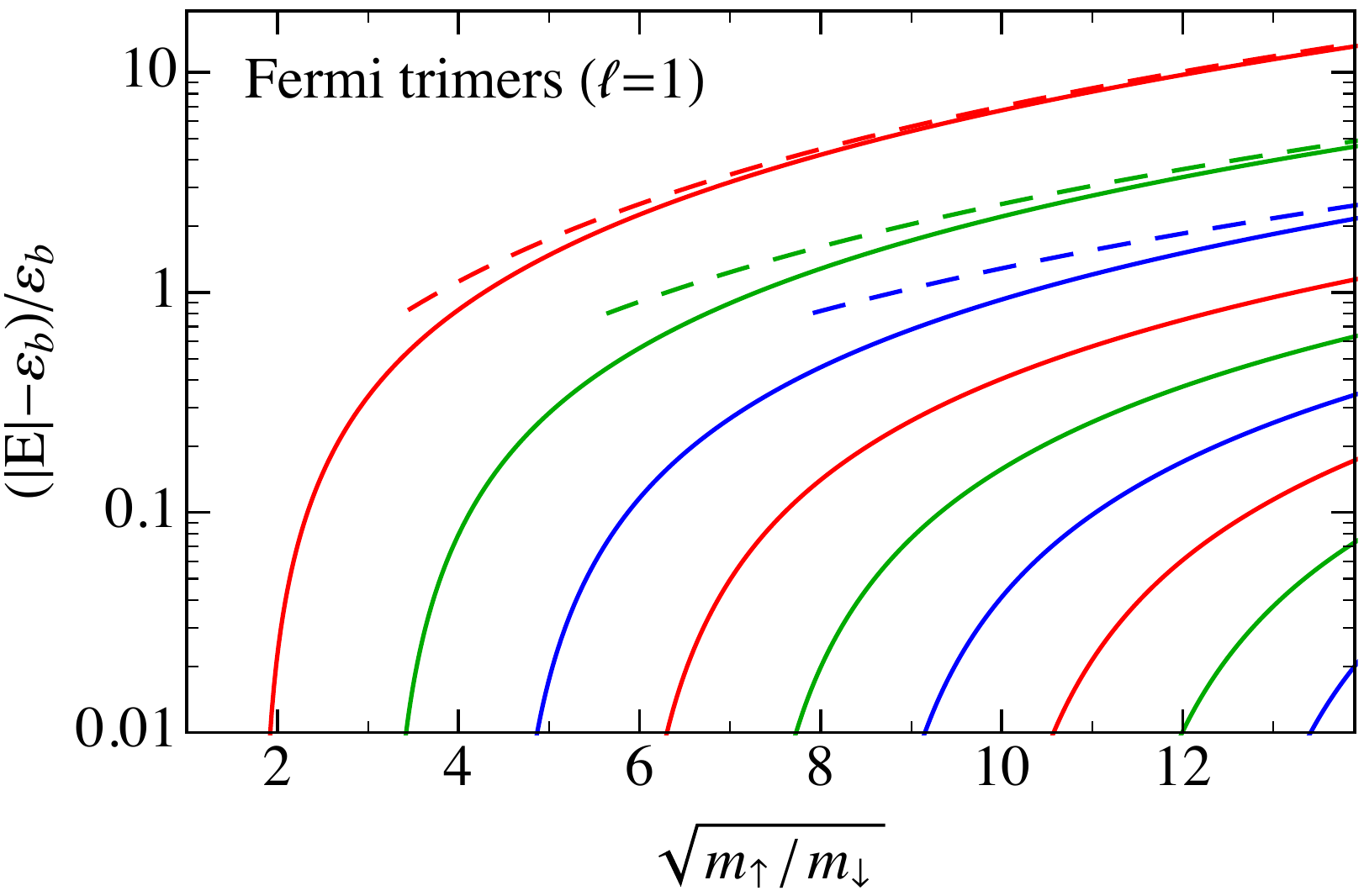}
\caption{The spectrum of $p$-wave trimers\cite{Pricoupenko2010} as a
  function of mass ratio, taken from Ref.~\refcite{Ngamp2013}. The
  dashed lines are the large mass ratio asymptotic hydrogen-like
  energy levels (\ref{eq:hydrogen}).
  \label{fig:spectrum}}
\end{figure}

Turning now to trimers, in Fig.~\ref{fig:spectrum} we show their
spectrum obtained from
\req{eq:stm}\cite{Pricoupenko2010,Ngamp2013}. To shed light on the
appearance of trimers at large mass ratio, it is instructive to turn
to the Born-Oppenheimer approximation (in the following discussion we
follow Ref.~\refcite{Ngamp2013}, but see also
Ref.~\refcite{Bellotti2013}): Assume that the wavefunction of the
light atom at position $\r$ adiabatically adjusts itself to the
positions of the heavy atoms, $\pm\R/2$. Then the wavefunction of the
light atom is
\begin{equation}
\psi_\mathbf{R}(\mathbf r) \propto 
K_0(\kappa_\mp(R)|\mathbf{r-R/2}|) \mp
K_0(\kappa_\mp(R)|\mathbf{r+R/2}|),
\label{eq:wave}
\end{equation}
where the upper (lower) sign describes even (odd) partial wave
scattering. The modified Bessel function of the second kind
$K_0(\kappa_\mp(R) r)$ is the decaying solution of the free
single-particle Schr{\"o}dinger equation with energy $\epsilon_\mp(R)
= -\kappa_\mp(R)^2 / 2m_\downarrow$. The energy of the light atom as a
function of separation of heavy atoms is determined from the
Bethe-Peierls boundary condition in 2D: $\left[\tilde r\psi'({\tilde
    r})/\psi\right]_{\tilde \r\rightarrow\0} = 1/\ln(\tilde r
/(2e^{-\gamma}a_{2\text{D}}))$ where $\tilde\r=\r\pm\R/2$. This leads
to the implicit equation
\begin{align}
\ln\left(-\frac{\epsilon_\mp(R)}{\eb}\right)
=\mp 2 K_0\left( \sqrt{-\frac{\epsilon_\mp(R)}{\eb}
  }\frac{R}{a_{2\text{D}}}\right).
\label{eq:BO}
\end{align}

\begin{figure}
\centering
\includegraphics[width=0.7\linewidth]{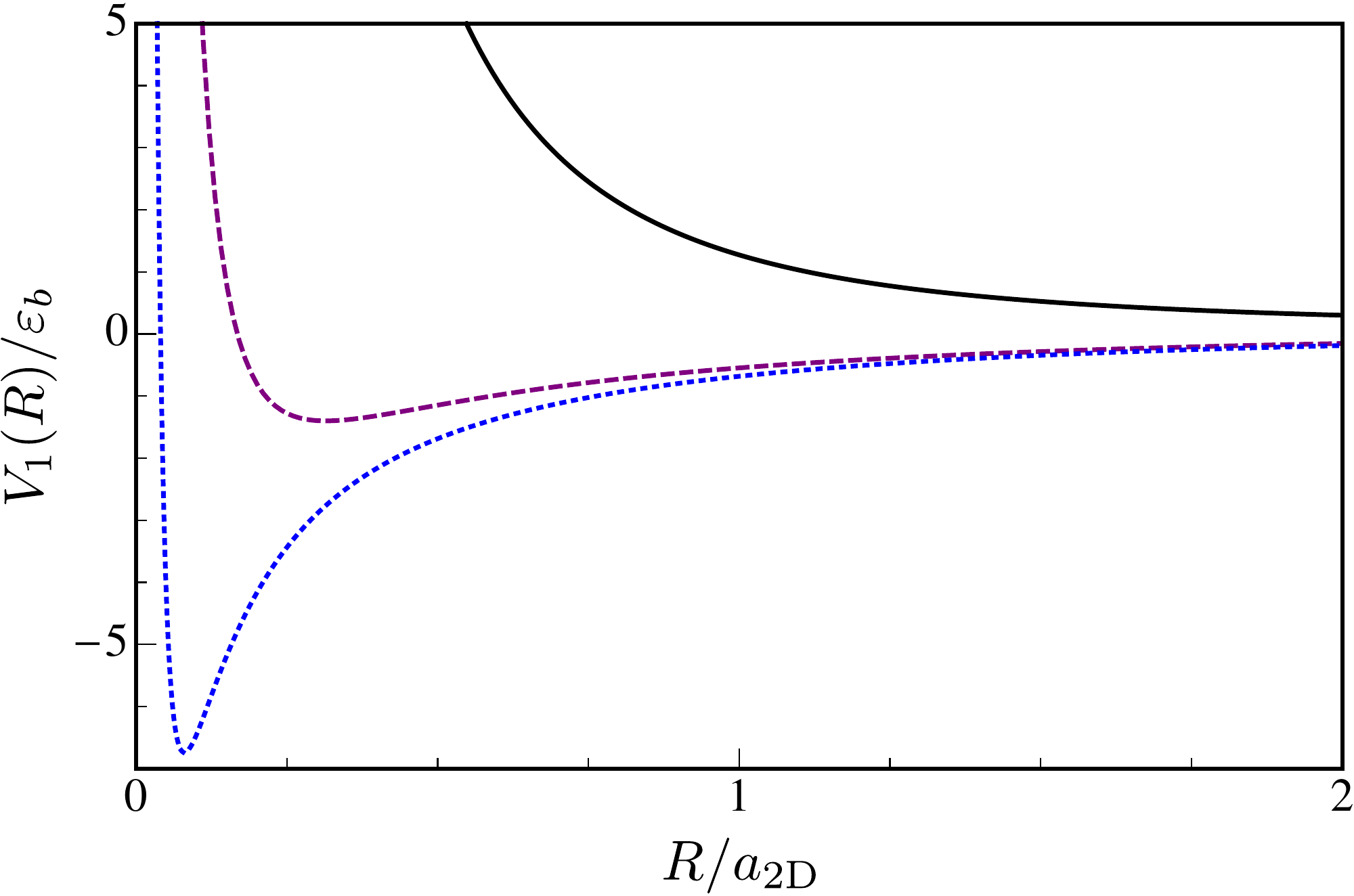}
\caption{Born-Oppenheimer effective $p$-wave potential between two
  heavy fermions mediated by a light atom. Top to bottom are the
  potentials for mass ratios $m_\up/m_\down=1$, 3.33 (critical mass
  ratio for the appearance of trimers), and 6.64 (the
  $^{40}$K-$^{6}$Li mass ratio).
  \label{fig:BO}}
\end{figure}

The energy levels of the light atom act as potential surfaces for the
motion of the heavy atoms. In the case of $p$-wave scattering, the
effective potential $V_p(R)=\epsilon_+(R)-\epsilon(\infty)+1/(m_\up
R^2)$ including the centrifugal barrier is shown in Fig.~\ref{fig:BO}.
The potential is measured from the limiting value of the potential at
large separation, $\epsilon(\infty)$, which reduces to $-\eb$ at large
mass ratios.  We see that when the mass ratio is small, the effective
potential is always repulsive. On the other hand, the potential
develops an attractive well as the mass ratio is increased and it is
in this well that trimers can form. For a large mass ratio, the odd
partial wave potential becomes $\epsilon_+(R) \approx -\frac{2 \eb
}{e^\gamma} \frac{a_{2\text{D}}}{R}$ at short distances, $R\ll
a_{2\text{D}}$. This potential is hydrogen-like and thus the spectrum
of the deepest bound trimers is~\cite{flugge1952} (appropriately
shifted by the dimer binding energy)
\begin{equation}
E_n = -\frac{m_\uparrow}{e^{2\gamma} m_\downarrow}
\frac{\eb}{2(n+1/2)^2} -\eb,
\quad
\label{eq:hydrogen}
\end{equation}
with integer quantum number $n\geq\ell$. In Fig.~\ref{fig:spectrum}
this asymptotic expression is shown to agree well with the exact
results for the deepest bound trimers. The expression
(\ref{eq:hydrogen}) makes it clear that deeply bound trimers in
different partial wave channels are quasi-degenerate\footnote{In fact,
  as the spectrum of even partial wave trimers consisting of two
  identical (non-interacting) heavy bosons and a light atom are
  determined from the same effective potential, these are also
  quasi-degenerate with the fermionic trimers arising in odd partial
  waves.}, as observed in Ref.~\refcite{Pricoupenko2010}. Remarkably,
at very short distances, the $1/R^2$ behavior of the centrifugal
barrier always dominates. This has the important consequence that the
trimers may be expected to be quite long-lived, as they are large (of
size $\ad$) and the constituent atoms do not approach each other
easily. This is completely unlike the 3D case, where the mediated
potential also goes as $1/R^2$ and the Efimov effect occurs at large
mass ratio.

The number of bound states at any given mass ratio may be
approximated\cite{Ngamp2013} by noting that it is proportional to the
number of nodes of the zero-energy wavefunction. The trimers exist in
the hydrogen-like part of the effective potential, $R\lesssim\ad$, and
in this regime the relative wavefunction of the heavy particles is
proportional to the Bessel function
$J_{2\ell}(2\sqrt{e^{-\gamma}(m_\up/m_\down)R/a_{2\text{D}}})$. The
wavefunction acquires an additional node each time the argument
increases by $\pi$, and consequently the number of trimers is
proportional to $\sqrt{m_\up/m_\down}$. This feature is clearly
observed in Figs.~\ref{fig:scat}(b) and \ref{fig:spectrum}. Using a
semi-classical approximation, Ref.~\refcite{Bellotti2013} found that
in the limit of a large mass ratio the number of bound states is
approximately 0.73$\sqrt{m_\up/m_\down}$.

\subsubsection{Bound states of three identical fermions and a light
  atom}
In fact, it is also possible for three identical fermionic atoms to
bind together owing to the attractive interaction mediated by a light
atom~\cite{Levinsen2013}. The critical mass ratio for the binding of
this tetramer is $m_\up/m_\down\approx5.0$, and like the trimer, the
tetramer binds in the $p$-wave state. It remains to be seen whether
more tetramers bind with increasing mass ratio, as in the case of
trimers, and whether they also form in higher partial waves. These
questions may presumably be answered within the Born-Oppenheimer
approximation. Interestingly, the tetramer is very close in energy to
a trimer plus a free atom, and this should lead to strong atom-trimer
interactions in the 2D heteronuclear Fermi gas for species close to
the critical mass ratio. A similar bound state has been predicted in
3D above a mass ratio of 9.5 (Ref.~\refcite{Blume2012}) and in 1D
above the mass ratio 2.0 (Ref.~\refcite{Mehta2014}). It is likely that
these states are continuously connected as the system is tuned between
the different geometries.

\subsection{Universal bound states in realistic experiments: Going
  beyond the 2D limit}

As discussed previously, realistic experiments on 2D Fermi gases
involve the presence of a tight confinement and the length scale
corresponding to this confinement always greatly exceeds the range of
the interatomic interactions. Thus, it is important to relate the
universal 2D few-body physics presented above to realistic
experiments, taking into account the 3D nature of the interactions.

In Sec.~\ref{sub:q2d}, the effects of confinement on the two-body
interaction were described. As should be clear from that discussion,
the 2D limit of the universal bound states described thus far
constitutes the regime where the 3D scattering length is negative and
much smaller than the confinement length, {\em i.e.}  $l_z/a\ll-1$; in
this limit we have a dimer whose binding energy $\eb$ is much smaller
than the level spacing in the harmonic trap, $\op$, and as the
energies of the universal bound states scale with $\eb$ we may be in a
regime where these are also negligible compared with $\op$.

It is natural to ask what happens to the bound trimers and tetramers
described above, once the confinement is relaxed. This question was
investigated in Ref.~\refcite{Levinsen2013} under the assumption that
both species of atoms are confined by a harmonic trap of the same
frequency. It was shown that the minimum energy $E$ in the problem of
$N$ spin-$\up$ atoms and a single spin-$\down$ atom in the center of
mass frame corresponds to a non-trivial solution of
\begin{align}
  \chi_{\k_2\ldots\k_N}^{n_0\ldots n_N} = &
  -\sum_{\k_1',n_0'n_1'}\frac{T_{n_0n_1}^{n_0'n_1'}(\k_0+\k_1,
    E_0+\epsilon_{\k_1n_1\up})}
  {E_0+\epsilon_{\k_1n_1\up}-\epsilon_{\k_0+\k_1-\k_1'
      n_0'\down}-\epsilon_{\k_1'n_1'\up}} \nn \\ & \hspace{5mm} \times
  \left\{\chi^{n_0'n_2n_1'n_3\ldots n_N}_{\k_1'\k_3\ldots\k_N}+
    \ldots+\chi^{n_0'n_Nn_2\ldots
      n_{N-1}n_1'}_{\k_2\ldots\k_{N-1}\k_1'}\right\}.
\label{eq:np1}
\end{align}
Here the single particle energies are $\epsilon_{\k
  n\sigma}=k^2/2m_\sigma+n\op$, $\k_1,\ldots,\k_N$ are the initial
momenta of the spin-$\up$ atoms, while $\k_0$ and $E_0\equiv
E-\sum_{i=1}^N\epsilon_{\k_in_i\up}$ are the initial momentum and
energy of the spin-$\down$ atom.  Since we consider scattering in the
center of mass frame of the 2D motion, we have $\k_0 = -\sum_i^N
\k_i$.  The energy is measured from the $N+1$ atom threshold
$(N+1)\op/2$. $T_{n_0n_1}^{n_0'n_1'}$ is related to the
quasi-two-dimensional $T$ matrix\footnote{ Specifically,
\begin{align*}
  T^{n_0'n_1'}_{n_0n_1}(\q,\epsilon) = &
  \sum_{n\,n_rn_r'\!\!}C^{n_0n_1}_{nn_r}(m_\down,m_\up)
  C^{n_0'n_1'}_{nn_r'}(m_\down,m_\up) \\ \nn & \times
  \sqrt{2\pi}l_zf_{n_r}f_{n_r'}\T\left(\epsilon-n\op+\frac{1}{2}\op-\frac{q^2}{2M}
  \right).
\end{align*}
The Clebsch-Gordan coefficients
$C^{n_0n_1}_{nn_r}(m_\down,m_\up)\equiv\langle n_0n_1|nn_r\rangle$
were obtained in Ref.~\refcite{Smirnov1962} and vanish unless
$n_0+n_1=n+n_r$.} Eq.~\eqref{eq:Tq2d} via a change of basis to the
relative and center of mass motion in the two-atom problem. The minus
sign on the \emph{r.h.s.}  arises from the antisymmetry of the vertex
$\chi$ under exchange of identical fermions.

\begin{figure}
\centering
\includegraphics[width=0.6\linewidth]{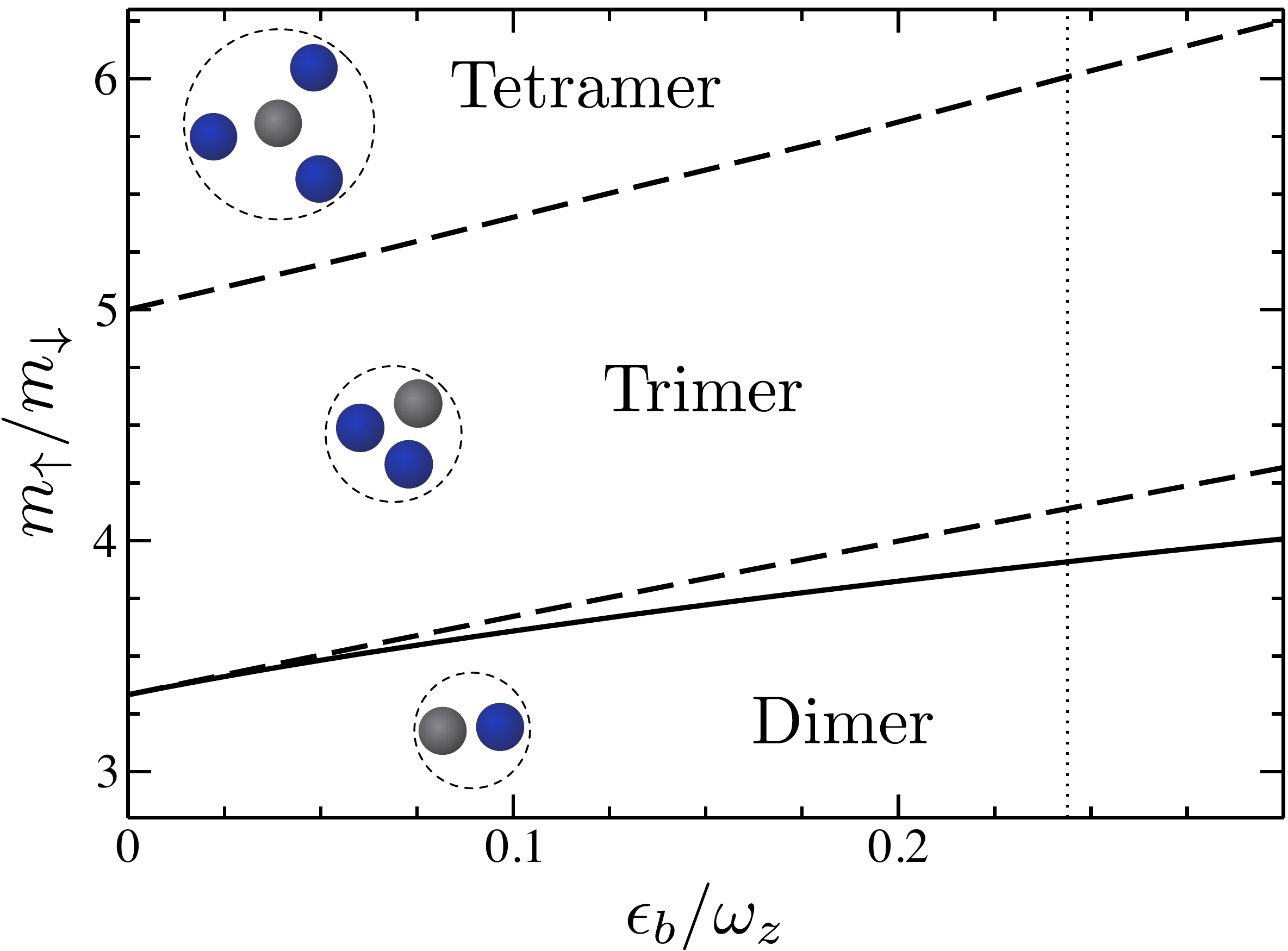}
\caption{The critical mass ratio for the formation of trimers and
  tetramers away from the 2D limit of $\eb/\op=0$, assuming the
  two species are confined by a harmonic potential characterized by
  the same frequency $\op$. The solid line is the result of exact
  calculations, while the dashed lines employ a two-channel model for
  the confinement. The vertical dotted line
  corresponds to the position of the 3D resonance. The figure is taken
  from Ref.~\refcite{Levinsen2013}.
  \label{fig:vacuumphase}}
\end{figure}

While Eq.~\eqref{eq:np1} is quite compact and in principle allows one
to capture the crossover from 2D to 3D physics, its numerical solution
quickly becomes prohibitive with increasing number of atoms.  Instead,
the departure from the 2D limit of the few-body bound states described
above may be considered in a perturbative expansion\cite{Levinsen2013}
in the parameter $\eb/\op$. This amounts to expanding the function
$\F$ in Eq.~\eqref{eq:Fcali} to linear order in $|E|/\op$ while
setting all harmonic oscillator quantum numbers to zero in
\req{eq:np1}, and follows from the antisymmetry of the vertex $\chi$
under exchange of any of the $N$ spin-$\up$ atoms. For definiteness we
write down here the resulting equation which determines the trimer
binding energy in the 3-body problem in the limit of strong
confinement:
\begin{align}
 & \frac{m_r}{2\pi}\left(
    \ln\left[
      \frac{-E+k_2^2/2\mad}{\eb}\right]-\ln(2)
    \frac{\eb+E-k_2^2/2\mad}{\op}\right) \chi_{\k_2}\nn \\ 
&=
  \sum_{\k_1}\frac{\chi_{\k_1}}
  {E-\epsilon_{\k_1\up}-\epsilon_{\k_2\up}-\epsilon_{\k_1+\k_2\down}}.
\end{align}

Figure~\ref{fig:vacuumphase} shows the behavior of the critical mass
ratio of the trimer and tetramer as the system is tuned away from the
strict 2D limit. We observe that the critical mass
ratio increases, consistent with the corresponding results in
3D~\cite{trimer,Blume2012}. This behavior is likely due to the
increased centrifugal barrier for $p$-wave pairing in
3D. Interestingly, at unitarity the critical mass ratio for tetramer
formation in this quasi-2D geometry is below the K-Li mass ratio.

\begin{figure}
\centering
\includegraphics[width=0.6\linewidth]{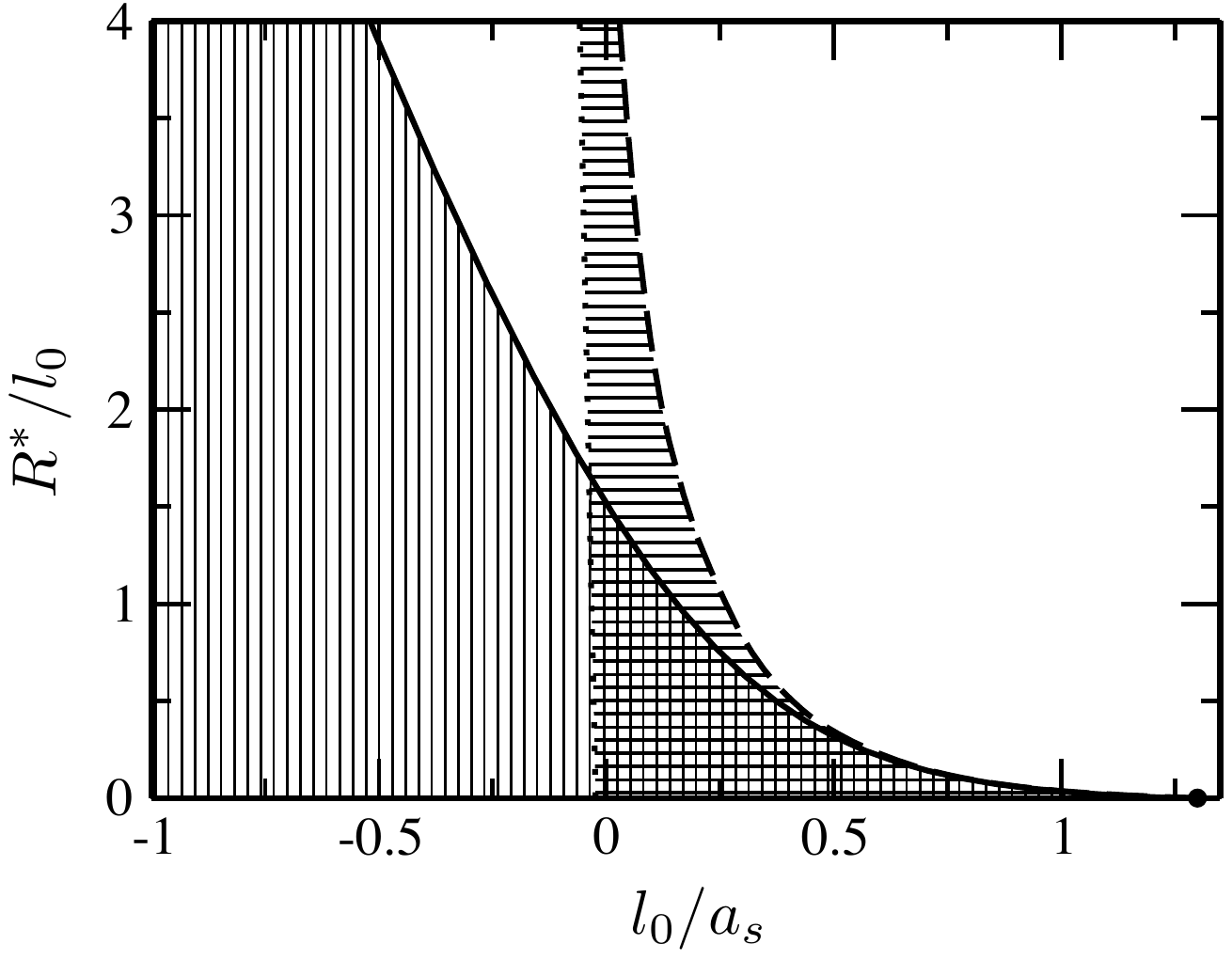}
\caption{Confinement induced trimers for the K-K-Li system: The solid
  line corresponds to the trimer formation threshold when both species
  are confined by a harmonic potential of the same frequency, while
  the dashed line assumes confinement of the heavy atom only. Three
  atoms bind to a trimer in the shaded
  regions. $l_0=1/\sqrt{m_\up\omega_\up}$ is the confinement length of
  the heavy atoms, while $R^*=-r_e/2>0$ is a length scale
  parameterizing the width of the Feshbach resonance ($r_e$ is the
  effective range). The figure is taken from
  Ref.~\refcite{Levinsen2009}.
  \label{fig:crit}}
\end{figure}

From the point of view of experiment, it is useful to determine the
precise conditions under which K and Li atoms will form such stable
few-body bound states. In the above we assumed that the confinement
frequencies of the two atomic species were identical, and while this
can be engineered using species dependent optical lattices this need
not be the case. However, even in the case where the frequencies are
species dependent, few-body properties may only be weakly affected by
this dependence; for instance, once the light atom oscillator length
greatly exceeds the two-body bound state, the light atom is
essentially confined by its interaction with the heavy atoms. This is
clearly illustrated in Fig.~\ref{fig:crit} which shows the critical
confinement length needed to confine two K atoms and a Li atom into a
trimer, and thus to make the atom-dimer interaction resonant in the
$p$-wave channel~\cite{Levinsen2009}. This result additionally takes
into account the fact that Li-K interspecies Feshbach resonances are
narrow in magnetic field width, associated with a weak coupling to the
closed channel molecular state. This latter effect tends to suppress
the attraction mediated by the light atom, and introduces an
additional complication in the quest to obtain these few-body states.

\subsection{Identical fermions with $p$-wave interactions --- Super
  Efimov states}

Finally, we mention that the three-body problem of identical fermions
in 2D interacting via a $p$-wave Feshbach resonance has also been
investigated~\cite{Levinsen2008,Moroz2013}. Ref.~\refcite{Levinsen2008}
studied the atom-dimer scattering problem and found that the
$T$-matrix at a momentum $\p$ is
$T(p)\propto\cos\left[\frac43\ln\ln\Lambda/p+\phi\right]$ where $\phi$
is a phase and $\Lambda$ a momentum cutoff. Remarkably, this
$T$-matrix displays a discrete scale invariance reminiscent of the
Efimov effect, with a three-body parameter set by the momentum
cutoff. However, unlike the standard Efimov effect, the scaling is
here characterized by a double exponential factor, which quickly
becomes enormous.  More recently, Ref.~\refcite{Moroz2013} studied the
bound state problem and found the existence of trimers with a doubly
exponential scaling of the binding energy, naming these super Efimov
states.

%% file: manybody.tex
\section{Ground state of the many-body system}\label{sec:bcsbec}

\subsection{The BCS-BEC crossover}

We now turn to the behavior of the many-body system, where one has a
finite density of spin-up and spin-down fermions, denoted $n_\up =
N_\up/A$ and $n_\down = N_\down/A$, respectively. At zero temperature,
in the absence of interactions, each type of fermion forms a filled
Fermi sea with radius in momentum space given by the Fermi wave vector
$k_{F\sigma} = \sqrt{4\pi n_\sigma}$ in 2D.  For an attractive
interspecies interaction, such as the short-range potential described
in Eq.~\eqref{eq:H}, a variety of pairing phenomena is expected to
occur in the Fermi system. In particular, for the case of equal
densities, $\kfup = \kfdown \equiv \kf$, the ground state can smoothly
evolve from the BCS regime of Cooper pairing to a Bose-Einstein
condensate (BEC) of tightly bound dimers with increasing interaction
strength. This constitutes the celebrated BCS-BEC crossover, which was
theoretically predicted several decades
ago~\cite{Leggett1980,eagles1969}, and first successfully realised in
3D cold-atom experiments in 2004
(Refs.~\refcite{regal2004,zwierlein2004}).

The different regimes of pairing are generally parameterized by the
dimensionless quantity $\kf l$, where $l$ is a typical length scale
that defines the strength of the interaction. For short-range $s$-wave
interactions in 3D, $l$ is simply the $s$-wave scattering length
$\as$, while in 2D it corresponds to $\ad$, which is related to the
size of the two-body bound state and is defined in
Sec.~\ref{sub:short}. Thus, in 2D, the BCS and BEC regimes correspond,
respectively, to $\kf \ad \gg 1$ and $\kf \ad \ll 1$, where the pair
size is much greater than the inter-particle spacing in the former
case, and much smaller in the latter. The dimensionless parameter $\kf
\ad$ automatically implies that there are two ways of achieving the
BCS-BEC crossover: by varying the interactions or by varying the
density. Note, however, that this is not the case for 3D contact
interactions, since a two-body bound state does not exist for
arbitrarily weak attraction, and the scattering length can change
sign. Thus, in 3D, one cannot traverse the entire crossover by varying
density alone.

This section will focus on the situation where the masses are equal,
i.e., $m_\up = m_\down \equiv m$. In any case, we do not expect the
qualitative picture of the BCS-BEC crossover to change for a small
mass imbalance. However, we can see from Fig.~\ref{fig:spectrum} in
Sec.~\ref{sec:few} that once $m_\up/m_\down \gtrsim 25$ (or equivently
when $m_\up/m_\down \lesssim 1/25$), the trimer state has a lower
energy than two dimers and therefore, in the BEC regime, the system
will prefer to form a mixed gas of trimers and light atoms. It remains
an open question what happens deep in the BCS regime, where the size
of the trimer becomes larger than the inter-particle distance.

\subsection{Mean-field description} \label{sub:MF} To gain insight
into the BCS-BEC crossover, it is instructive to employ a mean-field
approach to the problem.  We start by considering the full quasi-2D
problem as exists in experiment, and then we specialize to the 2D
limit in later sections.  The quasi-2D Fermi gas was first considered
within the mean-field approximation in Ref.~\refcite{MarT05}, but only
a few harmonic oscillator levels were included. Here, we follow the
approach in Ref.~\refcite{fischer2013}, which can in principle account
for an infinite number of levels.

Building on the formalism in Sec.~\ref{sec:2d}, the many-body
grand-canonical Hamiltonian in the quasi-2D geometry is (setting the
system area $A=1$):
\begin{align}
  \nonumber \hat{H}& = \sum_{\mathbf{k}, n,
    \sigma}(\epsilon_{\mathbf{k} n}-\mu)
  c^\dagger_{\mathbf{k} n \sigma} c_{\mathbf{k} n \sigma}\\
  & + \sum_{\substack{\mathbf{k}, n_1, n_2 \\ \mathbf{k}', n_3, n_4 \\
      \mathbf{q}}} \langle n_1 n_2 |\hat{g}| n_3 n_4 \rangle
  c^\dagger_{\mathbf{k} n_1 \uparrow}c^\dagger_{\mathbf{q}-\mathbf{k}
    n_2 \downarrow} c_{\mathbf{q}-\mathbf{k}' n_3
    \downarrow}c_{\mathbf{k}' n_4 \uparrow}\label{eq-fullham},
\end{align}
where $\epsilon_{\mathbf{k} n}=k^2/2m + n \op$ are the single particle
energies relative to the zero-point energy of the $n=0$ state.  Note
that since we have assumed that the masses and particle densities are
equal, the chemical potential must be the same for each spin $\sigma$,
i.e., $\mu_\up = \mu_\down \equiv \mu$.

The 3D attractive short-range interaction is set by the constant $g$
like in Sec.~\ref{sec:2d}.  In the many-body system, it is convenient
to work in the basis of the individual atoms rather than only
considering the relative pair motion as in the two-body problem.
However, since the interaction only depends on the relative motion,
the interaction matrix elements $\langle n_1 n_2|\hat{g}| n_3 n_4
\rangle$ are best determined by switching to relative and center of
mass harmonic oscillator quantum numbers, $\nu$ and $N$
respectively. This yields
\begin{align}
\label{eq-intmatelts}
\langle n_1 n_2|\hat{g}| n_3 n_4 \rangle &= g\sum_{N\nu \nu'}
f_\nu\langle n_1 n_2|N \nu \rangle
f_{\nu'}\langle N \nu'|n_3 n_4\rangle  \nonumber \\
&\equiv g\sum_N V_N^{n_1 n_2}V_N^{n_3 n_4},
\end{align}
where $f_\nu=\sum_{k_z}\tilde{\phi}_\nu(k_z)$, and $\tilde{\phi}_\nu$
is the Fourier transform of the $\nu$-th harmonic oscillator
eigenfunction.  These correspond to the $f$ coefficients defined
previously in Sec.~\ref{sec:2d}, but with momentum cut-off $\Lambda
\to \infty$, i.e., for even $\nu$, we obtain Eq.~\eqref{eq:funcn} with
$\lambda = 0$. In this case, $f_\nu$ reduces to the harmonic
oscillator wave function evaluated at $z=0$. The Clebsch-Gordan
coefficients in the matrix elements are given
by~\cite{Smirnov1962,ChaW67}
\begin{align} {\langle n_1 n_2|N \nu \rangle = \delta_{N+\nu,
      n_1+n_2}} \sqrt{\frac{N! \nu!}{2^{n_1+n_2}n_1! n_2!}}  \sum_{i+j
    = \nu} (-1)^{j} \binom{n_1}{i} \binom{n_2}{j} \>,
\end{align}
with $i=0,1, ... , n_1$ and $j=0,1, ... , n_2$.  Note that the
scattering process conserves parity since $\nu,\nu'$ must be even;
namely, if $n_1+n_2$ is even (odd), then the matrix element is only
non-zero when $n_3+n_4$ is also even (odd).  The 3D contact
interaction parameter $g$ can be written in terms of the quasi-2D
two-body binding energy $\eb$:
\begin{align}
\label{eq-inverseg}
-\frac{1}{g} = \sum_{\textbf{k}, n_1, n_2}\frac{f_{n_1+n_2}^2 |\langle
  n_1 n_2 | 0 \hspace*{2mm} n_1+n_2\rangle|^2}{\epsilon_{\textbf{k}
    n_1}+\epsilon_{\textbf{k} n_2}+\eb} \>.
\end{align}
Here, we simply take $N=0$ since $\eb$ is independent of the two-body
center of mass motion.  One can also connect $\eb$ to the 3D
scattering length $a_s$ using Eq.~\eqref{eq:q2d-bound} with $\lambda =
0$.

Following Ref.~\refcite{fischer2013}, we define the pairing order
parameter
\begin{align}
  \Delta_{\mathbf{q} N}=g\sum_{\mathbf{k}, n_1, n_2} V_N^{n_1 n_2} \langle
  c_{\mathbf{q}-\mathbf{k}n_2\downarrow}c_{\mathbf{k}n_1\uparrow}\rangle,
\end{align}
and assume that fluctuations around this are small, thus obtaining the
mean-field Hamiltonian,
\begin{align}
  \nonumber \hat{H}_\mathrm{MF} = & \sum_{\mathbf{k}, n,
    \sigma}(\epsilon_{\mathbf{k} n}-\mu) c^\dagger_{\mathbf{k} n
    \sigma} c_{\mathbf{k} n \sigma}\\ \nonumber & + \sum_{\mathbf{q},
    N}\bigg( \Delta_{\mathbf{q} N}\sum_{\mathbf{k}, n_1, n_2} V_N^{n_1
    n_2}c^\dagger_{\mathbf{k} n_1
    \uparrow}c^\dagger_{\mathbf{q}-\mathbf{k} n_2
    \downarrow}\\
  & + \Delta_{\mathbf{q} N}^\ast\sum_{\mathbf{k}', n_3, n_4} V_N^{n_3
    n_4}c_{\mathbf{q}-\mathbf{k}' n_3 \downarrow} c_{\mathbf{k}'n_4
    \uparrow} -\frac{|\Delta_{\mathbf{q} N}|^2}{g} \bigg).
\label{eq-mfham}
\end{align}
If we further assume that the ground state has a uniform order
parameter without nodes so that ${\Delta_{\mathbf{q}
    N}=\delta_{\mathbf{q} \vect{0}} \delta_{N 0} \Delta_0}$, then
Eq.~(\ref{eq-mfham}) only contains a single unknown parameter
$\Delta_0$.  Thus $\hat{H}_\mathrm{MF}$ can be diagonalized to yield
\begin{align}
\label{eq-quasiham}
\hat{H}_\mathrm{MF}=\sum_{\mathbf{k}, n} (\epsilon_{\mathbf{k}
  n}-\mu-E_{\mathbf{k} n}) -\frac{\Delta_0^2}{g} + \sum_{\mathbf{k},
  n, \sigma}E_{\textbf{k} n }\gamma^\dagger_{\textbf{k} n \sigma}
\gamma_{\textbf{k} n \sigma},
\end{align}
where $E_{\textbf{k} n }$ are the quasiparticle excitation
energies. The quasiparticle creation and annihilation operators are
respectively given by
\begin{align}
\label{eq-gammaup}
\gamma^\dagger_{\textbf{k} n \uparrow} & =\sum_{n'} (u_{\textbf{k}
n'n}c^\dagger_{\mathbf{k} n' \uparrow}+
v_{\textbf{k} n'n}c_{-\mathbf{k} n' \downarrow}) \\
\label{eq-gammadown}
\gamma_{-\textbf{k} n \downarrow} & =\sum_{n'} (u_{\textbf{k} n' n
}c_{-\mathbf{k} n' \downarrow}- v_{\textbf{k}
  n'n}c^\dagger_{\mathbf{k} n' \uparrow}) ,
\end{align}
where the amplitudes $u$, $v$ only depend on the magnitude $k \equiv
|\vect{k}|$ and satisfy $\sum_{n'} (|u_{\textbf{k} n'n}|^2 +
|v_{\textbf{k} n'n}|^2)=1$. Without loss of generality, we can choose
$u$, $v$ to be real.  Note that while the quasiparticles have a well
defined spin and momentum, they involve a superposition of different
harmonic oscillator levels. Since the ground state corresponds to the
vacuum state for the quasiparticles, the ground-state wave function
can be written
\begin{align}
|\Psi_\mathrm{MF}\rangle \propto \prod_{\vect{k}n\sigma} \gamma_{\vect{k}n\sigma} |0\rangle \ , 
\end{align}
where $|0\rangle$ is the vacuum state for the bare operators
$c_{\vect{k}n\sigma}$.  In the 2D limit where $\op \gg \mu, \eb$ and
we only have the lowest level $n=0$, we recover the standard BCS wave
function
\begin{align}
  \ket{\Psi_\mathrm{MF}} = \prod_{\k} \lba u_{\k 00} + v_{\k 00} c_{\k
    0 \up}^\dag c_{-\k 0 \down}^\dag \rba \ket{0} .
\end{align}

In general, we must minimize $\langle \hat{H}_\mathrm{MF} \rangle =
\sum_{\mathbf{k}, n} (\epsilon_{\mathbf{k} n}-\mu-E_{\mathbf{k} n})
-\frac{\Delta_0^2}{g}$ with respect to $\Delta_0$ at fixed $\mu$ to
obtain the ground state. For the 2D limit, this yields the usual form
$E_{\k} \equiv E_{\mathbf{k} 0} = \sqrt{(\epsilon_{\k}-\mu)^2 +
  \Delta^2}$, with $\ek \equiv \epsilon_{\k 0}$ and $\Delta \equiv
\Delta_0 V^{00}_0$. Using the density $2n_\sigma = - \del \langle
\hat{H}_\mathrm{MF} \rangle /\del \mu$ and the 2D Fermi energy $\ef =
\kf^2/2m$, we also obtain $\mu = \ef - \eb/2$ and $\Delta =
\sqrt{2\ef\eb}$, as derived
previously~\cite{Randeria1989,Randeria1990}. For the quasi-2D system,
one must use the general expression for the density
$n_\sigma=\sum_{\textbf{k}, n',n} |v_{\textbf{k} n'n}|^2$ and define
$\ef$ to be the chemical potential of an ideal Fermi gas with the same
density.

\begin{figure}
\centering
\includegraphics[width=1\linewidth]{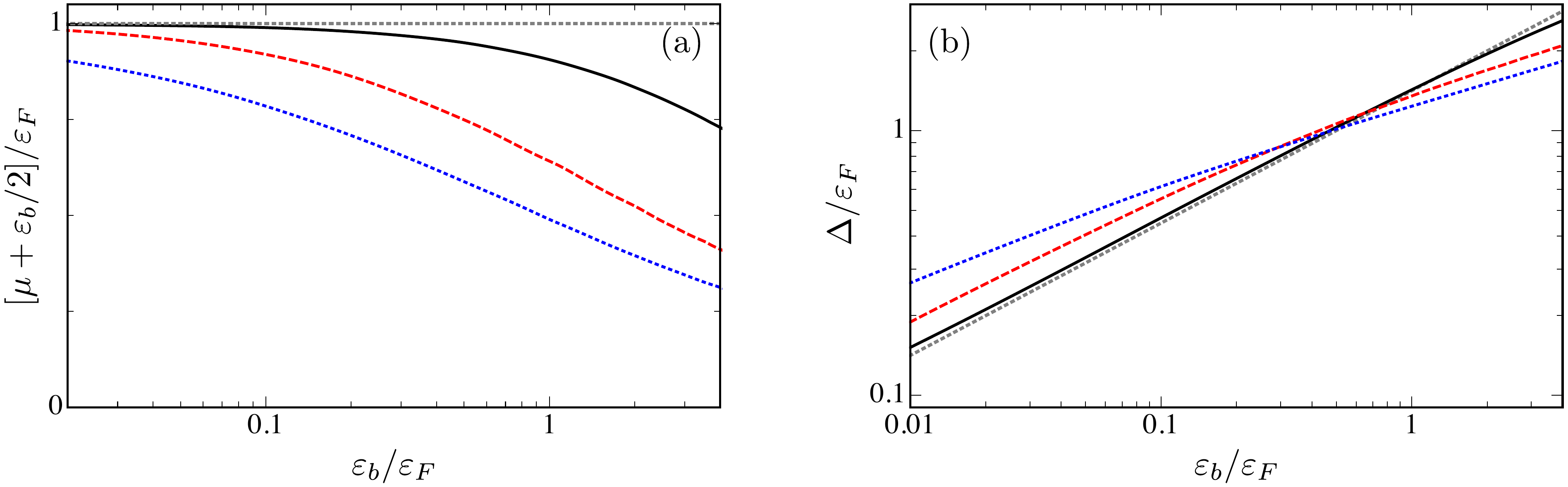}
\caption{Behaviour of the chemical potential (a) and the order
  parameter (b) as a function of the interaction parameter $\eb/\ef$
  for different confinement strengths. The solid (black), dashed
  (red), and dotted (blue) lines correspond to $\ef/\op = 0.1$, 0.5
  and 1, respectively.  The thin dotted lines are the results in the
  2D limit, $\mu=\ef-\eb/2$ and $\Delta = \sqrt{2\eb\ef}$.  In (a),
  the term from two-body binding, $-\eb/2$, has been subtracted from
  $\mu$ in order to expose the many-body corrections.
  The data is taken from Ref.~\refcite{fischer2013}.
  \label{fig:q2d-mu-gap}}
\end{figure}

As the interactions are varied in the quasi-2D Fermi gas, the chemical
potential always evolves from $\ef$ in the weakly interacting BCS
limit $\eb/\ef \ll 1$, to $-\eb/2$ in the BEC limit $\eb/\ef \gg
1$. However, the precise evolution in the BCS-BEC crossover is
dependent on the quasi-2D confinement frequency $\omega_z$ once we are
away from the 2D limit $\op \gg \ef, \eb$, and this has implications
for current 2D experiments, as we discuss later. If we take the limit
of weak confinement $\op \to 0$, there will be a crossover to 3D
pairing, albeit with a modified density of states due to the trapping
potential in the $z$-direction. For instance, in the BEC regime, we
should recover ``3D bosons'' with $\eb \simeq 1/m\as^2$ in the limit
$\eb \gg \op$ --- see the discussion in Sec.~\ref{sub:q2d}.  For
strong confinement, one might naively expect to observe 2D behaviour
once $\eb, \ef < \op$. However, the chemical potential can be
dramatically reduced from the 2D result even in this regime, as shown
in Fig.~\ref{fig:q2d-mu-gap} (a), and the deviation from 2D involves
multiple harmonic oscillator levels~\cite{fischer2013}.  Moreover,
once we approach $\ef \simeq \op$, the chemical potential is strongly
modified even when the interactions are weak, $\eb \ll \op$.  This is
analogous to the behaviour of the quasi-2D two-body $T$ matrix
\eqref{q2D-T}, which only resembles the 2D expression when the
collision energy $E \ll \op$, and is substantially different when $E
\simeq \op$. The pairing order parameter $\Delta$ is also modified by
the presence of $\op$, with $\Delta$ being increased in the BCS regime
with increasing $\ef/\op$, as seen in Fig.~\ref{fig:q2d-mu-gap}
(b). This suggests that pairing is enhanced by perturbing away from
the 2D limit and this therefore impacts the critical temperature for
superfluidity (see Sec.~\ref{sec:temp}).

\subsection{Perturbative regimes}\label{sub:perturb}

In order to go beyond mean-field theory, we restrict ourselves to the
2D limit $\op \gg \ef, \eb$, so that the different regimes are
completely parameterized by $\kf \ad$. In this case, the Hamiltonian
in \req{eq-fullham} reduces to
\begin{align}\label{eq:2dham}
  \hat{H}& = \sum_{\mathbf{k}, \sigma}\left(\ek -\mu \right)
  c^\dagger_{\mathbf{k} \sigma} c_{\mathbf{k} \sigma} + g
  \sum_{\mathbf{k}, \mathbf{k}', \mathbf{q}} c^\dagger_{\mathbf{k}
    \uparrow}c^\dagger_{\mathbf{q}-\mathbf{k} \downarrow}
  c_{\mathbf{q}-\mathbf{k}' \downarrow}c_{\mathbf{k}' \uparrow} ,
\end{align}
where $g$ is now an effective 2D contact interaction like in
\req{eq:H}, which gives rise to the required scattering length $\ad$
(or two-body binding energy $\eb$).  While mean-field theory provides
an appealingly simple and intuitive picture of the BCS-BEC crossover
in 2D, it is not expected to be quantitatively accurate and it at best
provides an upper bound on the energy, being in essence a variational
approach.  In particular, it substantially overestimates the effective
dimer-dimer interaction in the BEC regime.  Even in the BCS regime
where interactions are weak, it fails to capture the leading order
dependence of the energy on $1/\ln(\kf \ad)$ since it neglects the
interaction energy of the normal Fermi liquid phase. However, one can
extract accurate analytic expressions for the behaviour in the limits
$|\ln(\kf \ad)| \gg 1$ by performing a proper perturbative expansion
in the interaction.

In the regime of weak attraction $\ln(\kf \ad) \gg 1$, the gas behaves
as a Fermi liquid in the normal
state~\cite{engelbrecht1990,engelbrecht92,engelbrecht1992-2}. One can
show using perturbation theory in $g$ that the energy per particle in
this limit is \cite{Bloom1975,engelbrecht92}
\begin{align} \label{eq:weak-erg}
\frac{E}{N} = \frac{\ef}{2} \lba 1 -
  \frac{1}{\eta} + \frac{\mathcal{A}}{\eta^2} \rba ,
\end{align}
where $N = N_\up + N_\down$ and $\eta = \ln( \kf \ad)$. Of course, in
the ground state, the gas will be a paired superfluid rather than a
Fermi liquid, but the energy due to pairing scales as $\Delta^2/\ef
\sim \eb$ in this limit, which tends to zero faster than $\ef/\ln(\kf
\ad)$ as $\eb \to 0$.  While the structure of the perturbative
expansion is clear, there is some discrepancy in the literature
regarding the constant $\mathcal{A}$. A thorough calculation for the
repulsive Fermi gas using second-order perturbation
theory~\cite{engelbrecht92,engelbrecht1992-2} gives $\mathcal{A} = 3/4
- \ln(2) \simeq 0.06$. However, a recent QMC
calculation~\cite{Bertaina2011} finds a larger value: $\mathcal{A}
\simeq 0.17$.

In the opposite limit $\ln(\kf \ad) \ll -1$, the system can be
regarded as a weakly interacting gas of bosonic dimers. In this case,
the effective dimer-dimer interaction $g_d$ in the low-energy limit is
parameterized by the scattering length $\add \simeq 0.56 \ad$, as
noted in Sec.~\ref{sec:few}.  Moreover, the total energy of the system
can be written as $E = -\eb N_d + E_{d}$, where $N_d = N_\sigma = N/2$
and $E_d$ is the energy of a repulsive gas of $N_d$ bosons. We can
likewise introduce a boson chemical potential $\mu_d = 2 \mu +
\eb$. Note that in the limit $\add \to 0$ (or, equivalently, when $\eb
\to \infty$), we have $\mu_d \to 0^+$, as expected for a
non-interacting BEC. However, this behavior is not captured by BCS
mean-field theory, which predicts $\mu_d = 2\ef$. This corresponds to
an effective dimer-dimer interaction that only scales with the
density, as one might expect from a classical theory of interacting
dimers in 2D rather than an appropriately renormalized quantum one
(see Sec.~\ref{sec:dynamics}).

To extract the behaviour of the weakly repulsive Bose gas, we consider
the grand potential according to Bogoliubov theory:
\begin{align}
  \Omega = - \frac{\mu_d^2}{2g_d} - \frac{1}{2} \sum_{\vect{k}\neq 0}
  \lba \epsilon_{\vect{k}d} + \mu_d - \sqrt{\epsilon_{\vect{k}d}
    (\epsilon_{\vect{k}d}+2\mu_d)} \rba \>,
\end{align}
where $\epsilon_{\vect{k}d} = \frac{k^2}{2M}$ and $M = m_\up + m_\down
= 2m$.  After regularizing the momentum sum, we obtain (see, also,
Ref.~\refcite{MoraCastin2009})
\begin{align}
  \Omega = \frac{M \mu_d^2}{16 \pi }\lbc 1 - 2\ln\lba \frac{1}{M
    \add^2 \mu_d} \rba \rbc .
\end{align}
To relate this back to the Fermi system, we consider the density of
dimers:
\begin{align}
  n_d \equiv \frac{N_d}{A} = - \frac{\partial \Omega}{\partial \mu_d}
  = \frac{M \mu_d}{4 \pi} \ln\lba \frac{1}{M \add^2 \mu_d e} \rba .
\end{align}
Assuming that $\ln\lba\frac{1}{4\pi e n_d\add^2}\rba \gg 1$, this can
then be rearranged to obtain the leading order expression for $\mu_d$
in terms of the density $n_d$, i.e.,
\begin{align} \label{eq:mud}
\mu_d \simeq \frac{4\pi n_d}{M}
  \frac{1}{\ln\lba\frac{1}{4\pi e n_d\add^2}\rba} \lbc 1-
  \frac{\ln\ln\lba\frac{1}{4\pi e n_d\add^2}\rba}{\ln\lba\frac{1}{4\pi
      e n_d\add^2}\rba} \rbc .
\end{align}
This finally gives us the energy density $E/N_d = -\eb + E_{d}/N_d$,
with
\begin{align}
  \frac{E_d}{N_d} \simeq \frac{2\pi n_d}{M}
  \frac{1}{\ln\lba\frac{1}{4\pi e n_d\add^2}\rba} \lbc 1-
  \frac{\ln\ln\lba\frac{1}{4\pi e n_d\add^2}\rba}{\ln\lba\frac{1}{4\pi
      e n_d\add^2}\rba} - \frac{1}{2\ln\lba\frac{1}{4\pi e
      n_d\add^2}\rba} \rbc ,
\end{align}
where we have used the fact that $E_d = \Omega + \mu_d N_d$ at zero
temperature. This agrees with the expression used in
Ref.~\refcite{Bertaina2011}.

\subsection{Equation of state \label{sub:eos}}

\begin{figure}
\centering
\includegraphics[width=1\linewidth]{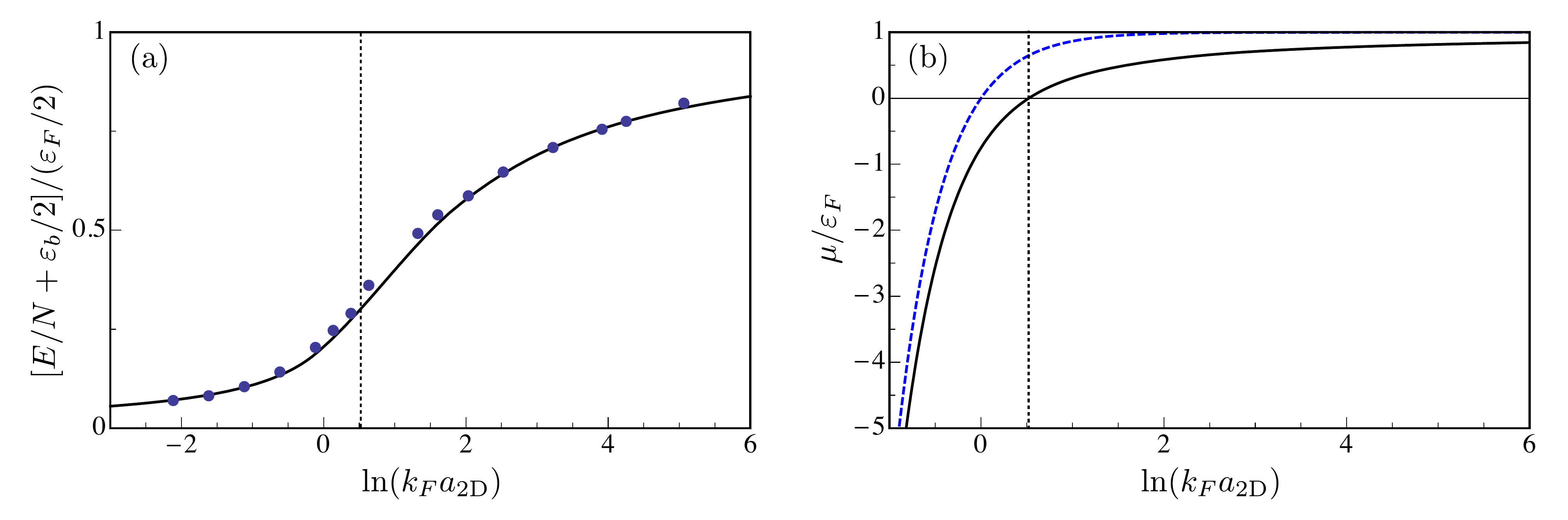}
\caption{
(a) Energy per particle in units of $\ef /2$ (the energy per particle
  in the non-interacting gas) as a function of interaction
  strength. The trivial
  two-body binding energy has
  been subtracted. The data points are the results of the QMC
  calculation \cite{Bertaina2011}, while the solid line is an
  interpolation between the known weak coupling results in the BCS and
  BEC limits.  (b) Chemical potential in units of the Fermi
  energy. The solid (black) line is an interpolation between the known
  limiting behaviors, while the dashed (blue) line is 
  the mean-field result $\mu = \ef - \eb/2$.
 In both plots, the vertical dotted line indicates the point at which
 the chemical potential is zero.  
 \label{fig:bertaina}}
\end{figure}

In order to obtain an accurate equation of state throughout the
BCS-BEC crossover, one must resort to numerical approaches such as the
fixed node diffusion QMC method mentioned previously
\cite{Bertaina2011}. The QMC result for the energy per particle is
displayed in Fig.~\ref{fig:bertaina}(a).\footnote{Note that we have
  used a different definition of $\ad$ compared with the original QMC
  paper -- see the discussion in Sec.~\ref{sub:short}.}  As expected,
the energy matches the known results in the BCS and BEC limits (taking
the QMC value for $\mathcal{A}$ in \req{eq:weak-erg}).  Indeed, by
interpolating between the known weak-coupling results in the BCS and
BEC regimes, we can obtain a reasonable curve for the energy
throughout the crossover that matches the QMC data. This also allows
us to easily extract other thermodynamic quantities such as the
chemical potential and the pressure --- one simply takes the
appropriate derivative of the expressions in the perturbative regimes
and then interpolates between the results. In particular, the chemical
potential in the BEC regime can be taken from \req{eq:mud}, while in
the BCS limit it corresponds to
\begin{align}
  \mu = \frac{\del E}{\del N} = \ef \lba 1 - \frac{1}{\eta} +
  \frac{4\mathcal{A}+1}{4 \eta^2} \rba .
\end{align}
Referring to Fig.~\ref{fig:bertaina}(b), we find that the interpolated
$\mu$ is lower than that from mean-field theory, but it still evolves
from $\ef$ to $-\eb/2$ with increasing attraction (decreasing
$\ln(\kf\ad)$). The point $\mu = 0$ may be regarded as the ``crossover
point'' that approximately separates Fermi and Bose regimes, as we
discuss in Sec.~\ref{sub:crossover}.

\begin{figure}
\centering
\includegraphics[width=0.7\linewidth]{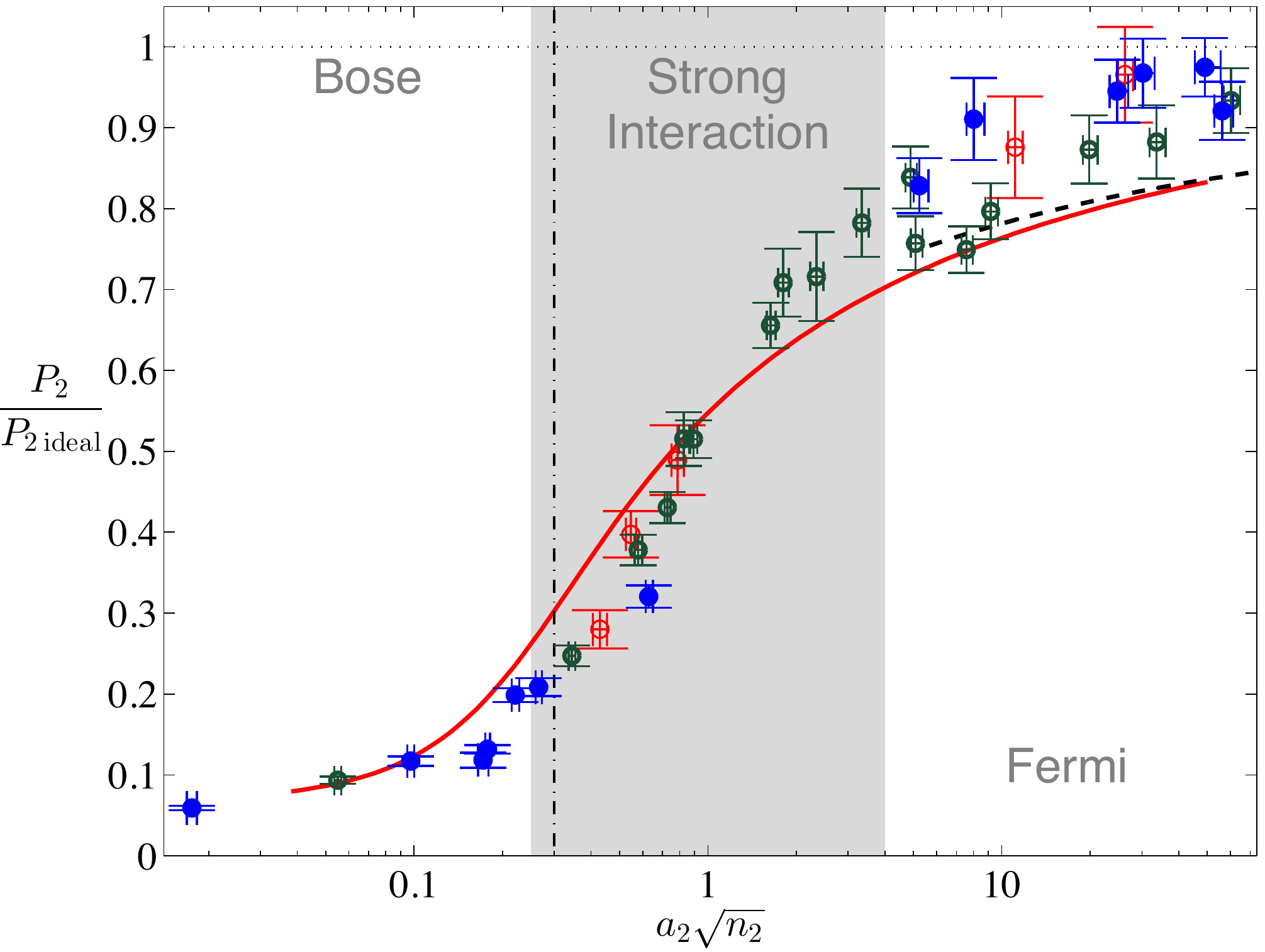}
\caption{The experimentally measured pressure $P_2$ in a quasi-2D
  Fermi gas \cite{Makhalov2014}. The pressure is scaled with respect
  to that of an ideal 2D Fermi gas, $P_{\rm 2 \, ideal}$.  Note that
  the experiment is not purely 2D since $\ef = 2\pi n_2/m \gtrsim
  0.5\op$, and $\eb > \op$ in the Bose limit. Therefore, the
  scattering length (denoted $a_2$) is not always simply related to
  $\eb$, and must be defined via the quasi-2D $T$ matrix.  The solid
  red curve is obtained from a fit to the 2D QMC
  data~\cite{Bertaina2011} while the dashed line approximately
  corresponds to the weak-coupling result in the Fermi
  regime. \newline\hspace{\textwidth} \tiny{Reprinted figure with
    permission from: V. Makhalov, K. Martiyanov, and A. Turlapov,
    \textit{Phys.\ Rev.\ Lett.} \textbf{112}, 045301 (2014). Copyright
    2014 by the American Physical Society.}
 \label{fig:pressure}}
\end{figure}

The pressure as a function of interaction has recently been measured
experimentally in a quasi-2D Fermi gas~\cite{Makhalov2014}.  As shown
in Fig~\ref{fig:pressure}, the comparison with the QMC prediction is
reasonable, aside from the BCS side of the crossover (or Fermi
regime), where the pressure is significantly higher. Indeed, this is
in the limit where the weak-coupling result becomes accurate and the
pressure should simply correspond to:
\begin{align}
  P \simeq \frac{(2 n_\up)^2 \pi}{2 m} \lba 1 - \frac{1}{\eta} +
  \frac{2 \mathcal{A}+1}{2 \eta^2} \rba .
\end{align}
However, the experiment was performed at finite temperature, while the
theory is for zero temperature. Indeed, a recent self-consistent
$T$-matrix (or Luttinger-Ward) calculation~\cite{Bauer2014} for the
normal state finds that the deviations in this regime are consistent
with a temperature of $T \simeq 0.15 T_F$, where $T_F =
\ef/k_B$. Another factor that complicates the analysis is the quasi-2D
nature of the gas. The experiment is never in the 2D limit since $\eb
>\op$ in the Bose regime and $\ef \gtrsim 0.5 \op$ throughout.  This
may account for the smaller but arguably more striking deviation from
the QMC prediction in the strongly interacting regime
(Fig~\ref{fig:pressure}). Here, the pressure is consistently below the
QMC curve, whereas one would generally expect the pressure to be
higher at finite temperature when $\kf\ad$ is fixed. However,
perturbing away from the 2D limit (which is equivalent to relaxing the
quasi-2D confinement) is expected to reduce the pressure in the plane
since the atoms can spread out in the transverse
direction~\cite{fischer2014}. This suggests that there are two
competing effects in the experiment: finite temperature tends to raise
the pressure, as evident in the Fermi regime, while quasi-2D effects
act to reduce it, particularly for strong interactions where both
$\eb/\op$ and $\ef/\op$ are sizeable.

\subsubsection{Contact}
Another important thermodynamic quantity is the contact density $C$,
which fixes the tail of the momentum distribution: $n(\k) \sim C/k^4$
as $k \to \infty$. This is related to the 2D equation of state by the
adiabatic theorem~\cite{Tan2008,Werner2012}
\begin{align}
  C = 2\pi m \frac{dE}{d\ln\ad} \> .
\end{align}
Since the contact determines the short-distance behavior of the gas
(i.e., it essentially gives the probability of finding a pair of $\up$
and $\down$ fermions close together), it can also be related to the
high-frequency and large-momentum limits of other correlation
functions in 2D such as the current response
function~\cite{Hofmann2011}.  Mean-field theory simply gives $C = m^2
\Delta^2$, which is consistent with the fact that $C$ should
monotonically increase with increasing attraction, and it yields the
correct two-body contact in the Bose limit.  However, the mean-field
result is not quantitatively accurate in the Fermi regime since it
does not capture the leading order dependence on the interaction, as
discussed in Sec.~\ref{sub:perturb}.

\begin{figure}
\centering 
\includegraphics[width=0.7\linewidth]{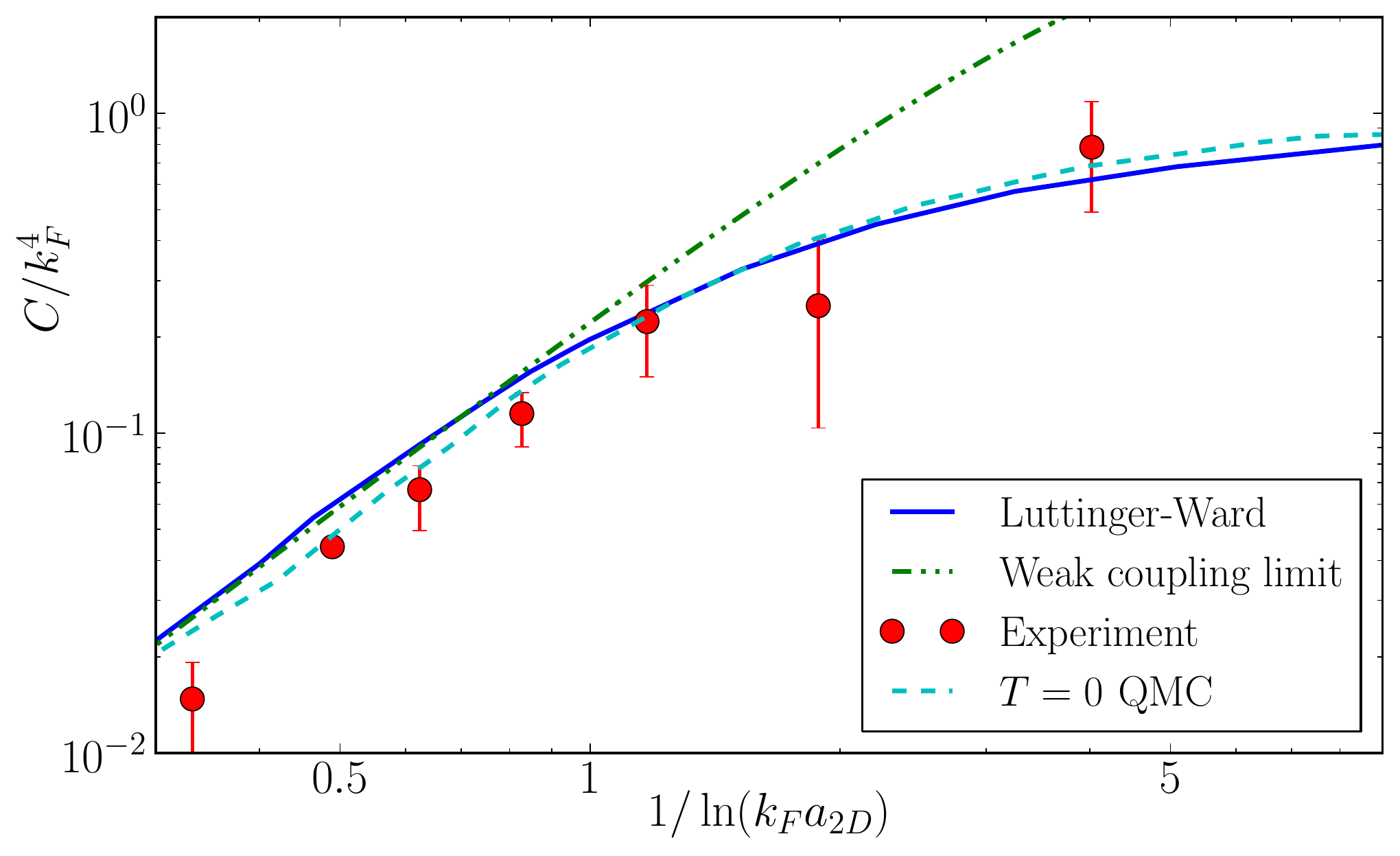}
\caption{Contact density $C$ in the regime $\ln(\kf\ad) > 0$. The
  filled circles correspond to the data in Ref.~\refcite{Frohlich2012}
  at $T/T_F = 0.27$, the dashed line is a fit to the 2D QMC
  result~\cite{Bertaina2011}, and the solid line is determined from a
  Luttinger-Ward, or self-consistent $T$-matrix, approach described in
  Ref.~\refcite{Bauer2014}. The $T=0$ weak-coupling limit is extracted
  from \req{eq:weak-erg}. The figure is adapted from
  Ref.~\refcite{Bauer2014}.
  \label{fig:contact}}
\end{figure}

The 2D contact may be experimentally determined from the
high-frequency tail of the RF spectrum~\cite{Langmack2012}, as well as
from the momentum profile. In contrast to the pressure, the contact
appears to be surprisingly insensitive to temperature in the
degenerate regime $T < T_F$. Figure~\ref{fig:contact} shows that there
is good agreement between the experimentally measured contact
density~\cite{Frohlich2012} at $T/T_F = 0.27$ and the $T=0$ QMC
result~\cite{Bertaina2011}. The contact determined using the
Luttinger-Ward approach at the same temperature ($T/T_F = 0.27$)
confirms that it is relatively unchanged for low
temperatures~\cite{Bauer2014}.
 
\subsection{The 2D crossover ``point'' \label{sub:crossover}}
Finally, we turn to the crossover from Fermi to Bose behavior in the
regime of strong interactions $|\ln(\kf\ad)| \lesssim 1$. From the
point of view of mean-field theory, the natural crossover point is at
$\mu=0$, since this marks a qualitative change in the quasiparticle
excitation spectrum $E_\k = \sqrt{(\ek -\mu)^2+\Delta^2}$. When
$\mu>0$, the minimum energy gap $\Delta$ occurs at finite momentum,
$k=\sqrt{2m \mu}$, corresponding to the remnants of a Fermi
surface. However, for $\mu<0$, the minimum gap occurs at $k=0$ and no
longer corresponds to $\Delta$. Indeed, for sufficiently strong
attraction, the gap in the single-particle spectrum becomes $-\eb/2$,
as expected.  Thus, the $\mu<0$ regime resembles the behavior of a gas
of bosonic dimers. According to mean-field theory, the point where
$\mu = 0$ corresponds to $\ln(\kf\ad) = 0$, and this is generally
viewed as playing a role analogous to the unitarity point $1/\as = 0$
in the 3D BCS-BEC crossover.\footnote{Note that the analogy between
  $1/a_s = 0$ in 3D and $\ln(\kf\ad) = 0$ in 2D is far from perfect,
  since we have $\mu > 0$ at $1/a_s=0$ and thus the unitarity point
  lies on the Fermi side of the crossover.}

However, if we go beyond mean-field theory and consider the more
accurate QMC calculations~\cite{Bertaina2011}, then we
find~\cite{Ngamp2013-2} that the point where $\mu=0$ instead occurs at
much weaker attraction, with $\ln(\kf\ad) \simeq 0.5$, as shown in
Fig.~\ref{fig:bertaina}. This suggests that the Fermi side of the
crossover occurs at larger $\ln(\kf\ad)$ than previously assumed, and
is consistent with the observation in QMC
simulations~\cite{Bertaina2011} that a variational wave function based
on dimers outperforms the one for a Fermi liquid once $\ln(\kf\ad)
\lesssim 1$.  Of course, it remains an open question how $\mu$ is
connected to the quasiparticle dispersion beyond the mean-field
approximation, but a negative $\mu$ already indicates strong
deviations from fermionic behavior.

\begin{figure}
\centering 
\includegraphics[width=0.7\linewidth]{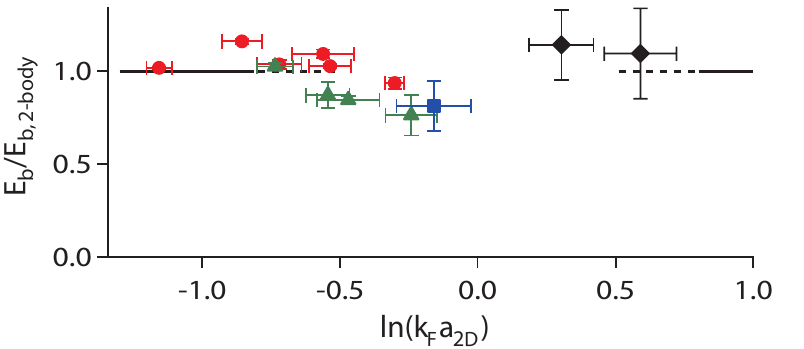}
\caption{Experimental measurement of the pairing gap $E_{b}$ in the 2D
  Fermi gas using RF spectroscopy, taken from
  Ref.~\refcite{Sommer2012}. The straight line corresponds to the
  mean-field result, which is simply the two-body binding energy
  (denoted here by $E_{b, {\rm 2\mhyphen body}}$).
  \newline\hspace{\textwidth} \tiny{Reprinted figure in (a) with
    permission from: A. T. Sommer, L. W. Cheuk, M. J. H. Ku,
    W. S. Bakr, and M. W. Zwierlein, \textit{Phys.\ Rev.\ Lett.}
    \textbf{108}, 045302 (2012). Copyright 2012 by the American
    Physical Society.}
\label{fig:pair-gap}}
\end{figure}

A recent experiment~\cite{Sommer2012} on pairing in the 2D Fermi gas
also suggests that the Bose regime extends beyond $\ln(\kf\ad) =
0$. Figure~\ref{fig:pair-gap} shows the pairing gap $E_{b} \equiv
E_{\k=\vect{0}} - \mu$, corresponding to the onset of the pairing peak
in the RF spectrum. At first glance, this measurement appears to
validate mean-field theory, which simply predicts $E_b = \eb$
throughout the crossover. However, this result is also expected for a
gas of dimers; thus an alternative explanation is that the experiment
only probes the Bose limit of the crossover, with $\ln(\kf\ad)
\lesssim 0.5$.

%% file: temp.tex
\section{Finite-temperature phenomenology}\label{sec:temp}

The behaviour of the 2D Fermi gas at finite temperature is even richer
than at zero temperature, since we have the possibility of superfluid
phase transitions and pairing without superfluidity. Above the
critical temperature $T_c$ for superfluidity, the normal phase is also
markedly different in the two perturbative limits, with a Fermi liquid
for $\ln(\kf\ad) \gg 1$ and a Bose liquid for $\ln(\kf\ad) \ll
-1$. This raises the question of whether the normal gas within the
crossover can display features intermediate between Fermi and Bose
behavior. In particular, there may exist a so-called ``pseudogap''
regime in the normal phase, where there is a suppression of spectral
weight at the Fermi surface that is reminiscent of a pairing
gap~\cite{Trivedi1995}. Such a phenomenon has been observed in the
quasi-2D cuprate superconductors, but its origin still remains a
mystery~\cite{Loktev2001}. By investigating its existence in
attractive Fermi gases, cold-atom experiments may help settle the
question of whether or not a pseudogap can be produced by pairing
alone, in principle.

In this section, we will review our current understanding of the
normal phase of the 2D Fermi gas, including the transition to
superfluidity at low temperatures. We will also briefly discuss how
the behavior is affected by the quasi-2D nature of the gas and the
in-plane trapping potential present in experiment. To simply the
equations, we set $\kb = 1$ in the following.

\subsection{Critical temperature for superfluid transition}
Two-dimensional gases are marginal in the sense that true long-range
order (i.e.\ condensation) only exists at $T=0$. Instead, the
superfluid phase at finite temperature exhibits quasi-long-range order
where the correlations decay algebraically~\cite{fisher1988}.
Increasing temperature further eventually results in a
Berezinskii-Kosterlitz-Thouless (BKT) transition to the normal
phase~\cite{fisher1988,holzmann2007,hadzibabic2006}.

\begin{figure}
\centering
\includegraphics[width=0.7\linewidth]{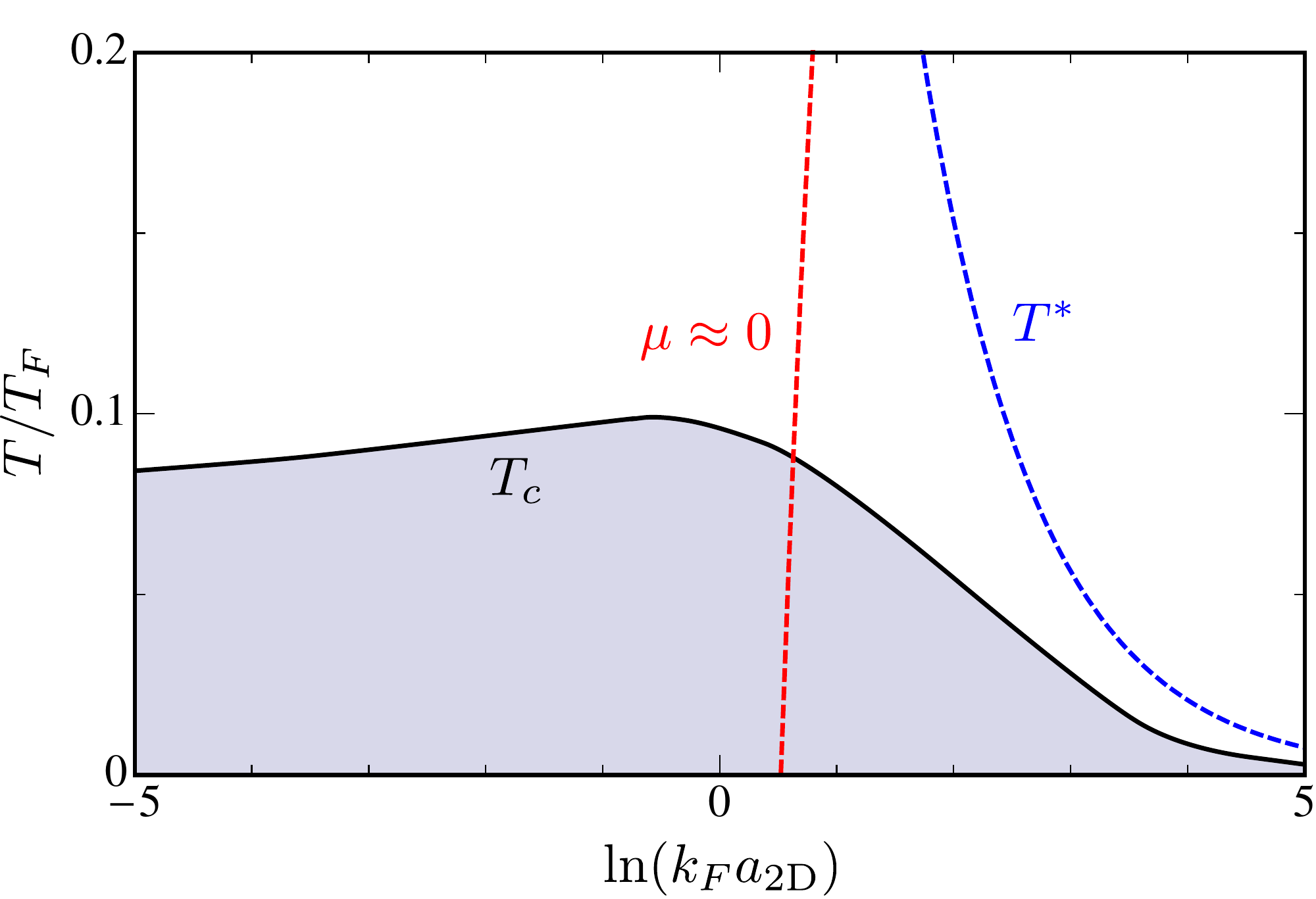}
\caption{Schematic 
  phase diagram throughout the BCS-Bose crossover.  The critical
  temperature for superfluidity is represented by the solid line, and
  corresponds to an interpolation between the known limits.  The
  dashed lines correspond to $\mu\approx 0$ and the onset of pairing
  $T^*$, which approximately bound the pseudogap region above
  $T_c$. The $\mu(T) \approx 0$ line is obtained by setting $T =
  \mu(0)$, while $T^*$ is estimated from the Thouless criterion
  \eqref{eq:thouless}.
\label{fig:sadestyle}}
\end{figure}

In the limit $\ln(\kf\ad) \ll -1$, the system corresponds to a weakly
interacting Bose gas and the BKT transition temperature
is~\cite{Petrov2003}:
\begin{align} \label{eq:thouless} \frac{T_c}{T_F} = \frac{1}{2} \lbc
  \ln\lba \frac{\mathcal{B}}{4\pi} \ln\lba \frac{4\pi}{\kf^2\ad^2}
  \rba \rba \rbc^{-1} ,
\end{align}
where $\mathcal{B} \simeq 380$. Note that $T_c \to 0$ in the limit
$\ln(\kf\ad) \to -\infty$, but since the dependence on $\ln(\kf\ad)$
is logarithmic, in practice we obtain $T_c/T_F \simeq 0.1$ for the
interaction regime accessible in experiment (see
Fig.~\ref{fig:sadestyle}).

In the BCS limit $\ln(\kf\ad) \gg 1$, the critical temperature is set
by the energy required to break pairs, which is the lowest energy
scale in the problem. Thus, one can estimate $T_c$ by taking the
mean-field Hamiltonian \eqref{eq-quasiham} and determining the point
at which $\Delta$ vanishes~\cite{BotS06}. From the resulting
linearized gap equation (or Thouless criterion) one
obtains~\cite{miyake83}
\begin{align}
\frac{T_c}{T_F} = \frac{2 e^{\gamma}}{\pi \kf\ad} .
\end{align}
A more thorough calculation that includes Gor’kov--Melik-Barkhudarov
corrections~\cite{Petrov2003} yields the BCS result above reduced by a
factor of $e$.

Referring to Fig.~\ref{fig:sadestyle}, we see that the results for
$T_c$ in the BCS and Bose limits can be smoothly interpolated,
suggesting that $T_c/T_F$ never exceeds 0.1. Note that $T_c$ has a
maximum in the regime $|\ln(\kf\ad)| <1$.  As yet, there is no
experimental observation of $T_c$ in the 2D Fermi gas.

\begin{figure}
\centering
\includegraphics[width=0.65\linewidth]{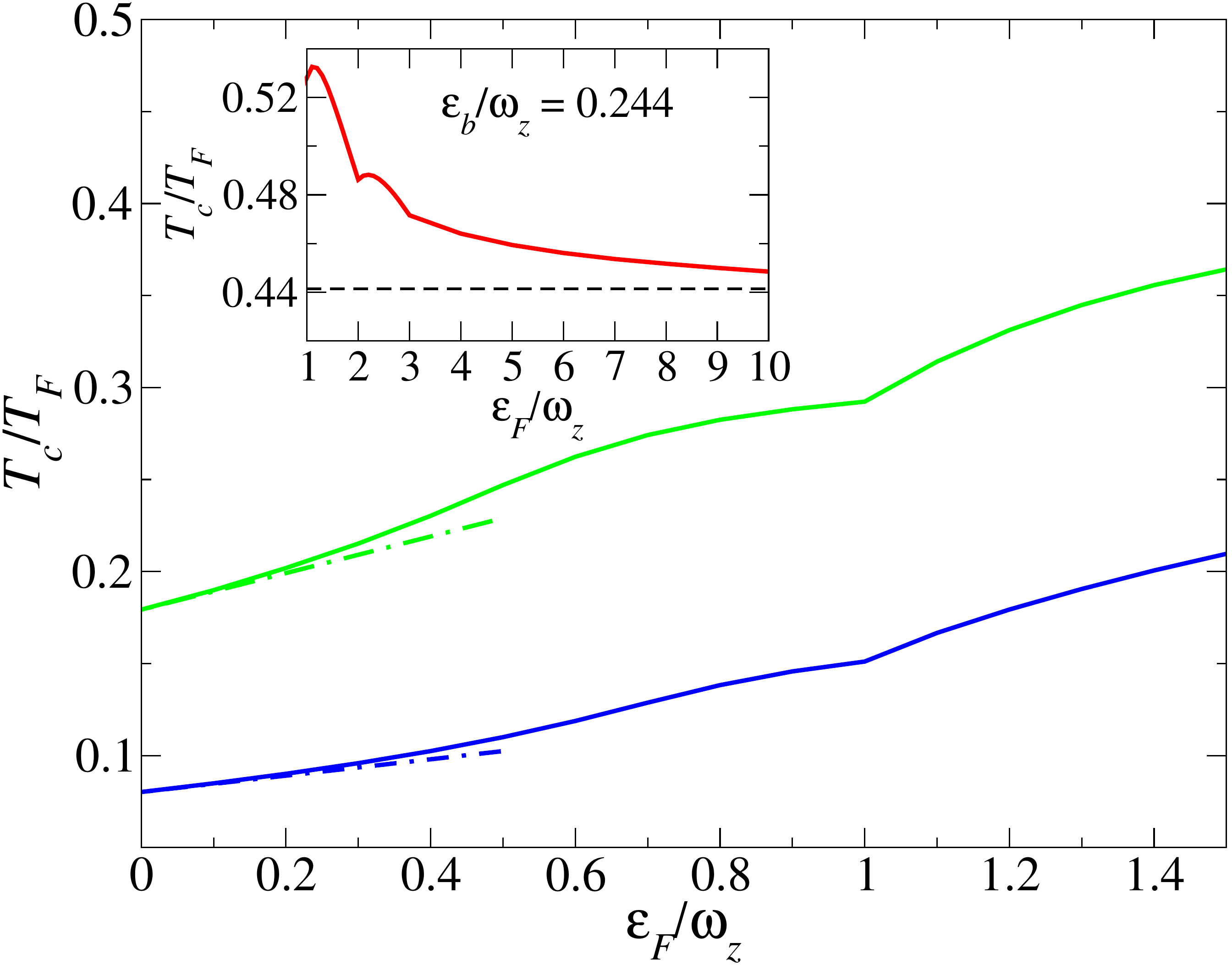}
\caption{Evolution of the mean-field critical temperature as the
  system is perturbed away from the 2D limit, taken from
  Ref.~\refcite{fischer2014}. The blue (bottom) and red (top) solid
  curves correspond to $\eb/\ef = 0.01$ and $\eb/\ef = 0.05$,
  respectively. The dash-dotted lines are the leading order behavior
  in $\ef/\op$.  Inset: The critical temperature at unitarity for
  large $\ef/\op$. The line tends towards the 3D result (dashed line)
  as $\ef/\op \to \infty$.
  \label{fig:q2d-tc}}
\end{figure}

\subsubsection{Quasi-2D case}
Given that experiments deal with quasi-2D Fermi gases, it is important
to understand the effect of a finite confinement length on $T_c$.
This is in general a challenging problem to address throughout the
BCS-Bose crossover, but it is possible to estimate the dependence on
$\ef/\op$ in the BCS limit. Using the mean-field approach for the
quasi-2D system described in Sec.~\ref{sub:MF}, one obtains a natural
generalization of the Thouless criterion to
quasi-2D:~\cite{fischer2014}
\begin{align}\label{eq:thouless-q2D}
  -\frac{1}{g}=\sum_{\vect{k}, n_1, n_2} (V_0^{n_1
    n_2})^2\frac{\tanh\left(\beta_c\xi_{ \mathbf{k} n_1 } /2
    \right)+\tanh\left(\beta_c\xi_{ \mathbf{k} n_2 } /2
    \right)}{2(\xi_{\vect{k} n_1}+\xi_{\vect{k} n_2})} ,
\end{align}
where $g$ is the 3D contact interaction, $\beta_c = 1/T_c$, and
$\xi_{\k n} = \epsilon_{\k n} - \mu$. Solving for $T_c$, we arrive at
the result plotted in Fig.~\ref{fig:q2d-tc}. While the mean-field
approach will overestimate $T_c$, it should be qualitatively accurate
in the BCS regime and we clearly see that $T_c/T_F$ increases as we
perturb away from 2D at fixed $\eb/\ef$. Indeed, the leading order
behavior in $\ef/\op$ is~\cite{fischer2014}
\begin{align}
  \frac{T_c}{T_F} = \frac{2 e^{\gamma}}{\pi \kf\ad} \lbc 1 +
  \frac{\ef}{\op} \ln \left(\frac{7+4 \sqrt{3}}{8}\right) \rbc .
\end{align}
This suggests that experiments will have a better chance of observing
$T_c$ and superfluidity if they are not purely 2D.  For intermediate
values of the confinement, we clearly see the presence of cusps at
integer values of $\ef/\op$, which correspond to discontinuities in
the density of states every time the Fermi energy crosses a harmonic
oscillator level.  In the limit $\op \to 0$, \req{eq:thouless-q2D}
yields the 3D expression for the Thouless criterion, as
expected.\footnote{The correct 3D expression is obtained by treating
  the confining potential in the $z$-direction within the local
  density approximation --- see, also, Sec.~\ref{sub:lda}.}  An
interesting possibility is that $T_c/T_F$ is maximized at intermediate
confinement strengths, where the geometry is between two and three
dimensions, but one would need to go beyond mean-field theory to
assess this.

\subsection{High temperature limit}

For high temperatures $T \gg T_F$, the gas is no longer quantum
degenerate and the behavior tends towards that of a classical
Boltzmann gas where the particle statistics are unimportant.  In this
limit, one may exploit the virial expansion described below, which has
the advantage of being a controlled approach at high temperatures
throughout the Fermi-Bose crossover.  As such, the virial expansion
can be used to investigate pairing phenomena at finite temperature and
thus provide a benchmark for both theory and experiment.

\subsubsection{Virial expansion}

In the following, we outline the basic idea of the virial expansion,
as applied to the uniform 2D Fermi gas.  Working in the grand
canonical ensemble, we define the virial coefficients $b_j$ such that
the grand potential $\Omega(T,\mu)$ is given by:
\begin{equation}
\Omega = -2T\lambda^{-2} \sum_{j\ge1} b_jz^j,
\label{eq:Pvirial1}
\end{equation}
where the thermal wavelength $\lambda = \sqrt{2\pi/mT}$, the fugacity
$z=e^{\beta \mu}$, and $\beta \equiv 1/T$.  In the high-temperature
limit, the thermodynamics of the system can be accurately described by
just the first few terms in the above power series. For a typical
Fermi gas, $z$ is the relevant expansion parameter, but this is not
the case in the Bose limit of the crossover where $k_F \ad \ll 1$.  In
this limit, $\mu \simeq -\eb/2$ at low temperatures so that $z \simeq
e^{-\beta \eb/2} \to 0$ as $T \to 0$, which naively suggests that the
virial expansion is valid at arbitrarily low temperatures.  However,
it can be shown that the coefficients $b_j$ also contain powers of
$e^{\beta \eb/2}$ that cancel the contribution from the binding energy
in $z$ when $j$ is even~\cite{Ngamp2013-2}. Thus, the relevant
expansion parameter in the Bose limit is instead $z^\text{(Bose)} =z
e^{\beta \eb/2}$, with corresponding coefficients $b_j^\text{(Bose)} =
e^{-j\beta \eb/2} b_j$.

The virial expansion effectively amounts to a cluster expansion,
whereby one determines the correlations between particles in a cluster
of a given size, and then increases the size of the cluster at each
order. For instance, $b_2$ only contains contributions from the one-
and two-body problems, $b_3$ further includes three-body scattering,
and so on.  As such, one can make use of the few-body results
described in Sec.~\ref{sec:few}.  For a recent review of the virial
expansion in cold gases, see Ref.~\refcite{LiuReview}.

\begin{figure}
\centering
\includegraphics[width=0.6\linewidth]{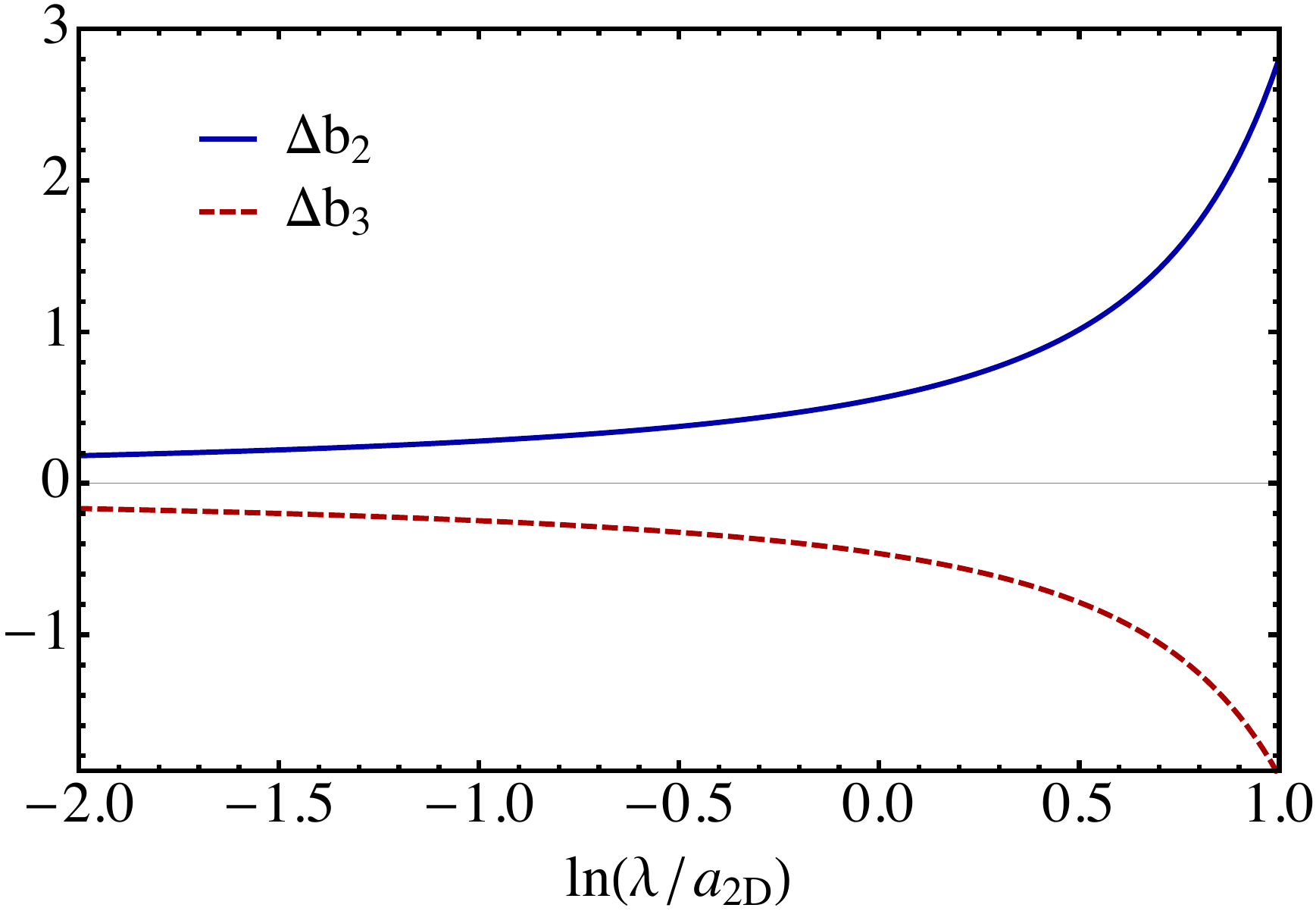}
\caption{The contribution from interactions to the second and third
  virial coefficients of the uniform 2D Fermi gas, taken from
  Ref.~\refcite{Ngamp2013-2}. The coefficients for the non-interacting
  gas are $b_j^{\rm (free)} = (-1)^{j-1} j^{-2}$ for $j\geq 1$, and
  $\Delta b_j \equiv b_j - b_j^{\rm (free)}$. Note that the virial
  coefficients are functions of $\ln(\lambda/\ad)$, or equivalently
  $\beta\eb$, only.  In the limit $\beta\eb \to \infty$, both $\Delta
  b_2$ and $\Delta b_3$ are dominated by the two-body bound state and
  thus they both go like $e^{\beta \eb}$ (but with different signs).
  \label{fig:virial}}
\end{figure}

The first calculation of the virial coefficients in a 2D Fermi gas was
for the trapped system~\cite{Liu2010}.  Indeed, one typically
determines each virial coefficient by solving the relevant few-body
problem in a harmonic trap. The coefficents for the trapped gas can be
straighforwardly mapped to those in the uniform case using the
relation: $b_j = j b_j^{\rm trap}$ (Ref.~\refcite{Ngamp2013-2}). The
lowest order coefficients are plotted in Fig.~\ref{fig:virial}. We see
that the correction to the second virial coefficient due to
interactions is attractive, as expected, since it lowers the grand
potential at fixed $\mu$ and $T$. However, this lowest order term is
expected to overestimate the attraction at lower temperatures and thus
the third-order correction acts to increase the energy.

One can also determine $b_j$ directly using a diagrammatic
approach~\cite{Leyronas2011} where the single-particle propagator $G$
is expanded in $z$, and this was first performed in 2D in
Ref.~\refcite{Ngamp2013-2}.  This approach also makes it
straightforward to determine the virial expansion for the spectral
function $A_\sigma(\k,\omega) = -2{\rm Im} G_\sigma(\k,\omega)$, which
is related to the probability of extracting an atom in state $\sigma$
with momentum $\k$ and frequency $\omega$. This allows one to
investigate pairing gaps in the spectrum at high temperature, as we
now discuss.

\subsection{Pseudogap}

The pseudogap regime is often synonymous with ``pairing above $T_c$''
in the cold-atom literature. However, such a scenario is trivially
achieved in a classical gas of diatomic molecules, where the gap in
the spectrum corresponds to the dimer binding energy.  To reproduce
the phenomenology of high-$T_c$ superconductors, one requires the
presence of a Fermi surface, since the pseudogap in these systems
manifests itself as a loss of spectral weight at the Fermi
surface~\cite{Loktev2001}. Indeed, it is not a priori obvious that
such a phenomenon can be replicated with an attractive Fermi gas: a
large attraction will surely lead to a pronounced pairing gap above
$T_c$, but it will also destroy the Fermi surface. It is therefore
reasonable to assume that any pseudogap regime must have $\mu >0$ in
addition to pairing, as schematically depicted in
Fig.~\ref{fig:sadestyle}.

\begin{figure}
\centering
\includegraphics[width=.98\linewidth]{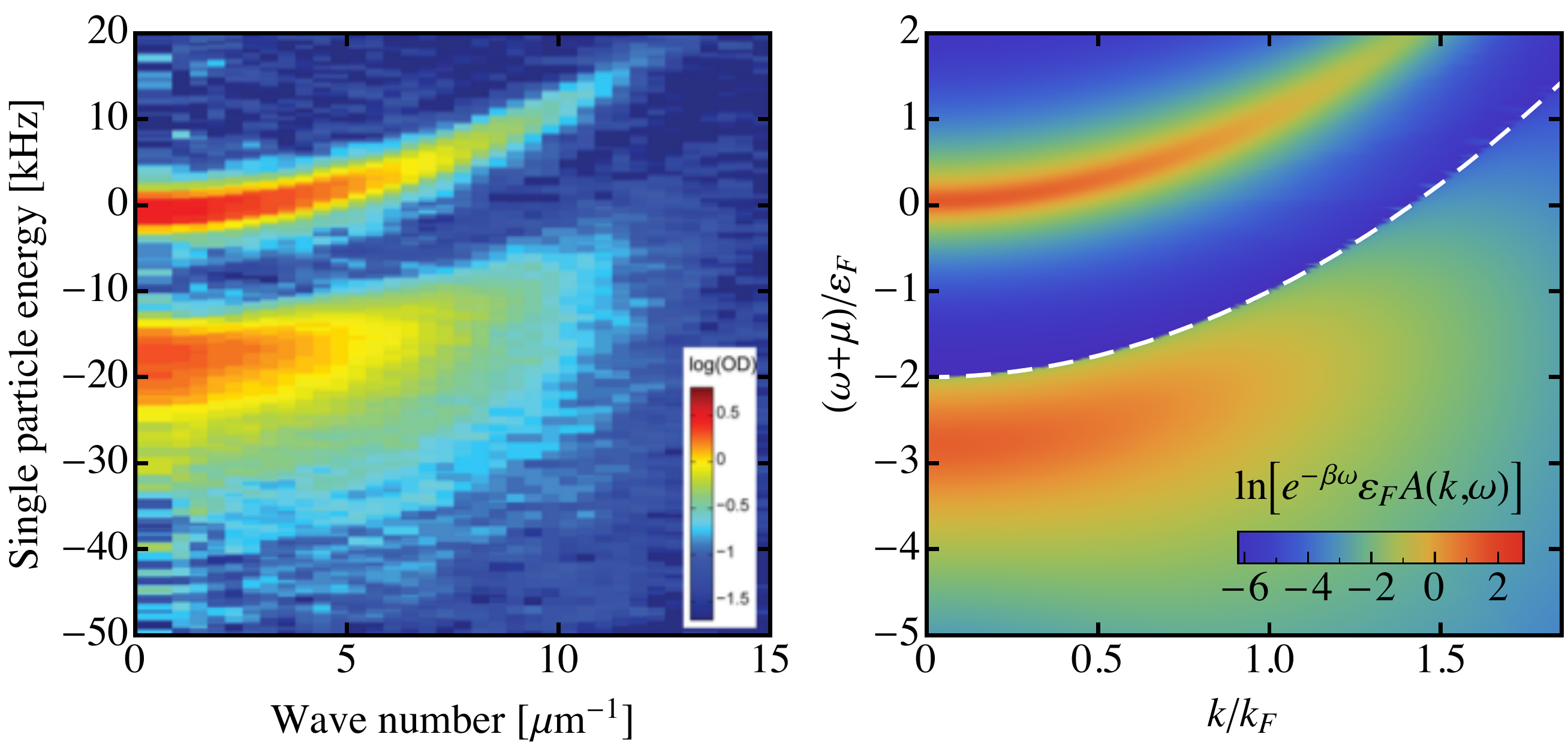}
\caption{The occupied part of the spectral function at $\ln(\kf\ad) =
  0$. (left) Measured momentum-resolved photoemission signal, taken
  from Ref.~\refcite{Feld2011}.  (right) Theoretical prediction at
  $T/T_F = 1$ from the virial expansion up to second
  order~\cite{Ngamp2013-2}.  The white dotted line marks the edge of
  the band of bound dimers (the incoherent part of the spectrum) and
  corresponds to the free atom dispersion shifted by the two-body
  binding energy, i.e., $\ek-\eb$. \newline\hspace{\textwidth} \tiny{
    (left) Reprinted by permission from Macmillan Publishers Ltd:
    \href{http://www.nature.com/nature/journal/v480/n7375/full/nature10627.html}
    {Nature {\bf 480}, 75 (2011)} copyright 2011.}
\label{fig:spectra}}
\end{figure}

The possibility of a pseudogap regime has been investigated in 3D
Fermi gases~\cite{he2005,chen2006,Gaebler2010,perali2011}, but its
existence is still under debate.  In 2D, the pseudogap regime is
expected to be much more pronounced than in 3D, since quantum
fluctuations suppress superfluid long-range order, and the system more
readily forms two-body bound states.  Already, a recent
measurement~\cite{Feld2011} of the spectral function in 2D has found
indications of a pairing gap above $T_c$.  However, a similar pairing
gap is found using the lowest order virial expansion of the spectral
function, which only includes two-body correlations, i.e., no Fermi
surface.~\cite{Ngamp2013-2,barth2014} In Fig.~\ref{fig:spectra}, the
agreement between experiment and theory suggests that the observed
pairing effectively arises from two-body physics only and therefore
does not correspond to a pseudogap. Furthermore, most of the
experimental measurments of the spectral function were apparently
performed in the regime where $\mu < 0$ (see Fig.~\ref{fig:sadestyle}
and Ref.~\refcite{Ngamp2013-2}).  Thus, it is likely that lower
temperatures and lower attraction are required to observe a pseudogap.
In particular, both non-self-consistent~\cite{Ohashi2013} and
self-consistent~\cite{Bauer2014} $T$-matrix approximations predict the
existence of a pseudogap in the regime $T/T_F \lesssim 0.2$ for
$\ln(\kf\ad) \simeq 1$.

\subsection{Equation of state in a trapped gas} \label{sub:lda}

The fact that the interaction parameter $\ln(\kf\ad)$ can be tuned by
varying the density has important consequences for the trapped 2D
gas. Specifically, it implies that $\ln(\kf\ad)$ decreases as we move
from the high-density region at the center of the trapped gas to the
low-density region at the edge.  Thus, we can in principle observe the
entire Fermi-Bose crossover in a single experiment.  This argument
relies on the local density approximation (LDA), where the in-plane
trapping potential can be incorporated into the chemical potential,
$\mu(r) = \mu - V(r)$, and thus each point in the trap corresponds to
a different $\ln\lba\kf(r)\ad\rba$.  For a harmonic potential with
frequency $\operp$, we require $\operp\ll T, \ef$ in order for LDA to
be valid.

\begin{figure}
\centering
\includegraphics[width=0.65\linewidth]{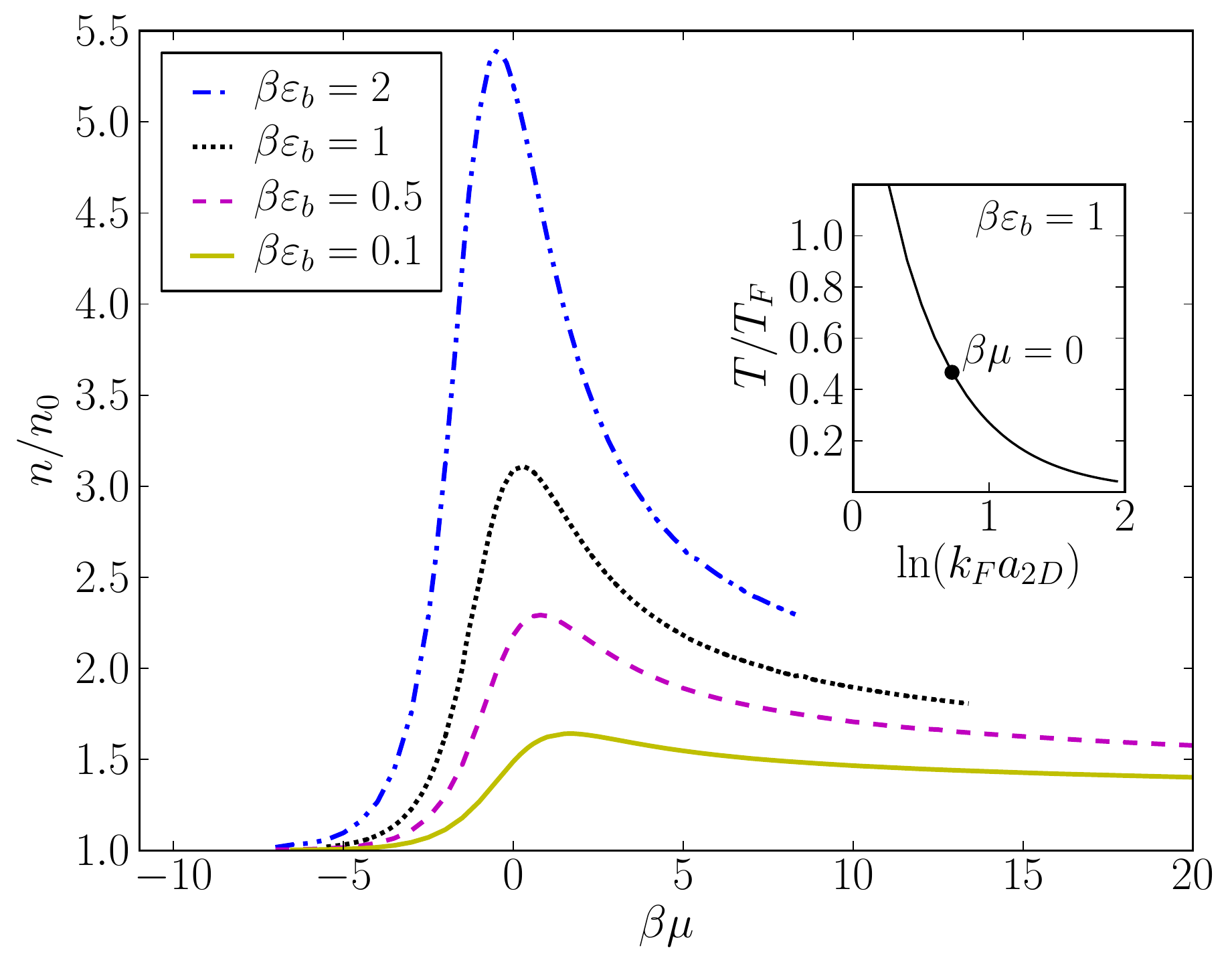}
\caption{The equation of state for the density at finite temperature,
  taken from Ref.~\refcite{Bauer2014}.  The density $n$ is normalized
  by $n_0(\beta\mu)$, the density of the non-interacting Fermi gas.
  The curves for large $\beta\eb$ are shown up to the critical value
  $\mu_c(\beta\eb)$ where the system is expected to enter the BKT
  phase. The inset shows a typical trajectory corresponding to fixed
  $\beta\eb$ in the phase space of $T/T_F$ versus $\ln(k_F\ad)$.
  Along this line, $\beta\mu$ increases with decreasing $T/T_F$.
 \label{fig:density}}
\end{figure}

One can make a direct connection with trapped-gas experiments by
considering the density $n(\beta\mu,\beta\eb)$ as a function of $\beta
\mu$ for different values of the interaction parameter $\beta \eb$
(see Fig.~\ref{fig:density}). Such an equation of state can be
straightforwardly extracted from the measured density profile in a
trap~\cite{ku12}.  To reveal the effects of interactions, we normalize
the density $n$ by that of the ideal Fermi gas, $n_0 = 2
\ln(1+e^{\beta \mu})/ \lambda^2$. In the high-temperature
(low-density) limit where $\beta \mu \to -\infty$, the behaviour
approaches that of an ideal Boltzmann gas, as expected.  However, with
decreasing temperature, $n/n_0$ eventually exhibits a maximum around
$\beta\mu \simeq 0$, implying that interactions are strongest at
intermediate rather than low temperatures.  This results from the fact
that decreasing $T/T_F$ at fixed $\beta\eb$ corresponds to an
increasing $\ln(k_F\ad)$.  Thus, we likewise expect the system to
approach a weakly interacting gas in the low temperature regime.  This
behavior is qualitatively different from that observed in 3D
\cite{ku12}, and is a direct consequence of the fact that one can
traverse the Fermi-Bose crossover in 2D by only varying the density.

%% file: polaron.tex
\section{The 2D polaron problem}\label{sec:polarized}

The properties of an impurity immersed in a quantum-mechanical medium
constitutes a fundamental problem in many-body physics. A classic
example in the solid state is the Fr{\"o}hlich polaron, an electron
moving in a crystal and interacting with the resulting bosonic lattice
vibrations. Due to the interactions, the system of impurity plus
lattice vibrations is better described in terms of a quasiparticle,
the polaron, which has modified effective mass, chemical potential,
charge, etc., compared with the free electron. The quasiparticle thus
encompasses both the electron and the cloud of excitations of the
medium.

In the context of two-component Fermi gases, the spin components may
be imbalanced straightforwardly, leading naturally to a polaron
problem in the limit of a large spin polarization, i.e., the problem
of a single spin-down impurity.  However, in contrast to the case of
the Fr{\"o}hlich polaron, the medium is now fermionic, and this can
strongly modify the character of the impurity quasiparticle, as we
discuss below. Furthermore, the properties of the polaron will
directly impact the topology of the whole phase diagram for the
spin-imbalanced Fermi gas.  It is well known that BCS pairing is very
sensitive to mismatched Fermi surfaces, and such a spin imbalance can
thus lead to more exotic superfluid phases.  For instance, the
formation of Cooper pairs at finite momentum may occur, giving rise to
the so-called Fulde-Ferrell-Larkin-Ovchinnikov (FFLO)
state~\cite{Fulde1964,larkin1965}.  For sufficiently large spin
imbalance, the system encounters the Chandrasekhar-Clogston limit and
ceases to display paired-fermion superfluidity. This limit has
recently been experimentally investigated in the strongly interacting
Fermi gas in both
3D\cite{Zwierlein2006,partridge2006,shin2008,Nascimbene} and 1D
\cite{Liao2010}, but it remains to be seen how the breakdown of
superfluidity occurs in the 2D Fermi gas.  For a further discussion of
the polarized Fermi gas in 3D, we refer the reader to, e.g., the
reviews of Refs.~\refcite{ChevyReview} and \refcite{sheehy2007}.

\begin{figure}
\centering
\includegraphics[width=0.7\linewidth]{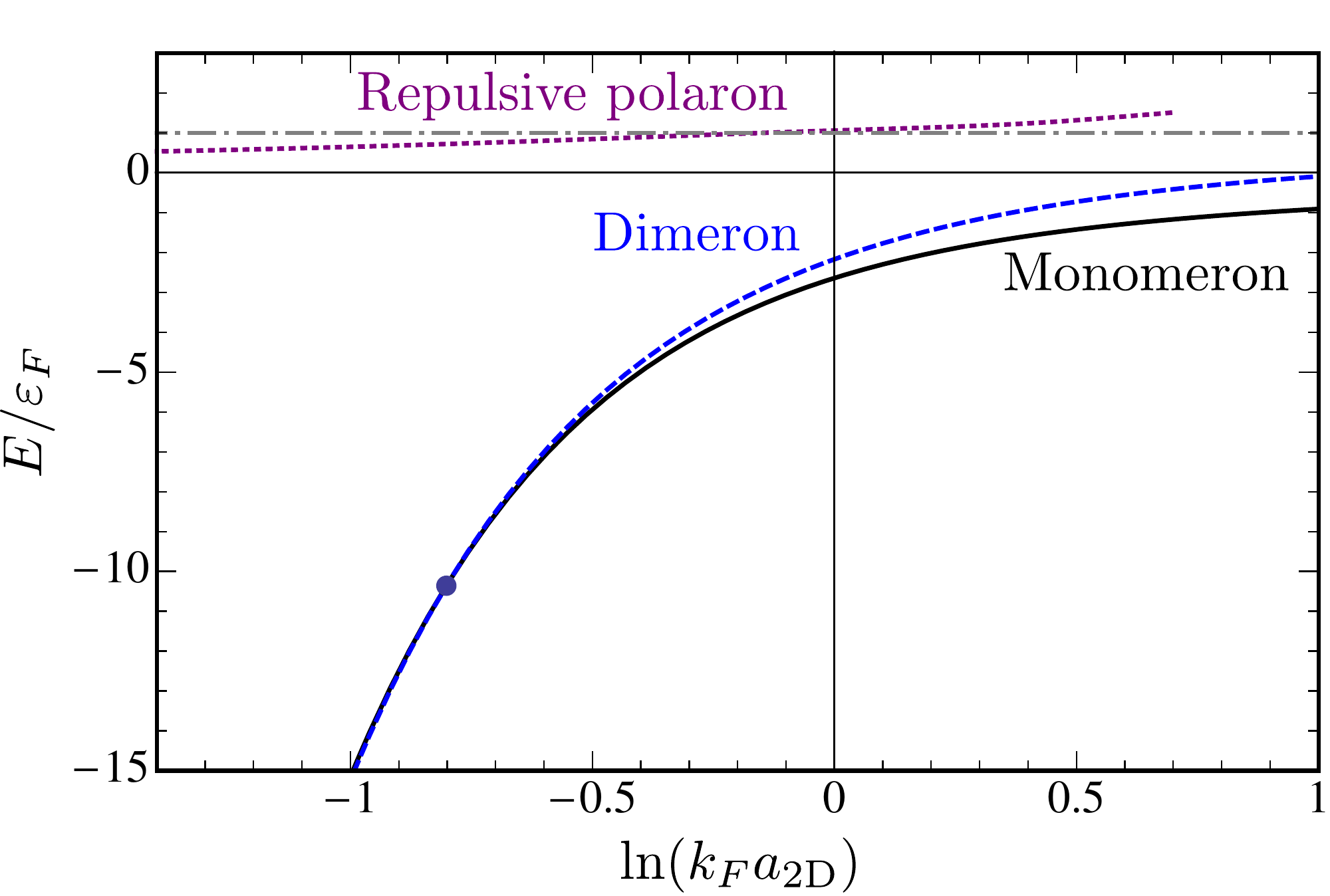}
\caption{The relevant quasiparticle branches in 2D for
  $m_\up=m_\down$: The monomeron (black, solid)\cite{Zollner2011}, the
  dimeron (blue, dashed)\cite{Parish2011}, and the repulsive polaron
  (purple, dotted)\cite{Schmidt2012,Ngamp2012}.  The filled circle
  marks the monomeron-dimeron transition in the ground state.  The
  dot-dashed marks the Fermi energy.  All quasiparticle energies
  displayed follow from variational wavefunctions limited to one
  particle-hole pair excitation (see text). \label{fig:polarons}}
\end{figure}

An important question concerns the nature of the ground state of a
spin-down impurity atom in a spin-up Fermi sea.  For weak attractive
interactions, the quasiparticle has properties similar to that of the
bare impurity and will be termed the ``monomeron''.\footnote{In this
  work we use the terminology monomeron, dimeron, trimeron, and
  tetrameron to denote the impurity bound to 0, 1, 2, and 3 majority
  atoms, respectively, in the presence of interactions with the Fermi
  sea. This replaces previous terminology (attractive polaron,
  molecule, dressed trimer, and dressed tetramer, respectively). As
  there is only one repulsive branch (see below) this is referred to
  as the repulsive polaron.}  However, as the interaction strength is
increased, the impurity can bind a majority particle to form a
two-body bound state dressed by particle-hole fluctuations of the
Fermi sea\cite{prokofiev2008,prokofiev2008_2,bruun2010}. This is
illustrated in Fig.~\ref{fig:polarons}, which shows the quasiparticle
branches for equal masses. Interestingly, in the fermionic problem,
the impurity can undergo a sharp transition in the ground state and
effectively change its statistics by binding fermions from the
majority fermions, an effect absent in the classic Fr{\"o}hlich
polaron example above.  Quasiparticles in a 2D Fermi gas have been
investigated in two experiments: Fermi polarons have been
observed\cite{Koschorreck2012}, while radio-frequency spectra of the
unpolarized Fermi gas have been interpreted in terms of
monomerons\cite{Zhang2012}.

The polaron problem also has relevance to the phenomenon of {\em
  itinerant ferromagnetism}. Here, a two-component equal-mass Fermi
gas with repulsive short-range interactions is predicted to
spontaneously undergo a transition to spin-polarized domains for
sufficiently strong repulsion.  This classic Stoner transition
received renewed interest when its observation was reported in a
recent MIT experiment on the repulsive branch in 3D Fermi
gases~\cite{MIT}.  Subsequently, however, it was shown that the
experiment had instead only observed a fast decay into
pairs~\cite{Pekker2011}; this realization led to the assertion that
the Fermi gas with strong short-range repulsive interactions can never
undergo a ferromagnetic transition\cite{MIT2}.  The central issue is
that strongly repulsive interactions can only be truly short-ranged if
the underlying potential is attractive, and thus any magnetic phase in
such a system will be metastable at best. As we describe below, the
properties of the repulsive polaron (see Fig.~\ref{fig:polarons}) are
crucial for determining whether saturated ferromagnetism may exist in
the 2D Fermi gas, and it, in fact, appears that the fast decay into
the attractive branch also precludes saturated ferromagnetism in
2D\cite{Ngamp2012}. For a recent review on polaron physics in
ultracold gases with an emphasis on the relation to itinerant
ferromagnetism, we refer the reader to Ref.~\refcite{Massignan2013}.

\subsection{Variational approach}

An intuitive way to describe the Fermi polaron theoretically is
through variational wavefunctions. The simplest is Chevy's
ansatz~\cite{chevy2006_2}:
\begin{align}
  \ket{P} & = \alpha_{0}^{(\vect{p})}
  c^\dag_{\vect{p}\downarrow} \fs
  + \sum_{\vect{k}\vect{q}} \alpha_{\vect{k}\vect{q}}^{(\vect{p})}
  c^\dag_{\vect{p}+\vect{q} - \vect{k}\downarrow}
  c^\dag_{\vect{k}\uparrow} c_{\vect{q}\uparrow} \fs.
\label{eq:p3}
\end{align}
Here and in the following we assume that $|\k|>\kf$ ($|\q|<\kf$)
describes a particle (hole). For simplicity, we define $\kf$ as the
Fermi momentum of the spin-$\up$ atoms. The wavefunction describes the
spin-$\down$ impurity as a quasiparticle at momentum $\p$ using two
terms: the first is simply the bare impurity on top of the
non-interacting majority Fermi sea, denoted by $\fs$, while the second
incorporates how the impurity can distort the Fermi sea by exciting a
particle out of it, leaving a hole behind.

The energy of the polaron state is obtained by minimizing the
expectation value $\bra{P}{\cal H}-E\ket{P}$ with respect to the
variational parameters $\alpha_0^{(\p)}$ and $\alpha_{\k\q}^{(\p)}$,
where ${\cal H}$ is the 2D Hamiltonian \eqref{eq:2dham}. This yields the
equation
\begin{equation}
  E-\epsilon_{\p\down}=\sum_\q\left[\frac1
g-\sum_\k\frac{1}{E-\ekup+\equp-\epsilon_{\p+\q-\k\down}}\right]^{-1}.
\label{eq:epol}
\end{equation}
Formally, the variational approach as introduced here only admits one
solution: the monomeron \cite{chevy2006_2}, which has energy less than
the impurity in vacuum.  However, the variational approach may be
extended to include metastable states where the energy is allowed to
have a finite imaginary part --- see Ref.~\refcite{Parish2013}.  In
this case, one also obtains a second solution, the ``repulsive
polaron'' \cite{Cui2010,Massignan2011}, which has an energy $E_{\rm
  rep}$ exceeding that of the impurity in vacuum and potentially even
exceeding the Fermi energy for strong interactions. The wavefunction
\eqref{eq:p3} may straightforwardly be extended by considering further
excitations; however the present approximation of one particle-hole
pair excitation gives a surprisingly good estimate of the energy and
the residue $Z=\left| \alpha_{0}^{(\vect{p})} \right|^2$.  This is due
to an approximate cancellation of higher order terms in the expansion
in particle-hole pairs~\cite{combescot2008}. A recent work in 3D has
demonstrated an impressive agreement between the variational approach
and experiment~\cite{InnsbruckPolaron}.

In addition to the states described by \req{eq:p3}, the impurity may
also (depending on the $\up$-$\down$ mass ratio) form dimeron,
trimeron, and tetrameron states by binding one or several majority
particles, in a natural analogy to the possible vacuum bound states
such as the dimer, trimer, and tetramer described in
Sec.~\ref{sec:few}. Remarkably, these states may be the ground states
even when they do not bind in vacuum. The variational wavefunctions
for such states can be generated in a similar fashion to \req{eq:p3}
above, but rather than displaying them here, we instead refer the
reader to the original works on the
dimeron\cite{combescot2009,punk2009,mora2009,Parish2011} and
trimeron\cite{Mathy2011,Parish2013}.

\subsection{The repulsive polaron and itinerant ferromagnetism}

Following the observation that recombination processes preclude
itinerant ferromagnetism in the 3D atomic Fermi gas,\cite{MIT2} it is
pertinent to ask the question whether the Stoner transition can take
place in a 2D Fermi gas.\cite{Conduit2010} The main difference between
the 2D and 3D Fermi gases with short-range interactions is that in 3D
the vacuum two-body bound state appears in the regime of strongest
interactions, $1/\kf\as=0$, whereas in 2D, the bound state only
approaches the continuum in the limit of weak attraction.  Thus, one
may speculate that the pairing mechanism that prevented the appearance
of itinerant ferromagnetism in 3D could be suppressed. Indeed, the
three-body recombination mechanism by which three atoms recombine into
an atom and a dimer takes completely different forms in
3D\cite{Petrov2002} and in 2D\cite{Ngamp2013}.  However, despite this
difference, the decay into the attractive branch is still strong
enough to exclude fully polarized itinerant ferromagnetism, as we now
discuss.

\begin{figure}
\centering
\includegraphics[width=0.7\linewidth]{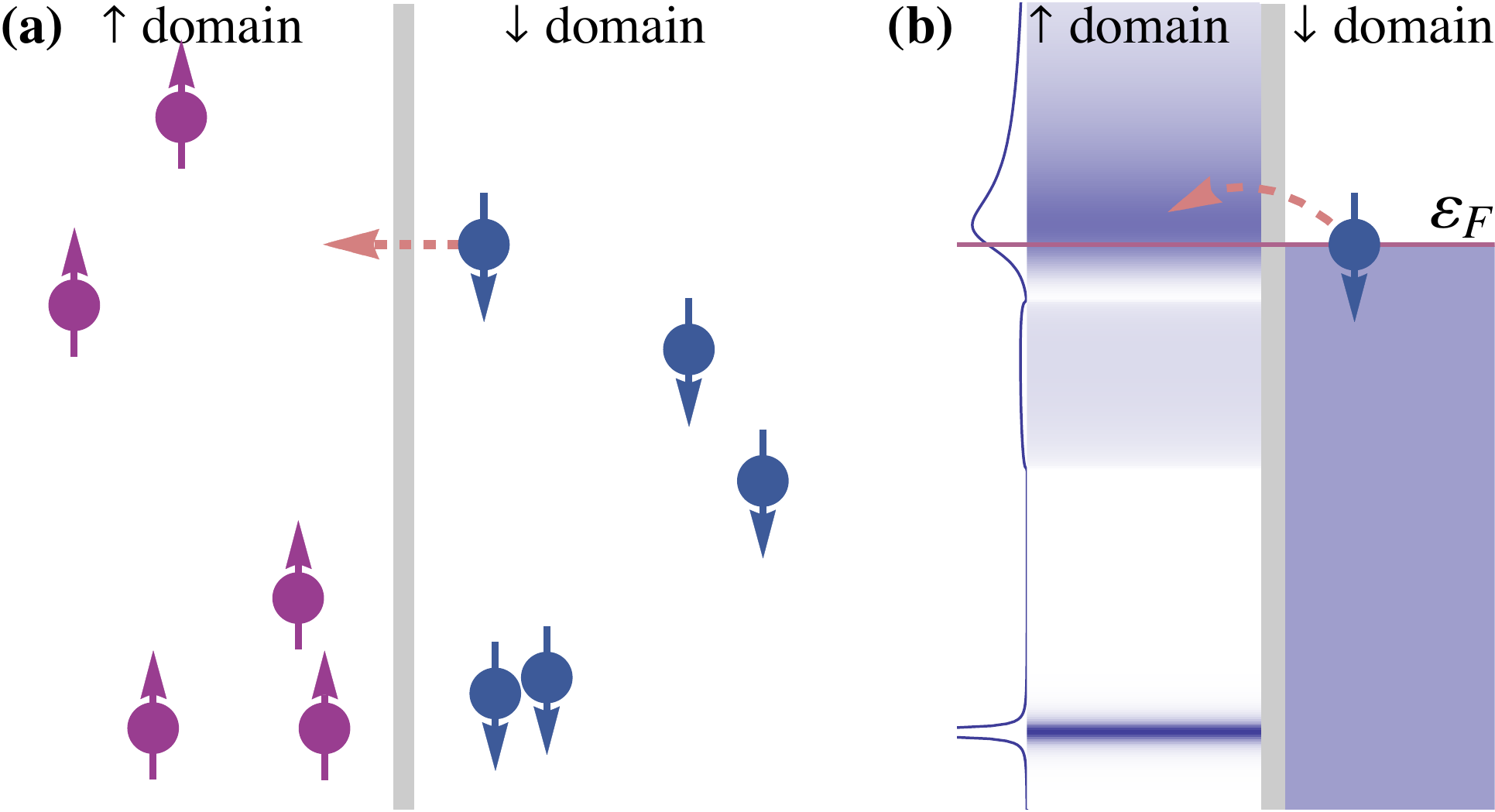}
\caption{Illustration of the stability condition for stable
  spin-polarized domains, taken from Ref.~\refcite{NgampThesis}.  (a)
  A spin-$\down$ atom can tunnel across the interface and become an
  impurity in the spin-$\up$ domain. (b) Density plot of the energy
  levels available to the fermion at $\ln(\kf\ad)=0.5$. The spectral
  function at $\k=\0$ of the impurity in the $\up$ domain is evaluated
  in the one particle-hole pair dressing approximation.
  \label{fig:domains}}
\end{figure}

Following Ref.~\refcite{Ngamp2012}, we investigate the stability of
{\em fully polarized} domains. To preserve $SU(2)$ symmetry and make a
direct connection with ferromagnetism, we confine the discussion to
equal-mass fermions. The fully polarized domains are illustrated in
Fig.~\ref{fig:domains}(a), and the central question is whether there
is an energy cost associated with moving a spin-$\down$ atom from its
domain to that of the spin-$\up$ atoms.  Assuming purely repulsive
interactions, this is the case if the energy of the dressed impurity
exceeds the Fermi energy in the spin-$\up$ region.  If one further
assumes mechanical equilibrium, where the pressures of the domains are
equal, then the Fermi energies $\ef^\up=\ef^\down$. Thus, referring to
Fig.~\ref{fig:polarons} and assuming that the impurity would tunnel
into the repulsive polaron state, the domains appear mechanically
stable if $\ln(\kf\ad)>-0.15$ and one concludes that itinerant
ferromagnetism is possible.

However, we must consider two other effects: The first is the finite
lifetime of the repulsive polaron, as the quasiparticle decay rate is
predicted to be a significant fraction of the Fermi energy in the
strongly interacting regime\cite{Ngamp2012} (the decay rate may also
be investigated as a pairing instability --- see
Ref.~\refcite{Pietila2012b}). This in turn leads to a large
uncertainty in the energy of the repulsive state, allowing atoms to
tunnel across the interface and depolarize the domains. Eventually, in
the weakly interacting regime $\ln(\kf\ad) \gg1$, the quasiparticle
decay rate becomes suppressed; however, as the tunneling probability
is proportional to the residue $Z$ of the corresponding quasiparticle,
and the residue of the repulsive branch is strongly suppressed in this
regime,\cite{Schmidt2012,Ngamp2012} the atoms will tunnel directly
into the attractive branch.  Combining the knowledge of the residue
and the lifetime of the repulsive polaron allows one to conclude that
even if spin polarized domains were to be artificially created, these
would not be dynamically stable\cite{Ngamp2012}.

The repulsive polaron has been observed in a recent
experiment~\cite{Koschorreck2012} and, in accordance with the theory,
no ferromagnetic transition was observed. In fact, the
experiment\footnote{In the experiment, the 2D scattering length was
  taken directly from the quasi-2D dimer binding energy, i.e.,
  $\ad^*=1/\sqrt{m\eb}$.  The relation between the present definition
  of $\ad$ and the one used in experiment is: $\ad=
  \ad^*\sqrt{\frac{\pi}{B}\frac{\eb}{\op}}e^{-\sqrt{\frac\pi2}{\mathcal
      F}_0(\eb/\op)}$, where ${\mathcal F}_0$ was introduced in
  Eq.~\eqref{eq:Fcali}. As argued in Ref.~\refcite{Levinsen2012}, the
  convention used for $\ad$ in this review yields a better agreement
  between the results of the quasi-2D experiments and the strict 2D
  theory presented here.\label{foot:a2d}} was limited to the regime
$-2.5<\ln(\kf\ad)<-1.3$, i.e., away from the limit where the
variational approach predicts $E_{\rm rep}>\ef$. In this regime, it
may be expected that $E_{\rm rep}\gtrsim0.3\ef$ (see
Fig.~\ref{fig:polarons}), whereas the experimentally observed energies
ranged from $10\%$ to $20\%$ of the Fermi energy. The discrepancy may
in part be due to the trap averaging\cite{Schmidt2012} and finite
temperature effects. However, in agreement with the
theory\cite{Ngamp2012}, the lifetime was severely suppressed,
preventing the detection of a coherent repulsive quasiparticle for
stronger interactions.

\subsection{Ground state of an impurity in a 2D Fermi gas}

In the following, we initially focus on the equal-mass case.  For a
single impurity attractively interacting with a 3D Fermi gas of
identical atoms, the existence of a sharp quasiparticle transition
from the monomeron to the dimeron state has been
predicted\cite{prokofiev2008,prokofiev2008_2,bruun2010}.  Such a
transition was recently observed experimentally for a finite density
of impurities~\cite{schirotzek2009}.  On the other hand, in the 1D
case, the exact Bethe ansatz solution\cite{McGuire1966} implies that
no such transition takes place. It is therefore natural to ask whether
a transition in the ground state occurs for an impurity in a 2D Fermi
gas, where quantum fluctuations are expected to be stronger than in
3D. The existence or otherwise of such a transition will impact the
overall phase diagram for the spin-imbalanced 2D Fermi
gas~\cite{He2008,conduit2008,Tempere2009,yin2014}.

\begin{figure}
\centering
\includegraphics[width=.49\linewidth]{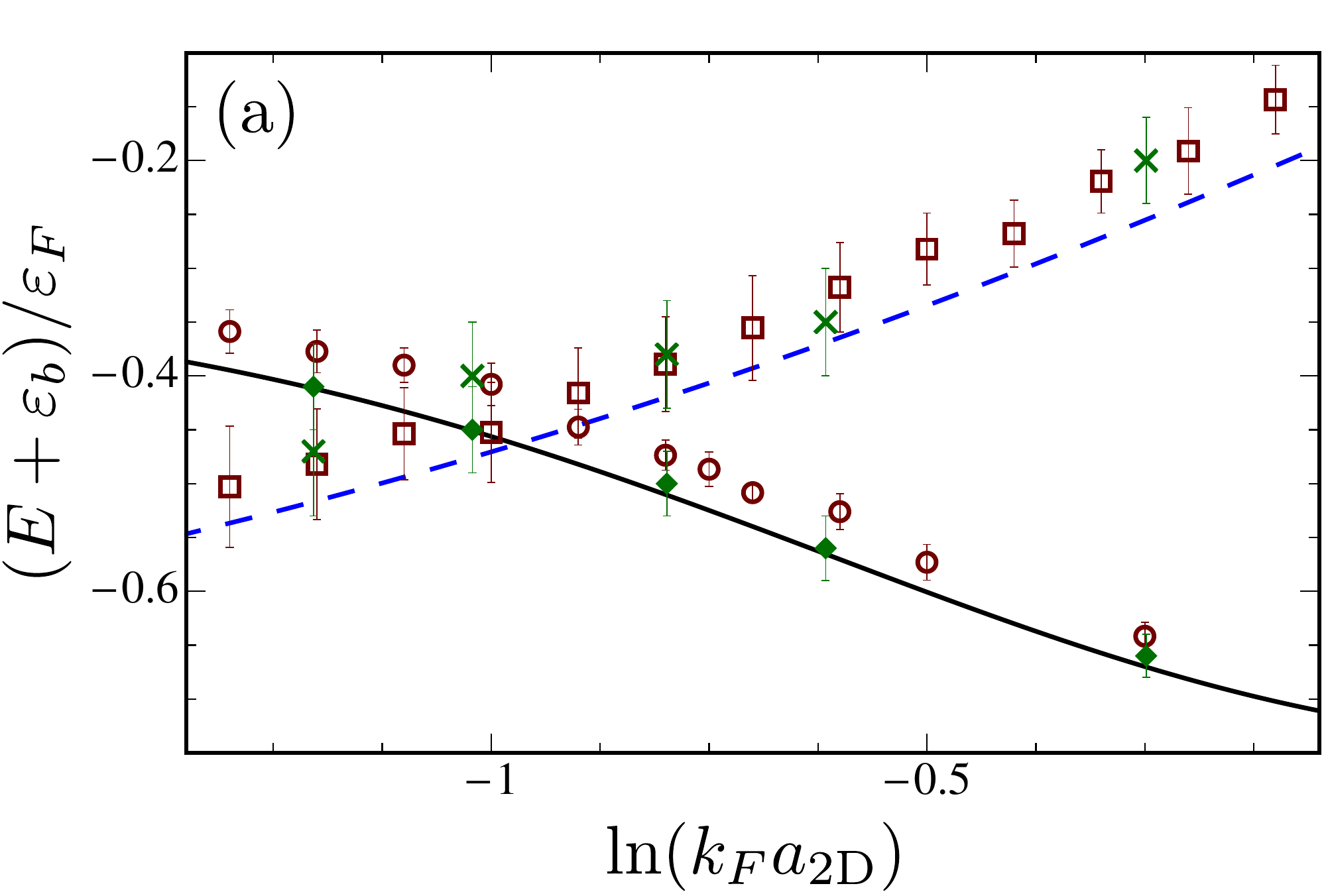}
\includegraphics[width=.49\linewidth]{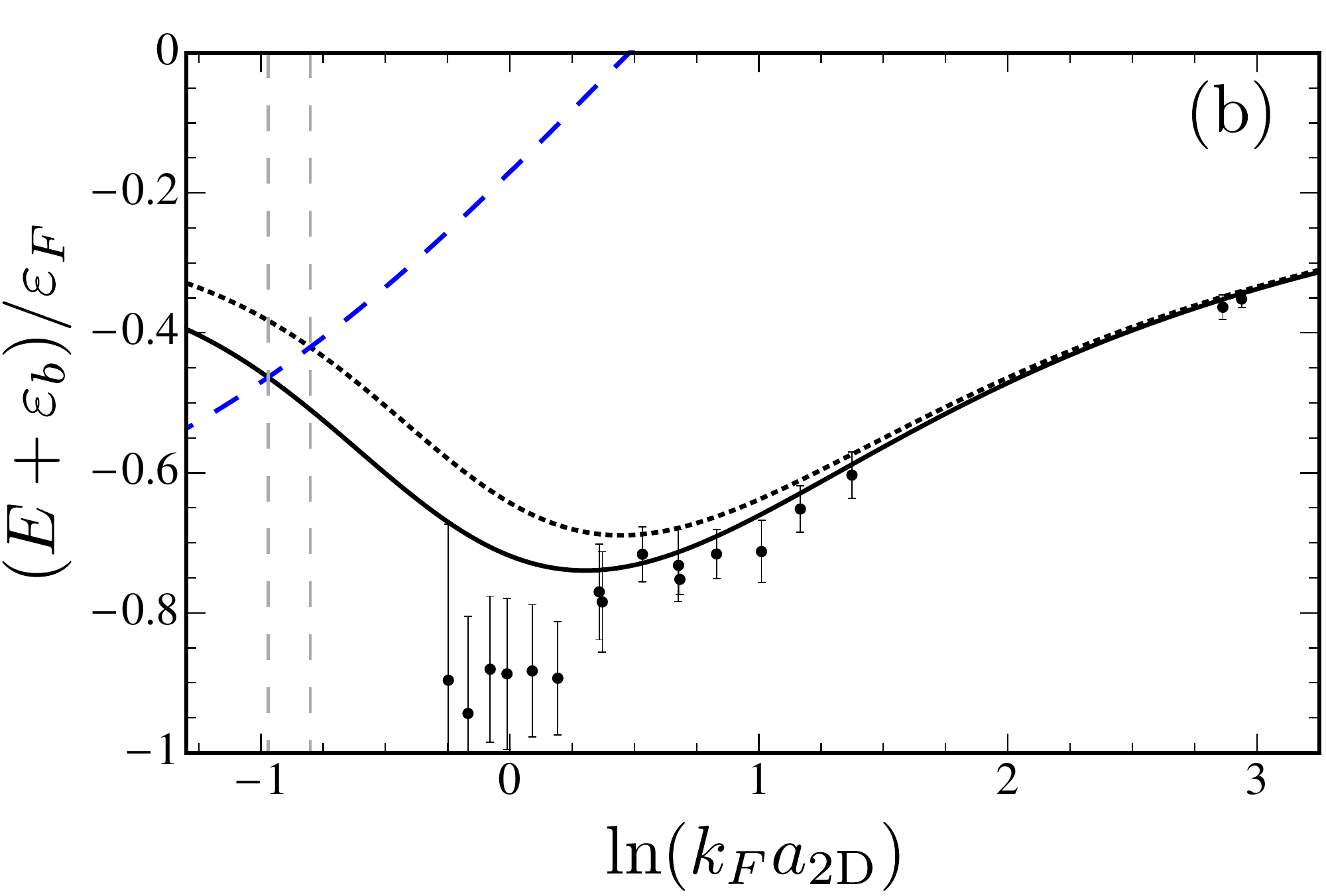}
\caption{ Energy of the impurity measured from the two-body binding
  energy. The dotted line corresponds to the energy of the monomeron
  within the Chevy ansatz\cite{Zollner2011}, Eq.~(\ref{eq:epol}),
  while the solid line is within the two particle-hole pair
  approximation\cite{Parish2013}. The dashed line is the energy of the
  dimeron within the one particle-hole pair
  approximation\cite{Parish2011}. (a) The data points are from the two
  recent Monte Carlo simulations: for the monomeron (dimeron) these
  are marked by red open squares (circles)\cite{Vlietinck2014} and by
  green diamonds (crosses)\cite{Kroiss2014}. Note that the variational
  approach provides an upper bound on the energy. (b) The ground state
  transition within the two approximations for the monomeron energy
  are illustrated by vertical dashed lines, while the experimental
  data\textsuperscript{\ref{foot:a2d}} is taken from
  Ref.~\refcite{Koschorreck2012}.}
\label{fig:2ph}
\end{figure}

The first work on this subject\cite{Zollner2011} did not find any
ground-state transition, the issue being that the authors did not
consider the monomeron and dimeron on an equal footing in terms of
particle-hole pair dressing of the variational wavefunctions. Later,
one of us\cite{Parish2011} included a particle-hole pair excitation in
the dimeron variational wavefunction to show that there is indeed a
ground state transition. We recently extended this analysis to argue
that, under a minimal set of assumptions, the critical interaction
strength for the monomeron-dimeron transition must lie in the
interval\cite{Parish2013}
\begin{align}
-0.97<\ln(\kf\ad)_{\rm{crit}}<-0.80.
\end{align}
The lower (upper) bound corresponds to comparing the dimeron dressed
by one particle-hole pair excitation with the monomeron dressed by two
(one) excitations. As seen in Fig.~\ref{fig:2ph}(a), our result agrees
with the critical interaction found in two recent diagrammatic Monte
Carlo studies: $\ln(\kf\ad)_{\rm crit}=-0.95(0.15)$
[Ref.~\refcite{Vlietinck2014}] and $\ln(\kf \ad)_{\rm crit} =
-1.1(0.2)$ [Ref.~\refcite{Kroiss2014}].
  
The monomeron was investigated in a recent
experiment\cite{Koschorreck2012}, and Fig.~\ref{fig:2ph}(b) shows that
for $\ln(\kf\ad)\geq0.3$ the comparison between theory and experiment
is excellent. For stronger attraction, the agreement becomes
progressively worse until at $\ln(\kf\ad) \simeq -0.6$ the measured
effective mass appears to diverge, which was taken to be a signature
of the monomeron-dimeron transition.\cite{Koschorreck2012} However, if
one extrapolates the measured residue to zero\cite{Koehl2012}, one
instead obtains a critical interaction strength of
$\ln(\kf\ad)_{\rm{crit}}=-0.88(0.20)$, which is in good agreement with
theory\cite{Parish2011,Parish2013}. As mentioned previously, the
experimental investigation of the polaron problem can be complicated
by temperature effects and trap averaging. In addition, one must
consider the fact that the high polarization limit typically
corresponds to a \emph{finite density} of spin-$\down$
impurities\cite{Lobo2006}.  Thus, we are faced with the question of
whether the single-impurity transitions are thermodynamically stable,
i.e., whether they are preempted by first-order transitions in the
thermodynamic limit. We have recently shown\cite{Parish2013} that a
first-order superfluid-normal phase transition preempts the
single-impurity transition at zero temperature, similarly to the
situation in 3D.\cite{Mathy2011} However, this result requires the
presence of a superfluid, and thus it is an open question whether
single-impurity transitions may exist at higher temperatures.

\begin{figure}
\centering
\includegraphics[width=1\linewidth]{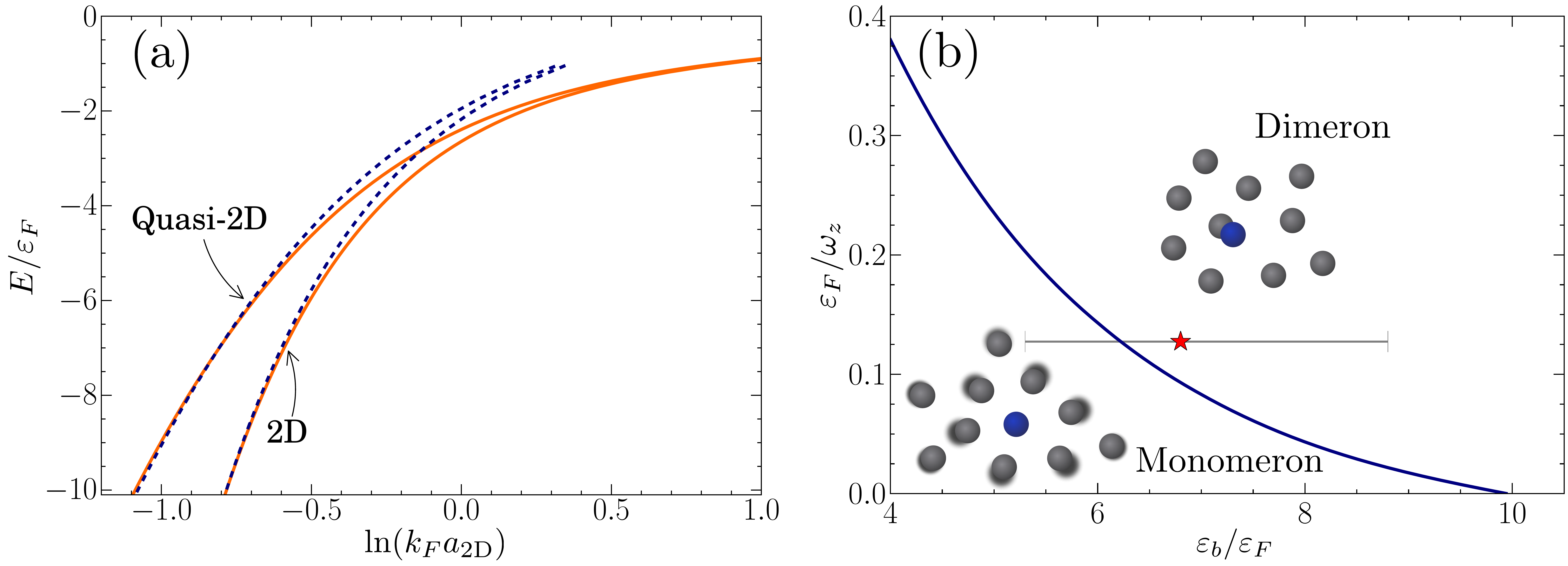}
\caption{The behavior of the quasi-2D polaron taken from
  Ref.~\refcite{Levinsen2012}.  (a) Monomeron (solid lines) and
  dimeron (dashed) energies in 2D and in quasi-2D at
  $\ef/\op=1/10$. (b) The single impurity phase diagram. The ground
  state of the impurity is a monomeron (dimeron) to the left (right)
  of the line. The theory lines are all within the one particle-hole
  pair approximation. The star shows the experimental transition point
  with error bars\cite{Koehl2012}.
  \label{fig:polgs}}
\end{figure}

In the present discussion, we have
mapped\textsuperscript{\ref{foot:a2d}} the results of the experiment
onto a pure 2D theory. However, let us now discuss the validity of
such an approach\cite{Levinsen2012}. The transverse confinement
applied in the experiment\cite{Koschorreck2012} was
$\op=2\pi\times78.5$kHz, while the Fermi energy of the majority
component was $2\pi\times10$kHz. This in turn means that the pure 2D
theory\cite{Parish2011,Parish2013} predicts the transition to occur
when $\eb\geq2\pi\times100$kHz, {\em i.e.} when the binding energy
exceeds the transverse confinement strength. In this regime, the
binding energy is strongly modified from the 2D prediction --- see
Fig.~\ref{fig:eb} and the discussion in Sec.~\ref{sec:2d}. On the
other hand, the ground state transition is governed by interactions
that take place at the typical energy scale $\sim\ef$. Since
$\ef\ll\op$ the low-energy quasi-2D theory described by
Eqs.~\eqref{eq:fcalapp} and \eqref{eq:ad} is still approximately
valid, explaining our choice of using a definition of $\ad$ which
derives from low-energy scattering rather than the binding energy.

The deviation from the pure 2D limit of the monomeron-dimeron
transition may be further investigated\cite{Levinsen2012} by including
harmonic oscillator levels in the variational wavefunction and using
the full quasi-2D Hamiltonian \eqref{eq-fullham}. The results of such
an analysis are shown in Fig.~\ref{fig:polgs}, where we see how the
transition point indeed changes very little in $\ln(\kf\ad)$ for
$\ef/\op \lesssim 1/10$, while on the other hand the change is rather large in
terms of the parameter $\eb/\ef$.

\begin{figure}
\centering
\includegraphics[width=.65\linewidth]{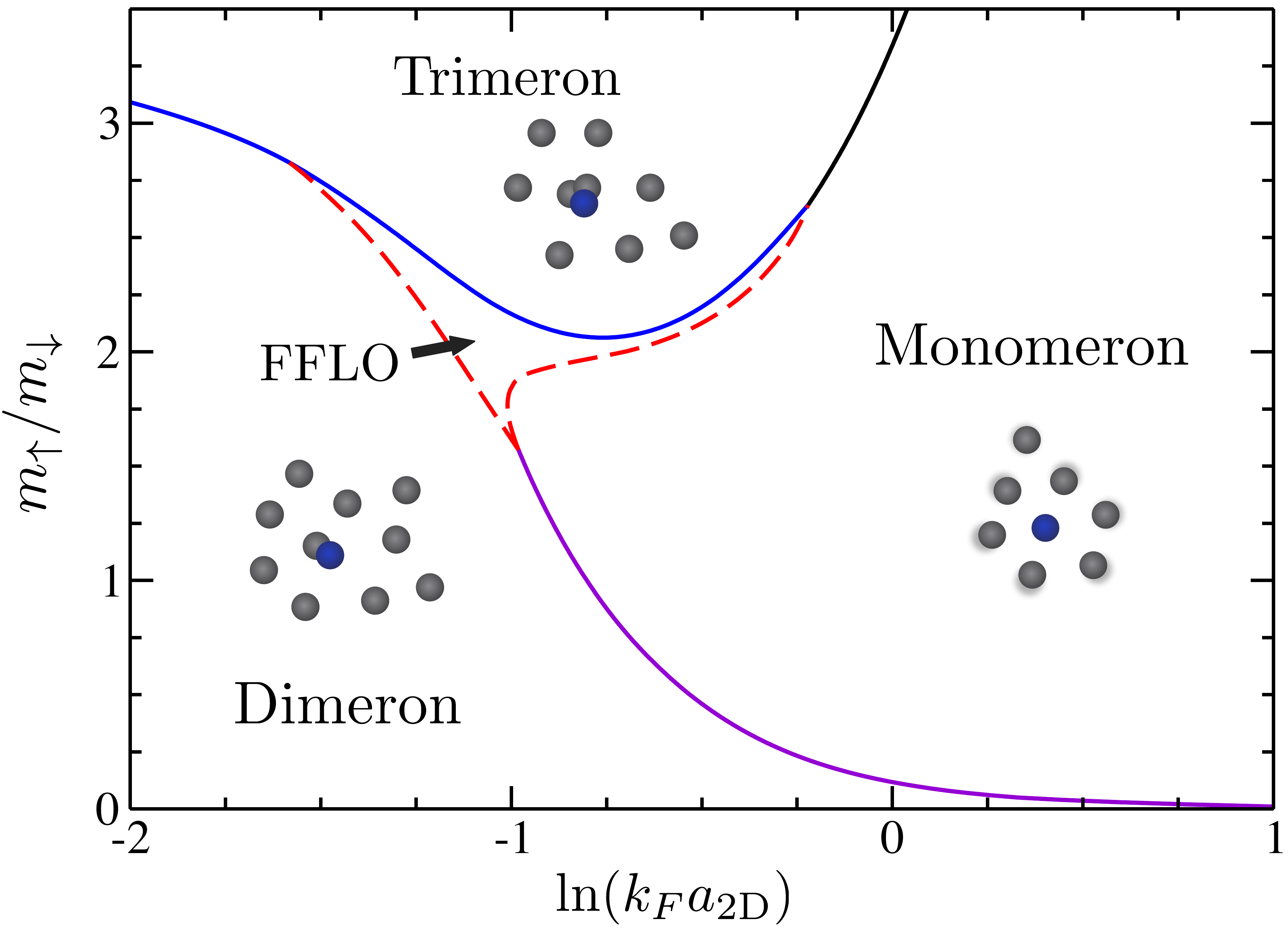}
\caption{Ground-state phase diagram for a single impurity atom of mass
  $m_\down$ immersed in a gas of fermions of mass $m_\up$, adapted
  from Ref.~\refcite{Parish2013}. The phase boundaries are derived
  within the one particle-hole pair dressing approximation. The
  single-impurity analog of the FFLO phase corresponds to a ground
  state dimeron at non-zero momentum.  }
\label{fig:2DFermi2}
\end{figure}

We finally turn to the mass-imbalanced system, where the single
impurity phase diagram\cite{Parish2013} as a function of interaction
strength takes the form displayed in Fig.~\ref{fig:2DFermi2}. In
Sec. \ref{sec:few} we discussed how in vacuum the $\up\up\down$
trimer appears\cite{Pricoupenko2010} when
$m_\up/m_\down=3.33$. Remarkably it is seen that the presence of a
Fermi sea favors trimer formation: within the approximation used, the
trimeron is predicted to be the ground state for mass ratios
$m_\up/m_\down\geq2.1$. This lower critical mass ratio may be
understood as a consequence of the kinetic energy cost involved in
forming a dimeron at rest: in the simplest approximation, the impurity
at momentum $+\k_F$ binds a majority atom at $-\k_F$ and, if the
impurity is sufficiently light, it may be energetically favorable to instead
form a dimeron at finite momentum or a trimeron. The same effect
was predicted in 3D\cite{Mathy2011}. While the trimeron is favored by
the Fermi sea, we found\cite{Parish2013} that the tetrameron appears
disfavored, i.e., the critical mass ratio for tetrameron formation
in the strongly interacting regime increases from its vacuum
value\cite{Levinsen2013} of $m_\up/m_\down=5.0$.

The possibility of a dimeron at finite momentum is of considerable
interest, since it is a single-particle analog of the FFLO phase ---
for a small but finite density of impurities this has been shown to
lead to a spatially-modulated
superfluid\cite{Parish2011b,goldbart2011}. We see in
Fig.~\ref{fig:2DFermi2} that the FFLO dimeron occupies a considerable
part of the phase diagram, making it possible that FFLO physics may be
observed in the strongly spin-imbalanced 2D Fermi gas.

%% file: dynamics.tex
\section{Dynamics}\label{sec:dynamics}
Dynamical properties provide a powerful probe into the nature of
interactions in strongly correlated quantum systems.  For instance, it
has been predicted that the harmonically trapped 2D quantum gas
features an $SO(2,1)$ dynamical scaling symmetry due to the
(classical) scale invariance of the uniform gas with contact
interactions. A consequence of this symmetry is the existence of an
undamped monopole breathing mode with frequency exactly twice that of
the trap\cite{Pitaevskii1996}. While true in the absence of
interactions, the scale invariance which exists at the classical level
is broken by the procedure of renormalization\cite{Pitaevskii1996},
the so-called {\em quantum anomaly}\cite{Olshanii2010}. Thus, the
shift of the breathing mode frequency probes the breaking of scale
invariance in the interacting 2D quantum gas. Another dynamical
phenomen is that of spin diffusion, the process that evens out
differences in spin polarization across the gas. Here the diffusivity
in the strongly interacting and degenerate regime is naturally of
order $\hbar/m$ and an interesting possibility is that there is a
universal lower bound set by quantum mechanics.

As of now, there have been experiments on the collective modes in a
harmonic trap\cite{Vogt2012} and on spin
transport\cite{Koschorreck2013}. The results of the experiments have
indicated several surprising features of the 2D Fermi gas: an undamped
breathing mode with a frequency compatible with no shift from the
classical (non-interacting) result; a quadrupole mode strongly damped
even in the weakly interacting regime; and a transverse spin
diffusivity three orders of magnitude smaller than in any other
system. The strong damping of the quadrupole mode may be
explained\cite{Baur2013}, at least in part\cite{Chiacchiera2013}, by
the anisotropy of the trapping potential used. However at first sight
the other two features appear contradictory, as the results of the
breathing mode experiment indicate that the effect of interactions is
much weaker than expected by theories, while the spin diffusivity
experiment indicates the opposite. Ultimately, further experiments as
well as possibly finite temperature QMC calculations will likely be
needed to shed light on the discrepancy.

In this section we assume a purely 2D geometry, such that the
transverse confinement frequency $\op$ drops out of the problem. The
experiments described here are indeed all in the regime where $T\leq
T_F\lesssim 0.1\op$, so this approximation is reasonable.

\subsection{Classical scale invariance, a hidden $SO(2,1)$ symmetry,
  and the breathing mode}

It has been predicted\cite{Pitaevskii1996} that a 2D quantum gas in a
harmonic transverse trapping potential features an undamped monopole
breathing mode with frequency exactly twice that of the trap. The
origin of this surprising result is the (classical) scale invariance
of the Hamiltonian with a short-range $\delta$-function interaction.
Define the 2$N$ dimensional vector
$\vect{X}=(\boldsymbol{\rho}_1,\cdots,\boldsymbol{\rho}_N)$, with
$\boldsymbol{\rho}_i$ the positions of the atoms $i=1,\cdots,N$, and
the hyperradius $X\equiv |\vect{X}|$. In real space, both terms in the
Hamiltonian
\begin{align}
  H_0=-\frac{\Delta^2_\vect{X}}{2m}+
  g\sum_{i<j}\delta(\boldsymbol{\rho}_i-\boldsymbol{\rho}_j)
\end{align}
scale as $\lambda^{-2}$ under the scale transformation $\vect{X}\to
\lambda\vect{X}$, and consequently the Hamiltonian is scale
invariant. While this is true classically, the procedure of
renormalization of the quantum theory introduces a scale, the 2D
scattering length $\ad$, as discussed in Section \ref{sec:2d}; the
absence of a scale in the classical theory and the introduction of one
through renormalization is known as a quantum anomaly. The scale
invariance is still approximately valid in the limits $\ad\to0$ and
$\ad\to\infty$ where the following results apply.\footnote{Whereas the
  scale invariance in 2D is only exact in the trivial non-interacting
  limits, it is, in fact, quantum mechanically exact for the 3D
  unitary Fermi gas \cite{Werner2006} as well as for the 1D gas in the
  Tonks limit, both strongly interacting systems.}

The presence of a harmonic trapping potential
\begin{align}
  H_{\rm{trap}}=\frac12 m\omega_0  X^2
\end{align}
in the 2D plane obviously breaks the scale invariance as it scales as
$\lambda^2$ under $\vect{X}\to\lambda\vect{X}$. However, it leads to a
very interesting algebra\cite{Pitaevskii1996}: using the usual
commutation relations for $\vect{X}$ and $\vect P\equiv
i\del_\vect{X}$, one may easily show
$\left[H_{\rm{trap}},H\right]=i\omega_0 ^2Q$. Here $Q\equiv
\frac12(\vect{P}\cdot \vect{X}+\vect{X}\cdot \vect{P})$,
$e^{-\ln(\lambda)Q}$ is the generator of scale
transformations\cite{Nishida2007}, and $H=H_0+ H_{\rm{trap}}$ is the
total Hamiltonian. Then defining the operators
\begin{align}
L_1=\frac1{2\omega_0 }(H_0-H_{\rm{trap}}), \hspace{5mm}
L_2=Q/2, \hspace{5mm}
L_3=\frac1{2\omega_0 }(H_0+H_{\rm{trap}}),
\end{align}
these satisfy
\begin{align}
\left[L_1,L_2\right]=-iL_3, \hspace{5mm}
\left[L_2,L_3\right]=iL_1, \hspace{5mm}
\left[L_3,L_1\right]=iL_2,
\end{align}
which is the algebra of the Lorentz group in 2D, $SO(2,1)$. As usual,
one may then define raising and lowering operators
$L_\pm=\frac1{\sqrt2}(L_1\pm iL_2)$. From the commutation relations
$\left[H,L_\pm\right]=\pm2\omega_0 L_\pm$ it follows that if
$|\Psi_g\rangle$ is the ground state with energy $E_g$, the state
$L_+|\Psi_g\rangle$ has energy $E_g+2\omega_0 $ while
$L_-|\Psi_g\rangle=0$. Thus the repeated action of $L_+$ generates a
tower of states, separated by $2\omega_0 $ and these may be identified
with the breathing modes of the system. For instance, if the system is
initially in a stationary state with a constant trap frequency
$\omega_0 $ at time $t<0$, the trap frequency is slightly perturbed
during the interval $0<t<t_f$, and returns to its initial value at
time $t>t_f$, one finds\cite{Werner2006} that the final state scale
oscillates around unity with frequency $2\omega_0 $. That is, the
lowest breathing mode has been excited.

\begin{figure}[h]
\centering
\includegraphics[width=0.7\linewidth]{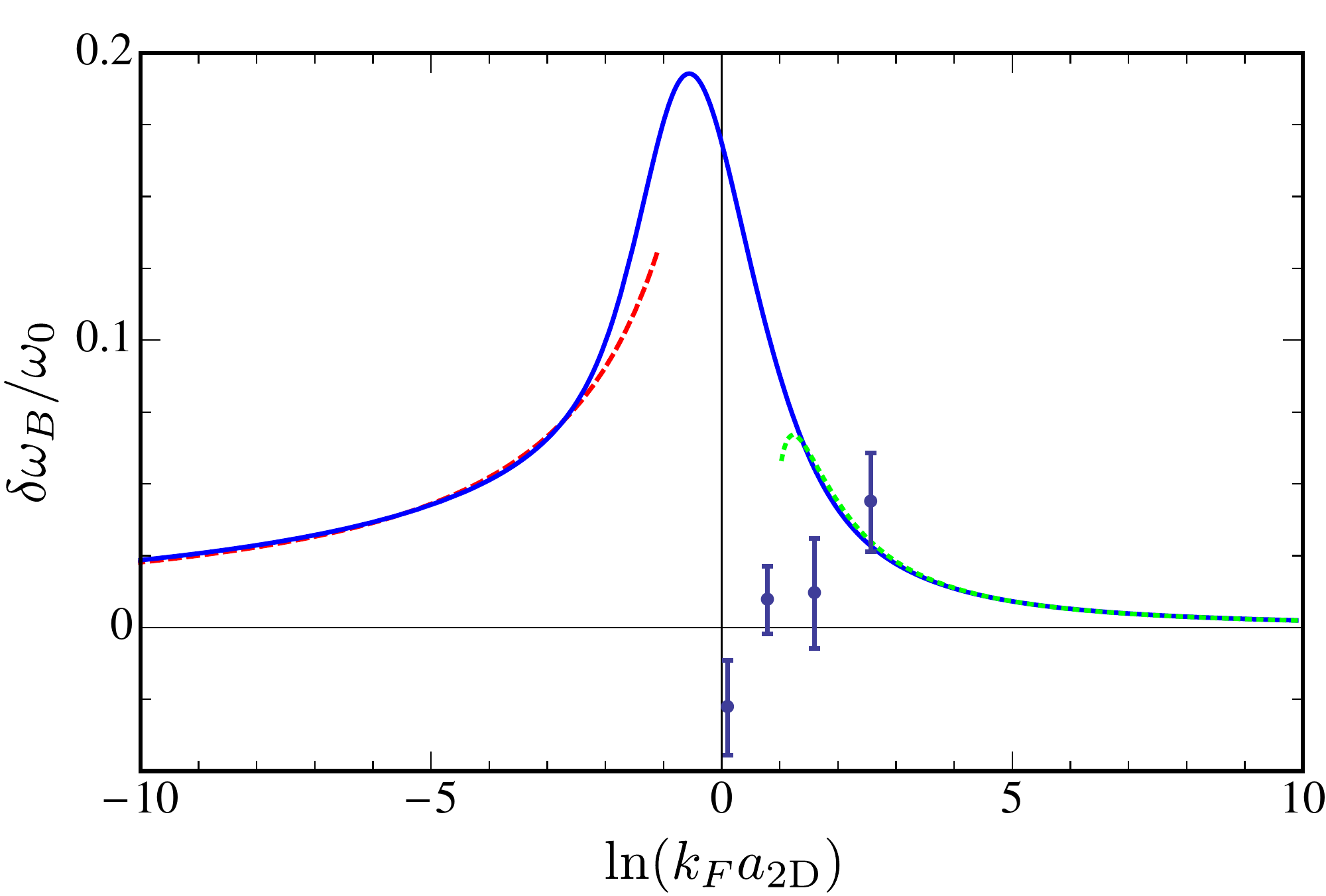}
\caption{Shift of the breathing mode as a function of interaction
  parameter. The solid line (full theory), dashed (BEC limit), and dotted
  (BCS limit) are the theory curves from
  Ref.~\refcite{Hofmann2012} while the data points are from the
  experiment\cite{Vogt2012}. 
  \label{fig:breathe}}
\end{figure}

In the above scale invariant (and non-interacting) regimes, the
breathing mode is undamped and its frequency is independent of
amplitude. On the other hand, in the interacting quantum system, the
breathing mode is shifted to $\omega_B= 2\omega_0+\delta\omega_B$ from
its non-interacting value, as discussed in the Bose case in
Ref.~\refcite{Olshanii2010}. In the 2D Fermi gas,\footnote{In fact,
  within mean-field theory, the breathing mode has frequency
  $2\omega_0$ in the entire BCS-BEC crossover\cite{Taylor2012}.} this
shift has been modelled\cite{Hofmann2012} (see also
Refs.~\refcite{Taylor2012,Gao2012,Moroz2012}) by assuming a
hydrodynamic description of the strongly interacting regime, and a
polytrope $P\sim n^{\gamma+1}$ for the dependence of pressure on
density. These assumptions allow for a solution of the linearized
hydrodynamic equations, and in turn for the breathing mode frequency,
$\omega_B=\omega_0\sqrt{2+2\gamma}$. $\gamma$ itself was obtained by
comparing with the zero-temperature QMC data\cite{Bertaina2011}
discussed in Section \ref{sec:bcsbec}. The resulting frequency shift
is shown in Fig.~\ref{fig:breathe}, and is seen to be of order $10\%$
in the regime of strong interactions, $\ln(\kf\ad)\sim0$.

Experimentally, the breathing mode was investigated \cite{Vogt2012}
using a procedure essentially as described above: The two-dimensional
confinement was adiabatically lowered from the initial configuration
and then abrubtly returned to its original configuration. After a
variable wait time, the confinement was switched off and the density
distribution was revealed by an absorption image after time of
flight. The experiment investigated a large range of interaction
strengths, $0\lesssim\ln\kf \ad\lesssim500$. Surprisingly, the results
of the experiment were consistent with the scale invariant assumption
above, i.e., no significant frequency shift was observed, even in the
regime of strong interactions (see Fig.~\ref{fig:breathe}).  The
results beg the question whether the zero-temperature equation of
state\cite{Bertaina2011} is appropriate for the comparison with the
experiment at $T/T_F=0.4$, i.e., whether the apparent scale invariance
arises due to finite temperature effects\cite{Hofmann2012}. Indeed, in
the high temperature limit the shift of the breathing mode may be
analyzed by combining the virial expansion of the equation of state
with a variational method in the hydrodynamic
regime\cite{Chafin2013}. The results of this analysis are consistent
with the experiment and the theoretical curve in
Fig.~\ref{fig:breathe} in the regime of validity of the approach,
$\ln(\kf\ad)\gtrsim1.75$.

The damping of the breathing mode is related to the bulk viscosity. In
particular both the bulk viscosity and the damping are expected to
vanish in the normal phase in the regime where the $SO(2,1)$ symmetry
is exact, as first pointed out in the context of the unitary Fermi
gas\cite{Son2007}. Using a sum rule, the bulk viscosity has been
argued\cite{Taylor2012} to vanish in the weakly interacting limits
$|\ln(\kf\ad)|\gg1$. However, in the intermediate strongly interacting
regime one expects a non-vanishing bulk viscosity and related damping
of the breathing mode. Therefore it is surprising that the experiment
\cite{Vogt2012} measures a damping consistent with vanishing bulk
viscosity across the entire interaction range.

\subsection{Quadrupole mode}

\begin{figure}[h]
\centering
\includegraphics[width=\linewidth]{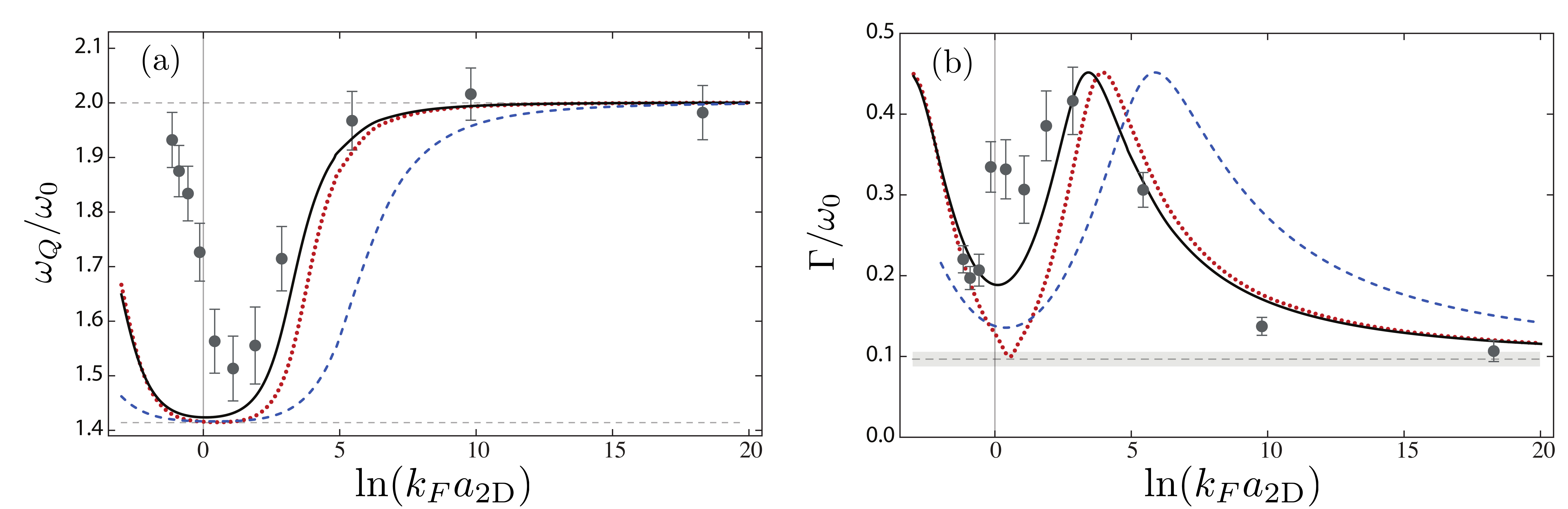}
\caption{(a) Quadrupole frequency and (b) damping as a function of
  interaction parameter. The lines are the theoretical
  curves\cite{Baur2013} in the Boltzmann limit (blue, dashed), with
  Pauli blocking only (black, solid), and with additional medium
  effects (red, dotted). The black dots are the results of the
  experiment\cite{Vogt2012}. The trap frequency here is
  $\omega_0=\sqrt{\omega_x\omega_y}$. In (b) a constant shift has been
  applied to the theory to account for systematic effects.
  \newline\hspace{\textwidth} \tiny{Figure adapted with permission
    from Ref.~\refcite{Baur2013}. Copyrighted by the American Physical
    Society.}
  \label{fig:quad}}
\end{figure}

In addition to the breathing mode, the experiment\cite{Vogt2012}
considered the quadrupole mode, corresponding to an excitation with
velocity field ${\bf x}-{\bf y}$ oscillating with frequency
$\omega_Q$. The excitation procedure was similar to the monopole mode
described above: the radial trap was adiabatically made elliptical,
followed by an abrubt return to the original configuration, a short
free oscillation, and an absorption image after time of flight. The
results of the experiment are shown in Fig.~\ref{fig:quad}(a). Two
regimes are immediately identifiable: the collisionless regime, where
$\ln(\kf\ad)\gg1$ and $\omega_Q\approx 2\omega_0$, and the
hydrodynamic regime, where $\omega_Q\approx\sqrt2\omega_0$. The theory
curves\cite{Baur2013} (see also Ref.~\refcite{Wu2012}) are calculated
using kinetic theory and correctly identify the onset of the
hydrodynamic regime. The theory is not expected to be valid when
$\ln(\kf\ad)\lesssim0.5$, where the (zero-temperature) chemical
potential is negative\cite{Ngamp2013-2} (see the discussion in
Sec.~\ref{sub:eos}) and pairing becomes significant.

The damping of the quadrupole mode is shown in Fig.~\ref{fig:quad}(b):
it is seen that this is a sizeable fraction of the trap frequency, and
curiously this is the case even in the weakly interacting regime
$\ln(\kf\ad)\gg1$ --- in fact, the experiment\cite{Vogt2012} showed
that the large damping persists up to very large $\ln(\kf\ad)\sim500$,
far into the collisionless regime where the damping is expected to
vanish, and kinetic theory is valid. Ref.~\refcite{Baur2013} argued
that the large damping in the weakly interacting regime arises mainly
from systematic effects, which generate an approximately constant
damping across all interaction strengths (indeed a smaller constant
damping was also observed in the breathing mode experiment described
above). The anisotropy of the trap before time of flight may then
account for the remaining discrepancy in this
limit\cite{Baur2013}. However, a recent analysis using a realistic
trapping potential and including even the effect of gravity concluded
that the damping of the quadrupole mode in the weakly interacting
regime could not be explained by the specific geometry of the trap
\cite{Chiacchiera2013}. In the strongly interacting hydrodynamic
regime, the damping may be related to the shear viscosity of the
gas\cite{Bruun2012,Schafer2012,Enss2012}.

\subsection{Spin diffusion}

An interesting application of fermionic quantum gases is the study of
spin transport. These systems may provide particularly clean
experimental realizations compared with, e.g., $^3$He-$^4$He
solutions, since in the quantum gases the interactions are tunable and
spin states may be manipulated in a coherent manner by radio-frequency
pulses. One basic transport process is that of spin diffusion,
recently investigated in the context of ultracold atomic
gases\cite{Sommer2011,Koschorreck2013}. This process acts to even out
differences in polarization. Writing the magnetization as a product of
the magnitude and the direction, $\bm{\mathcal M}={\mathcal
  M}\hat{\bold e}$, there are two contributions to the magnetization
gradient $\nabla \bm{\mathcal M}=(\nabla {\mathcal M})\hat{\bold
  e}+{\mathcal M}\nabla\hat{\bold e}$. The first of these,
longitudinal diffusion, acts between regions of different magnitude of
magnetization, while the second, transverse diffusion, acts between
regions of different orientation.

In the light of the proposed quantum limit of the ratio of shear
viscosity to entropy density\cite{Policastro2001}, it is interesting
to ask the question whether quantum mechanics provides a lower bound
for other transport phenomena such as spin diffusion in a strongly
interacting Fermi gas. As decoherence is introduced by collisions, the
resulting spin diffusivity may be expected to go as the collision
speed of two atoms multiplied by the mean free path. In the degenerate
regime, the former may be taken to be $\hbar \kf/m$, while the mean
free path is $1/n\sigma\sim 1/\kf$; the density $n\sim \kf^2$ and the
cross section in the degenerate regime takes its strongest value
allowed by quantum mechanics, {\em i.e.}  $\kf^{-1}$.  Thus the
diffusivity may be expected to be of order $\hbar/m$ in the degenerate
regime and indeed the lowest spin diffusivity for longitudinal spin
currents has been measured to be $6.3\hbar/m$ in a 3D quantum
degenerate Fermi gas at unitarity\cite{Sommer2011}. In general this
argument is too simple; for instance it neglects the effect of Pauli
blocking which causes the longitudinal spin diffusivity in the Fermi
liquid to diverge as $1/T^2$ at low temperature.  Note that for the
purpose of this discussion we have displayed $\hbar$ explicitly.

\begin{figure}[h]
\centering
\includegraphics[width=0.7\linewidth]{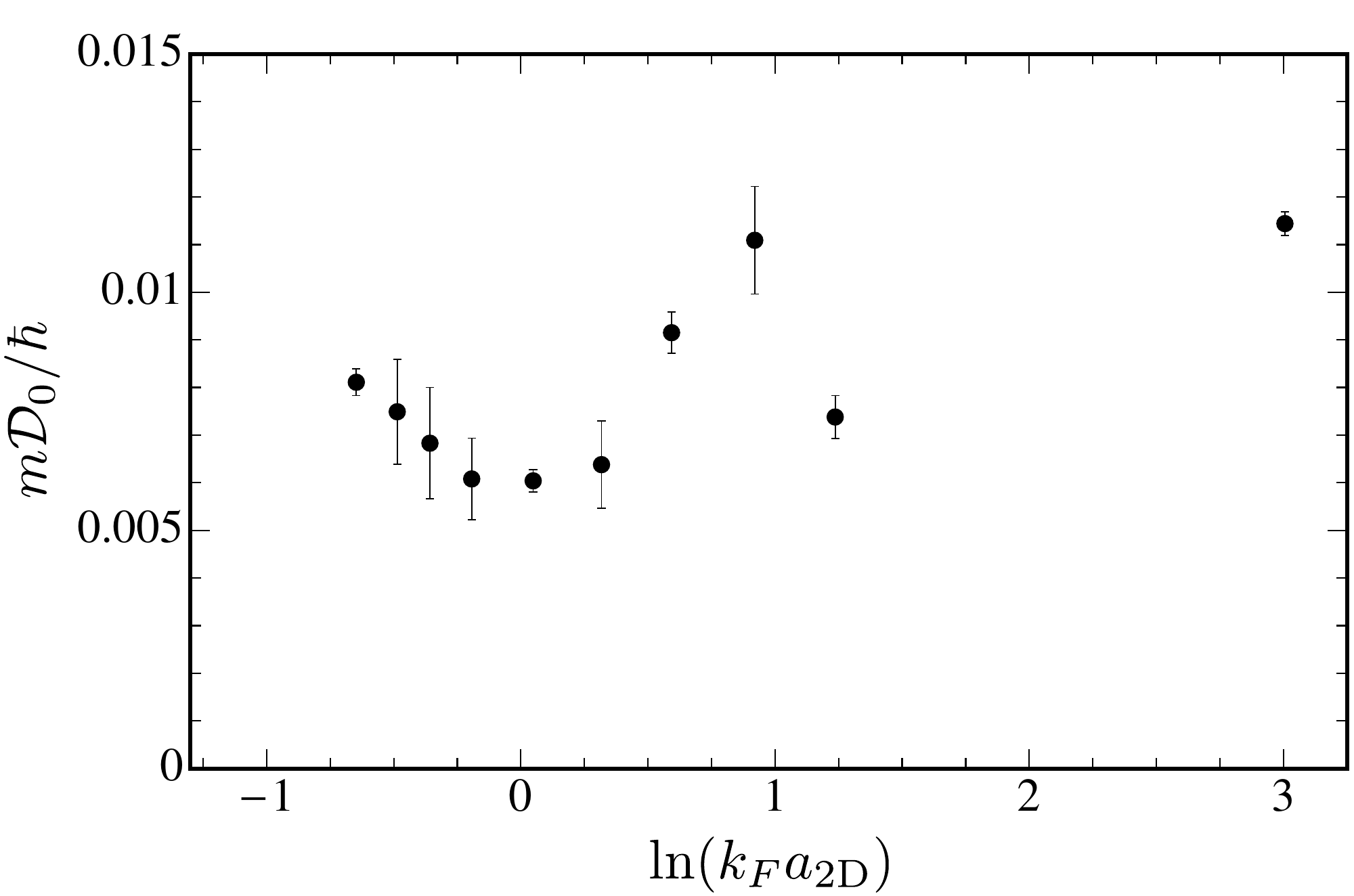}
\caption{Transverse spin diffusivity measured in
  experiment\cite{Koschorreck2013} across the strongly interacting
  regime. \label{fig:spin}}
\end{figure}

Surprisingly, a recent experiment~\cite{Koschorreck2013} has found a
transverse spin diffusivity in the strongly interacting regime that is
orders of magnitude smaller than $\hbar/m$. Starting from a fully
polarized 2D gas of $^{40}$K atoms, the experiment used a spin-echo
technique consisting of three consecutive radio-frequency pulses:
First, a $\pi/2$ pulse was applied to rotate the spin into a coherent
superposition of $\up$ and $\down$ states. A magnetic field gradient
ensured a transverse spin wave due to the difference in gyromagnetic
ratio of the two spin states, thus lifting the spin polarization and
allowing the different spin states to collide and diffuse. Trivial
dephasing due to the magnetic field gradient was reversed by the
application of a $\pi$ pulse after a time $\tau$. This ensured that
the spin state would refocus at time $2\tau$ in the absence of
decoherence, in which case the final $\pi/2$ pulse would rotate the
spin state back to the original one.  The experimental observable was
the final magnetization $\langle M\rangle\equiv
(N_\up-N_\down)/(N_\up+N_\down)$, and by measuring this for different
spin evolution times $2\tau$, the transverse spin diffusivity was
extracted.\footnote{${\cal D}_0$ was obtained from the magnetization
  using the time evolution $\langle M_z\rangle\propto e^{-(2/3){\cal
      D}_0 (\delta\gamma B')^2\tau^3}$ with $\delta\gamma$ the
  difference in gyromagnetic ratio between the two spin states and
  $B'=\del B_z/\del x$ the magnetic field gradient.} The results are
shown in Fig.~\ref{fig:spin}, and it is seen that ${\cal D}_0$ has a
shallow minimum around $\ln(\kf\ad)=0$, with values as low as
$0.006\hbar/m$.

\begin{figure}
\centering
\includegraphics[width=0.7\linewidth]{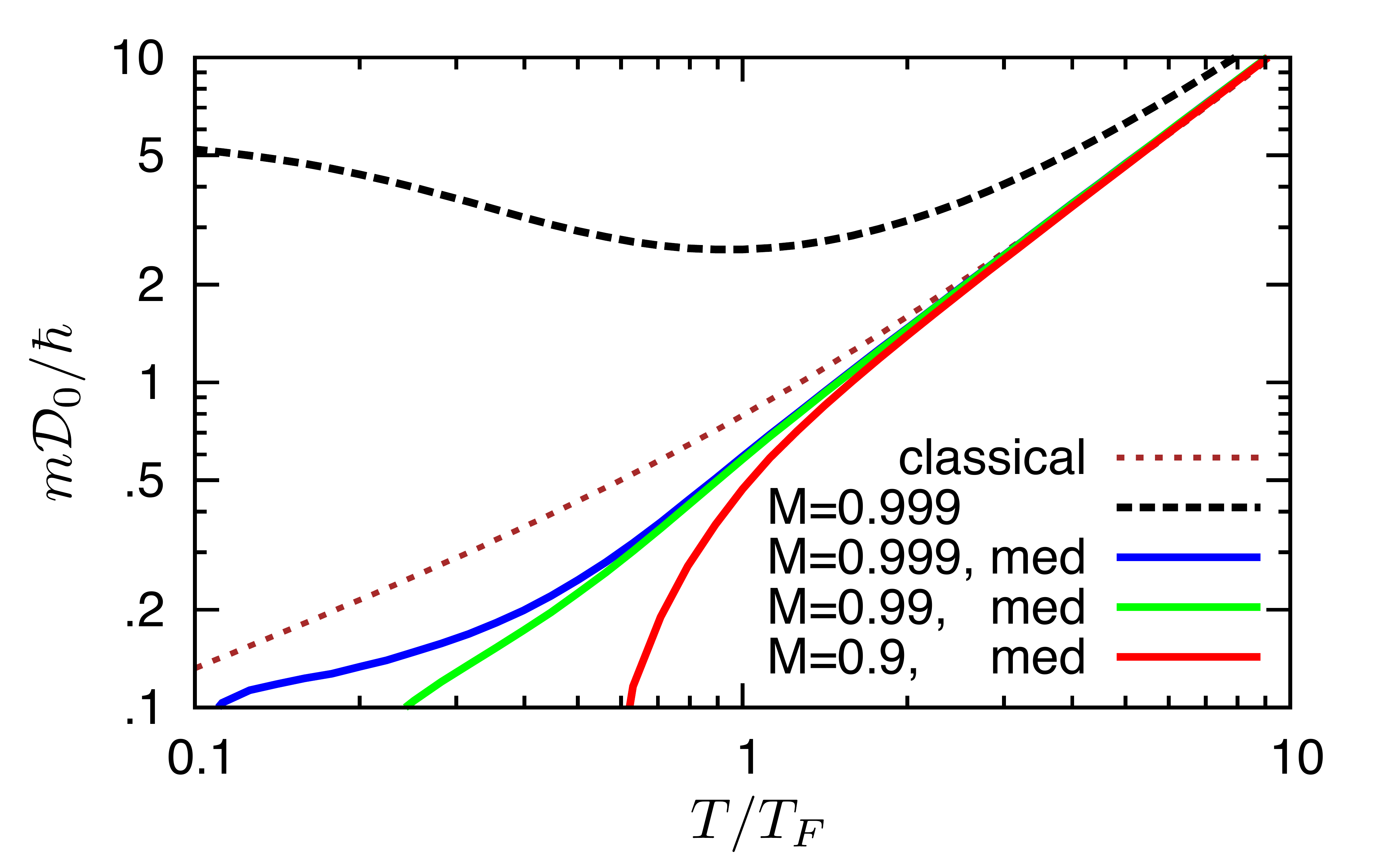}
\caption{${\cal D}_0$ as a function of temperature at $\ln(\kf\ad)=0$
  for various magnetizations\cite{Enss2013}. The dashed (black) line
  includes Pauli blocking only, while the dotted line is the
  high-temperature Boltzmann limit.  For comparison, the
  experiment\cite{Koschorreck2013} was carried out at $T/T_F=0.24(3)$.
  \newline\hspace{\textwidth} \tiny{Reprinted figure with permission
    from: T. Enss, \textit{Phys.\ Rev.\ A} \textbf{88}, 033630
    (2013). Copyright 2013 by the American Physical Society.}
\label{fig:enss}}
\end{figure}

The transverse spin diffusivity in the 2D Fermi gas was recently
investigated using a kinetic theory based on a many-body $T$
matrix\cite{Enss2013}. Indeed, it was found that medium effects could
substantially suppress the spin diffusion below $\hbar/m$, see
Fig.~\ref{fig:enss}. As shown in the figure, the theory also predicts
that at temperatures below $T_F$, the transverse spin diffusivity is
quite sensitive to magnetization and gets suppressed as the
magnetization decreases. The origin of the suppression lies in the
enhanced cross section in the many-body system close to the Thouless
pole and, as discussed in Sec.~\ref{sec:bcsbec}, the Thouless pole
overestimates the critical temperature. However, it is likely that the
theory captures the correct qualitative behavior; thus this feature
may have implications for the interpretation of the experiment, which
assumes a constant transverse spin diffusivity over the timescale
$2\tau$.

%% file: conclusion.tex
\section{Discussion and outlook}\label{sec:outlook}

Low-dimensional Fermi gases are expected to feature stronger
correlations and larger quantum fluctuations than their 3D
counterparts.  Yet, some of the first experiments on the 2D Fermi gas
appear to have observed the opposite.  The monopole breathing mode
apparently displays no shift from the predicted value in the absence
of interactions, indicating a scale invariant system. \cite{Vogt2012}
Likewise, the energy of the repulsive branch of the polarized Fermi
gas was found to be much smaller than that predicted theoretically
\cite{Koschorreck2012}.  This could imply one of two things: either
our expectation of strong correlations in the 2D Fermi gas is
incorrect, or there are additional factors present in 2D experiments
that need to be taken into account. For instance, the apparent scale
invariance may be influenced by finite temperature, the quasi-2D
nature of the gas, or even the in-plane trapping potential and trap
averaging.  Thus, a detailed theoretical understanding of these
effects is important.

Indeed, a major challenge currently facing experiments on 2D Fermi
gases is to achieve ultracold temperatures under strong
confinement. As such, superfluidity in the 2D Fermi gas has not yet
been realized experimentally.  Given that the BKT transition has
already been observed in 2D Bose gases\cite{hadzibabic2006}, it is
likely that superfluidity in the Bose regime of the crossover in Fermi
gases will soon be realized.  It may prove more difficult to observe
the superfluid phase in the BCS regime because of the reduced $T_c$ in
this limit.  However, as one of us has recently
argued\cite{fischer2014}, the quasi-2D nature of the gas could turn
out to be advantageous here, since mean-field theory predicts that
$T_c/\ef$ is increased as the confinement is relaxed at fixed
$\ef/\eb$ and the Fermi system is tuned away from 2D.  This raises the
tantalizing possibility of $T_c$ being maximal in the regime
intermediate between 2D and 3D.
 
Thus far, cold-atom experiments have only just begun to explore the
behavior of fermions in 2D.  Even above $T_c$, a pseudogap regime has
not yet been conclusively observed: while a gap in the spectra has
been nicely demonstrated,\cite{Feld2011} it seems likely that this is
due to two-body effects only, and any apparent reduction of the gap at
finite temperature is due to thermal broadening\cite{Ngamp2013-2}.
Thus, the interaction strength {\em vs} temperature phase diagram
requires further investigation.  In the future, we expect an
increasing array of tuning ``knobs'' to be added to the exploration of
2D Fermi gases.  There is the prospect of varying the spin imbalance
and achieving superfluid-normal transitions at zero temperature.
Moreover, heteronuclear Fermi-Fermi mixtures promise a fascinating new
playground, where novel bound states become possible as the mass ratio
is increased.  Ultimately, one would like to fully uncover the
fundamental differences between 2D and other dimensions.